\documentclass[twocolumn]{aastex61hack}
\usepackage{natbib}
\newcommand{\HI}{\ensuremath{\mbox{\ion{H}{1}}}}
\newcommand{\HII}{\ensuremath{\mbox{\ion{H}{2}}}}
\newcommand{\HeI}{\ensuremath{\mbox{\ion{He}{1}}}}
\newcommand{\MgII}{\ensuremath{\mbox{\ion{Mg}{2}}}}
\newcommand{\CIV}{\ensuremath{\mbox{\ion{C}{4}}}}
\newcommand{\CaII}{\ensuremath{\mbox{\ion{Ca}{2}}}}

\newcommand{\OII}{\ensuremath{[\mbox{\ion{O}{2}}]}}
\newcommand{\OIII}{\ensuremath{[\mbox{\ion{O}{3}}]}}
\newcommand{\OVI}{\ensuremath{\mbox{\ion{O}{6}}}}

\newcommand{\OIId}{\ensuremath{[\mbox{\ion{O}{2}}]} $\lambda\lambda$3726,3728}
\newcommand{\OIIId}{\ensuremath{[\mbox{\ion{O}{3}}]} $\lambda\lambda$4958,5006}
\newcommand{\HeIline}{\ensuremath{\mbox{\ion{He}{1}}} $\lambda$3187}

\newcommand{\NHI}{\ensuremath{N(\mbox{\ion{H}{1}})}\relax}
\newcommand{\logNHI}{\ensuremath{\log N(\mbox{\ion{H}{1}})}\relax}

\newcommand{\Rvir}{\ensuremath{R_{\rm vir}}}
\newcommand{\vesc}{\ensuremath{v_{\rm esc}}}
\newcommand{\logMstar}{\ensuremath{\log M_\star}}
\newcommand{\logMhalo}{\ensuremath{\log M_{\rm h}}}
\newcommand{\xh}{\ensuremath{[{\rm X/H}]}\relax}
\newcommand{\z}{$z$}

\providecommand{\zabs}{\ensuremath{z_{\rm abs}}}

\newcommand{\msun}{\ensuremath{{\rm M}_\odot}}

\newcommand{\column}{cm$^{-2}$}

\newcommand{\kms}{km\,s$^{-1}$}
\newcommand{\hub}{km\,s$^{-1}$\,Mpc$^{-1}$}

\newcommand{\hst}{{\em HST}}

\newcommand{\abssample}{36}
\newcommand{\qsosample}{23}
\newcommand{\ifuabssample}{19}
\newcommand{\ifuqsosample}{11}
\newcommand{\ifucutgalsample}{14} 
\newcommand{\ifugalsample}{23} 
 
\newcommand{\ifunonsamplet}{seven} 
\newcommand{\hostsperabs}{one} 

\defcitealias{lehner2013}{{\rm L13}}
\defcitealias{hm1996}{{\rm HM05}}
\defcitealias{hm2012}{{\rm HM12}}
\defcitealias{tpw2017}{Tumlinson, Peeples, \& Werk 2017}

\shorttitle{BASIC I: A Dual Population of Low-metallicity Absorbers} 
\shortauthors{Berg et al.}

\begin{document}

\title{The Bimodal Absorption System Imaging Campaign (BASIC) I. A Dual Population of Low-metallicity Absorbers at $\z\ <1$}

\correspondingauthor{Michelle A. Berg}
\email{michelle.berg@austin.utexas.edu}

\author[0000-0002-8518-6638]{Michelle A. Berg}
\affiliation{Department of Physics and Astronomy, University of Notre Dame, Notre Dame, IN 46556, USA}
\affiliation{Department of Astronomy, The University of Texas at Austin, Austin, TX 78712, USA}

\author[0000-0001-9158-0829]{Nicolas Lehner}
\affiliation{Department of Physics and Astronomy, University of Notre Dame, Notre Dame, IN 46556, USA}

\author[0000-0002-2591-3792]{J. Christopher Howk}
\affiliation{Department of Physics and Astronomy, University of Notre Dame, Notre Dame, IN 46556, USA}

\author[0000-0002-7893-1054]{John M. O'Meara}
\affiliation{W. M. Keck Observatory 65-1120 Mamalahoa Hwy. Kamuela, HI 96743, USA}

\author[0000-0002-0668-5560]{Joop Schaye}
\affiliation{Leiden Observatory, Leiden University, P.O. Box 9513, 2300 RA, Leiden, The Netherlands}

\author[0000-0001-5892-6760]{Lorrie A. Straka}
\affiliation{Leiden Observatory, Leiden University, P.O. Box 9513, 2300 RA, Leiden, The Netherlands}

\author[0000-0001-5810-5225]{Kathy L.~Cooksey}
\affiliation{Department of Physics \& Astronomy, University of Hawai`i at Hilo, Hilo, HI 96720, USA}

\author[0000-0002-1218-640X]{Todd M. Tripp}
\affiliation{Department of Astronomy, University of Massachusetts, 710 North Pleasant Street, Amherst, MA 01003-9305, USA}

\author[0000-0002-7738-6875]{J. Xavier Prochaska}
\affiliation{Department of Astronomy and Astrophysics, University of California, Santa Cruz, CA 95064, USA}
\affiliation{Kavli Institute for the Physics and Mathematics of the Universe (WIP), 5-1-5 Kashiwanoha, Kashiwa, 277-8583, Japan}

\author[0000-0003-4754-6863]{Benjamin D. Oppenheimer}
\affiliation{CASA, Department of Astrophysical and Planetary Sciences, University of Colorado, 389 UCB, Boulder, CO 80309, USA}

\author[0000-0001-9487-8583]{Sean D. Johnson}
\affiliation{Department of Astronomy, University of Michigan, Ann Arbor, MI 48109, USA}

\author[0000-0003-3938-8762]{Sowgat Muzahid}
\affiliation{IUCAA, Post Bag 04, Ganeshkhind, Pune-411007, India}
\affiliation{Leibniz-Institute for Astrophysics Potsdam (AIP), An der Sternwarte 16, D-14482 Potsdam, Germany}

\author[0000-0002-3120-7173]{Rongmon Bordoloi}
\affiliation{Department of Physics, North Carolina State University, 2401 Stinson Drive, Raleigh, NC 27695, USA}

\author[0000-0002-0355-0134]{Jessica K. Werk}
\affiliation{Department of Astronomy, University of Washington, Box 351580, Seattle, WA 98195, USA}

\author[0000-0003-0724-4115]{Andrew J. Fox}
\affiliation{AURA for ESA, Space Telescope Science Institute, 3700 San Martin Drive, Baltimore, MD 21218, USA}

\author[0000-0002-3097-5381]{Neal Katz}
\affiliation{Department of Astronomy, University of Massachusetts, 710 North Pleasant Street, Amherst, MA 01003-9305, USA}

\author[0000-0001-5020-9994]{Martin Wendt}
\affiliation{Institut f$\ddot{u}$r Physik und Astronomie, Universit$\ddot{a}$t Potsdam, Karl-Liebknecht-Str 24/25, D-14476 Golm, Germany}
\affiliation{Leibniz-Institute for Astrophysics Potsdam (AIP), An der Sternwarte 16, D-14482 Potsdam, Germany}

\author[0000-0003-1455-8788]{Molly S. Peeples}
\affiliation{Space Telescope Science Institute, Baltimore, MD 21218, USA}
\affiliation{Department of Physics and Astronomy, Johns Hopkins University, Baltimore, MD 21218, USA}

\author[0000-0003-3381-9795]{Joseph Ribaudo}
\affiliation{Department of Engineering and Physics, Providence College, Providence, RI, 02918, USA}

\author[0000-0002-7982-412X]{Jason Tumlinson}
\affiliation{Space Telescope Science Institute, Baltimore, MD 21218, USA}
\affiliation{Department of Physics and Astronomy, Johns Hopkins University, Baltimore, MD 21218, USA}


\begin{abstract} 
The bimodal absorption system imaging campaign (BASIC) aims to characterize the galaxy environments of a sample of \abssample\ \HI-selected partial Lyman limit systems (pLLSs) and Lyman limit systems (LLSs) in \qsosample\ QSO fields at $z\la 1$. These pLLSs/LLSs provide a unique sample of absorbers with unbiased and well-constrained metallicities, allowing us to explore the origins of metal-rich and low-metallicity circumgalactic medium (CGM) at $z<1$. Here we present Keck/KCWI and VLT/MUSE observations of \ifuqsosample\ of these QSO fields (\ifuabssample\ pLLSs) that we combine with \hst/ACS imaging to identify and characterize the absorber-associated galaxies at $0.16 \la z \la 0.84$. We find \ifugalsample\ unique absorber-associated galaxies, with an average of \hostsperabs\ associated galaxy per absorber. For \ifunonsamplet\ absorbers, all with $<10\%$ solar metallicities, we find no associated galaxies with $\logMstar \gtrsim 9.0$ within $\rho/\Rvir$ and $|\Delta v|/\vesc$ $\le$ 1.5 with respect to the absorber. We do not find any strong correlations between the metallicities or \HI\ column densities of the gas and most of the galaxy properties, except for the stellar mass of the galaxies: the low-metallicity ($\xh \le -1.4$) systems have a probability of $0.39^{+0.16}_{-0.15}$ for having a host galaxy with $\logMstar \ge 9.0$ within $\rho/\Rvir \le 1.5$, while the higher metallicity absorbers have a probability of $0.78^{+0.10}_{-0.13}$. This implies metal-enriched pLLSs/LLSs at $z<1$ are typically associated with the CGM of galaxies with $\logMstar > 9.0$, whereas low-metallicity pLLSs/LLSs are found in more diverse locations, with one population arising in the CGM of galaxies and another more broadly distributed in overdense regions of the universe. Using absorbers not associated with galaxies, we estimate the unweighted geometric mean metallicity of the intergalactic medium to be $\xh \la -2.1$ at $z<1$, which is lower than previously estimated.
\end{abstract}


\section{Introduction} \label{sec:intro}

The circumgalactic medium (CGM) is the interface between the interstellar medium (ISM) and the intergalactic medium (IGM). As such, the gas that flows into and out of the CGM has the potential to shape the evolution of the galaxy and possibly reflects this evolution. If the gas can cool and fall onto the galaxy, then it can be used to form new stars, but if the galaxy is unable to refuel from the CGM, then the current supply of ISM gas will quickly be used up (e.g., \citealt{mb2004,tacconi2010}). Whether or not a galaxy can accrete new gas is crucial in determining its evolutionary state (star-forming or quiescent). Therefore, characterizing and comparing the flows around different types of galaxies and galaxy environments is critical to understanding the evolutionary path of a galaxy.

In the low redshift universe, surveys using the Cosmic Origins Spectrograph (COS) and exquisite imaging capabilities on the {\em Hubble Space Telescope} ({\em HST}), large-scale surveys of galaxies (e.g., SDSS; \citealt{york2000}), and large ground-based telescopes\textemdash especially those with integral-field unit (IFU) instruments\textemdash have all contributed to transforming our knowledge of the CGM in recent years. When using QSO absorption-line spectroscopy to characterize the gaseous surroundings of galaxies, three main sample selections have been employed: 1) absorbers are pre-selected based on their properties such as the presence of a specific ion like \OVI, \CIV, \MgII, or \HI\ (e.g., \citealt{steidel1992,nielsen2013,schroetter2016,schroetter2019,zabl2019,lofthouse2020,lundgren2021,wendt2021}); 2) galaxies are pre-selected based on some of their properties (e.g., mass or luminosity) at some distances from observable QSOs, and then the absorbers in the QSO spectra at similar redshift are characterized (e.g., \citealt{bowen1995,bowen2002,chen2010,tumlinson2011,bordoloi2014,werk2014,borthakur2016,heckman2017,ho2017,johnson2017,keeney2017,berg2018,chen2018,berg2019,pointon2019,lehner2020}); and 3) fields are pre-selected to have UV-bright QSOs and the galaxies are surveyed to some magnitude limit without knowledge of the absorbers observed in the QSO spectra (e.g., \citealt{cooksey2008,prochaska2011,johnson2013,tejos2014,burchett2016,burchett2019,bielby2019,prochaska2019,chen2020,muzahid2021}).

These different sample selections all have pros and cons and are complementary. CGM surveys based on field selection are less subject to galaxy or absorber biases (although there can be subtle effects due to intervening absorbers if using UV brightness as a basis for selection). Unfortunately, these surveys are typically not large enough to have a representative sample of massive galaxies, and the observations may also be too shallow to be sensitive to dwarf galaxies (but see e.g., \citealt{johnson2017,chen2020}). However, completeness and depth are becoming less of an issue with deep IFU observations. Galaxy selection has the particular advantage of providing statistical samples of some galaxy types in a redshift range that would otherwise be difficult to get from a blind survey (e.g., massive luminous red galaxies, LRGs, \citealt{chen2018,berg2019} or galaxies hosting an active galactic nucleus, AGN, \citealt{berg2018}). Similarly, absorber selection is particularly useful for assembling statistically useful samples of some specific types of absorbers (e.g., strong \MgII- or \HI-selected absorbers), which are rarer.\footnote{While these sample selections have different limitations and advantages, we emphasize that clearly defining the way a sample is selected is critical to understanding any possible biases or limitations in absorber-galaxy surveys. For example, in the context of galaxy-centric surveys, some types of galaxies or environments (e.g., the IGM) can be missed. Some CGM properties may also only be connected to certain types of galaxies. Similarly, a metal selection in absorber-centric surveys can be inherently metallicity-biased.}

In this paper, we present the first results from our absorber-selected survey: the bimodal absorption system imaging campaign (BASIC). With BASIC, we aim to characterize the galaxy environments associated with a sample of \abssample\ \HI-selected absorbers detected toward \qsosample\ background QSOs. The absorbers have a column density range of $16.0 \le \logNHI < 19.0$ [\column], a redshift range of $0.1 \la \zabs \la 1$ with the average $\zabs = 0.47$, and display a metallicity range of $-3 \la \xh \la 0$.\footnote{Here we use the traditional square-bracket notation $\xh \equiv \log N_{\rm X}/N_{\rm H} - \log ({\rm X/H})_\sun $, where X is an $\alpha$-element, unless otherwise stated.} This sample is mostly drawn from \citet[][hereafter \citetalias{lehner2013}]{lehner2013} with updated \NHI\ measurements and metallicity estimates from the COS CGM Compendium (CCC, \citealt{lehner2018,lehner2019,wotta2019}). Absorbers within this \HI\ column density range are known as partial Lyman limit systems (pLLSs, $16.0 \le \logNHI < 17.2$) and Lyman limit systems (LLSs, $17.2 \le \logNHI < 19.0$). At $z\la 1$, the baryon overdensity ($\delta \equiv \rho/(\Omega_b\rho_c)$) of these types of absorbers is of order $10^2$--$10^3$ \citep{wotta2016}, in line with the overdensities of gaseous galaxy halos observed in simulations and calculated analytically \citep{schaye2001}. In cosmological simulations, these absorbers correspond to cold ($T \sim10^4$\,K) dense gas clumps and can represent new or recycled fuel for star formation \citep{oppenheimer2010,vandevoort2012,fg2016,hafen2019}. The metallicity of the absorbers can help distinguish between possible origins \citep{vs2012,hafen2017,suresh2019}.

Owing to the \HI-selection, \HI\ column density range, and sensitivity of the COS data that were used to analyze these absorbers, there is no metallicity bias in our sample (i.e., both high- and low-metallicity gas can be discovered if it is present). The \HI\ column densities of the pLLSs and LLSs in our sample are also all very well constrained (with typical errors $\la 0.05$--0.1\,dex, \citetalias{lehner2013}; \citealt{lehner2018}), limiting the uncertainty in the ionization modeling used to estimate the metallicity. As shown in CCC \citep{lehner2018,lehner2019,wotta2019}, this type of absorber selection has revealed a population of pLLSs and LLSs at $z\la 1$ that have a remarkably wide range of metallicities not observed in higher \HI\ column density absorbers (see also \citealt{ribaudo2011b}; \citealt{tripp2011}; \citetalias{lehner2013}; \citealt{wotta2016}).\footnote{Similar properties are also observed at higher redshifts (e.g., \citealt{fumagalli2016,lehner2016,lehner2022}).} Absorbers with $\logNHI >19.0$ at $z\la 1$ have metallicities mostly in the range $-1.4\la \xh \la 0$ (\citetalias{lehner2013}; \citealt{lehner2019,wotta2019}, and see also, e.g., \citealt{battisti2012,rafelski2012,som2015,quiret2016}). In contrast, over the same redshift interval, absorbers with $15.1 \la \logNHI \la 18.0$ exhibit a wide range of metallicities from $\xh \la -3$ to super-solar at $\xh \ga 0$, and 40\%--50\% of these absorbers have metallicities of $\xh \la -1.4$ \citep{lehner2019,wotta2019}. Perhaps even more surprising is that $\approx$20\% of the pLLSs/LLSs at $z\la 1$ have metallicities of $\xh \la -2$, implying this gas had little or no additional chemical enrichment over $\approx$6 Gyr when compared with comparable higher-redshift absorbers \citep{fumagalli2016,lehner2016}. In cosmological zoom simulations, LLS/pLLS gas around galaxies with $\la 1\%$ solar metallicity at low redshift is typically found flowing into galaxies from the IGM \citep{hafen2017}. Therefore, the plausible source for these low-metallicity absorbers that have not been processed recently in galaxies is the IGM. On the other hand, metal-enriched absorbers can have many origins including galaxy outflows, AGN outflows, tidal stripping from satellites, or recycled inflowing material from ancient outflows. 

The discovery of this wide range of metallicity in pLLSs and LLSs (as well as lower \NHI\ absorbers) at $z\la 1$ has raised a key question: what types of galaxies and environments house these different absorbers? BASIC aims to directly answer this question by identifying and characterizing the galaxies associated with a sample of pLLSs and LLSs that exhibit a wide range of metallicities. In this first paper in the series, we focus on \ifuqsosample\ QSO fields (\ifuabssample\ absorbers) for which we have obtained IFU observations using Keck/KCWI and VLT/MUSE and \hst\ imaging with the Advanced Camera for Surveys (ACS). Due to our ability to more completely identify associated galaxies in these fields in a magnitude-limited manner, we will use them to estimate the fraction of associated galaxies we may be missing in the other fields where we only have long-slit spectroscopic observations in Paper~II.

This paper is organized as follows. In Section~\ref{sec:abs_descr}, we provide more details on the absorber selection for the survey. In Section~\ref{sec:obs}, we describe the ground- and space-based observations of the fields. Section~\ref{sec:galprop} lays out the methods we use to derive the associated galaxy properties, and in Section~\ref{sec:results}, we present a summary of the associated galaxy properties. In Section~\ref{sec:discussion}, we discuss the implications of our results and compare them to other surveys and models. A summary of our main results are listed in Section~\ref{sec:conclusion}. Throughout this paper we adopt the cosmology from \citet{planck2016}, notably $H_0 = 67.7$ \hub, $\Omega_m (z=0) = 0.309$, and $\Omega_\Lambda(z=0) = 0.691$. We report air wavelength for lines with $\lambda > 3000$\AA. All distances are in units of proper kpc. Metallicities are reported relative to the solar abundance from \citet{asplund2009}.

\section{Description of the \HI-Selected Absorber Sample}\label{sec:abs_descr}

\subsection{Sample Selection}\label{sec:abssamp}

For the present paper, we focus on \ifuabssample\ absorbers observed toward \ifuqsosample\ QSOs for which we have IFU observations of the fields in addition to \hst\ imaging. We list the QSO fields in our BASIC-IFU sample in Table~\ref{tab:fieldinfo}, which were selected as follows. The sample of galaxy fields probed by BASIC is based on an \HI-selected sample of absorbers that have $16.0 \le \logNHI < 19.0$ (i.e., they are pLLSs and LLSs). These absorbers were found blindly in UV QSO spectra that were obtained for other scientific reasons. Therefore, these absorbers are not pre-selected based on knowledge of their metal content or their galaxy environment.

Our sample of pLLSs/LLSs at $z\la 1$ is originally drawn from \citetalias{lehner2013}, but with the updated absorber search from \citet{lehner2018} used for CCC. CCC is an archival \HI-selected absorption line study of 261 new absorbers at $z\lesssim1$ with $15.0 < \logNHI < 19.0$ found in QSO UV spectra, along with stronger absorbers compiled from the literature (see references therein). This updated search resulted in five more pLLSs found along the \citetalias{lehner2013} sightlines. For our BASIC survey, we have removed one of the original \citetalias{lehner2013} fields (PG1216+069) due to the presence of several bright foreground stars that impaired reliable galaxy identification. We included the HE1003+0149 field in its place, which has four \HI-selected absorbers instead of one toward PG1216+069.\footnote{The number of absorbers along this sightline was a factor in its selection to maximize the efficiency of the IFU observations.} We also do not include the TON153 sightline in this sample due to the selection of that system on previously identified \MgII\ absorption (\citealt{kacprzak2012}; \citetalias{lehner2013}). All of the QSOs were observed with \hst/COS using the G130M and/or G160M grating (R $\approx$ 17,000). In total, the BASIC survey comprises \abssample\ pLLS and LLS absorbers in \qsosample\ QSO fields.

\begin{deluxetable*}{lcccc}
\tablecaption{Field Information \label{tab:fieldinfo}}
\tablehead{\colhead{QSO Field} & \colhead{J2000 Name} & \colhead{$z_{\rm em}$} & \colhead{Telescope/Instrument} & \colhead{Integration}}
\startdata
HE0153$-$4520 & J015513.20$-$450611.9 & 0.451 & VLT/MUSE & 7200 \\
PHL1377 & J023507.38$-$040205.6 & 1.437 & VLT/MUSE & 10800 \\
PKS0405$-$123 & J040748.42$-$121136.3 & 0.572 & VLT/MUSE & 35100 \\
HE0439$-$5254 & J044011.90$-$524818.0 & 1.053 & VLT/MUSE & 9000 \\
PKS0552$-$640 & J055224.49$-$640210.7 & 0.680 & VLT/MUSE & 7200 \\
HE1003+0149 & J100535.25+013445.5 & 1.080 & VLT/MUSE & 7200 \\
PG1338+416 & J134100.78+412314.0 & 1.217 & Keck/KCWI & 1100 \\
J1419+4207 & J141910.20+420746.9 & 0.874 & Keck/KCWI & 900 \\
J1435+3604 & J143511.53+360437.2 & 0.429 & Keck/KCWI & 900 \\
PG1522+101 & J152424.58+095829.7 & 1.328 & VLT/MUSE & 7200 \\
J1619+3342 & J161916.54+334238.4 & 0.471 & Keck/KCWI & 900, 1200\tablenotemark{a} \\
\enddata
\tablecomments{The QSO names in the first column are taken from \citetalias{lehner2013}. The second column lists the QSO names as reported in the COS CGM Compendium \citep{lehner2018,lehner2019,wotta2019}. In the last column, we list the average integration for each pointing in seconds.}\tablenotetext{a}{This field was observed twice with different integration times.}
\end{deluxetable*}

\subsection{Metallicity of the Absorbers}\label{sec:absmet}

We do not re-estimate the column densities or metallicities ($\xh \equiv \log N_{\rm X}/N_{\rm H} - \log ({\rm X/H})_\sun $) used in this work, but it is useful to remind the reader about the key points and caveats of the CCC analysis \citep{lehner2018,lehner2019,wotta2019}. The column densities were estimated in individual components on the basis that they could be resolved in several ionic transitions and in the \HI\ absorption at the \hst/COS G130M and G160M resolution. The \HI\ and metal-line column densities were also estimated over the same velocity interval defined by the velocity width of the weak, typically unsaturated \HI\ Lyman series transition. The latter is critical since the strength of the metal and \HI\ absorption, and hence ionization or metallicity, may drastically change with velocity as demonstrated by \citet[][see also \citealt{kacprzak2019,zahedy2019,haislmaier2021}]{lehner2019}. 

As empirically shown by \citet{lehner2018}, all of the pLLSs and LLSs (and specifically the absorbers in BASIC) are predominantly ionized with $x(\HII) > 90\%$. Therefore, large ionization corrections are required to determine the metallicity of the absorbers. The good correspondence between the velocities at the peak optical depths of the \HI\ and of the ions used to determine the metallicity suggests they are co-spatial and motivates the use of a single-phase ionization model for estimating ionization corrections (but see e.g., \citealt{marra2022}). A Bayesian formalism and Markov chain Monte Carlo (MCMC) techniques with grids of photoionization models were used to determine the ionization corrections necessary to derive gas-phase metal abundances \citep{lehner2019,wotta2019}. The photoionization models were constructed using {\tt Cloudy} (version C13.02; see \citealt{ferland2013}), assuming a uniform slab geometry in thermal and ionization equilibrium, with the slab illuminated by a Haardt-Madau EUV background (EUVB) radiation field from quasars and galaxies \citep{hm1996,hm2012}. \citetalias{hm1996} (as implemented in {\tt Cloudy}) was adopted as the fiducial radiation field for CCC, and hence BASIC. To derive the metallicities, low to intermediate ions that fit a single-phase photoionization model are used, and the metallicities are based on $\alpha$-elements (e.g., O, Si, Mg) and carbon (where C/$\alpha$ was allowed to vary in the models). Typically all low and intermediate ions available for an absorber are used in the models, allowing the model to find the best-fit metallicity encompassing multiple $\alpha$-element ion column densities. \citet{lehner2019} details when and why certain ions are not used in the modeling and how this affects the metallicity values.

It is important to consider how the need for large ionization corrections impacts the metallicities of the absorbers. All of the errors reported for the metallicities in this paper are based on the 68\% confidence intervals (CI) from the Bayesian MCMC analysis exploring the full ionization parameter range (\NHI, $z$, metallicity, density, and C/$\alpha$) summarized in Table~3 of \citet{wotta2019}, assuming the \citetalias{hm1996} EUVB; this accounts for the statistical errors. An important systematic error results from the uncertainty in the EUVB. As explored in detail in \citet[][see in particular their Figures~9--10 and Table~7]{wotta2019} and \citet{haislmaier2021}, adopting a different EUVB can systematically change the metallicities. For example, the difference between \citet{hm2012} and \citetalias{hm1996} leads to an average systematic increase in the metallicity of $+0.38 \pm 0.17$ dex for the pLLSs and $+0.16 \pm 0.12$ dex for LLSs at $z\la 1$ (see \citealt{gibson2022} for the explanation of this behavior). Another systematic can be introduced if the gas is assumed to have multiple phases \citep{howk2009}. Considering the three absorbers studied in detail by \citet{haislmaier2021} that are in common with CCC, when multiphase effects are accounted for in the metallicity calculation the overall agreement is well within 0.2 dex. We do not report the systematic error from the EUVB that often can dominate the error budget in the figures or tables in this paper because it would be directly related to the assumed EUVB of \citetalias{hm1996} adopted in CCC. However, we emphasize that the metallicity is still accurate enough to separate low-metallicity absorbers from metal-enriched absorbers (see below for absorber definitions). These systematic errors are comparable to the uncertainties in chemical abundances determined from emission lines in galaxy spectra (see Section~\ref{sec:gmet}). 

\begin{figure}[h]
    \epsscale{1.17}
    \plotone{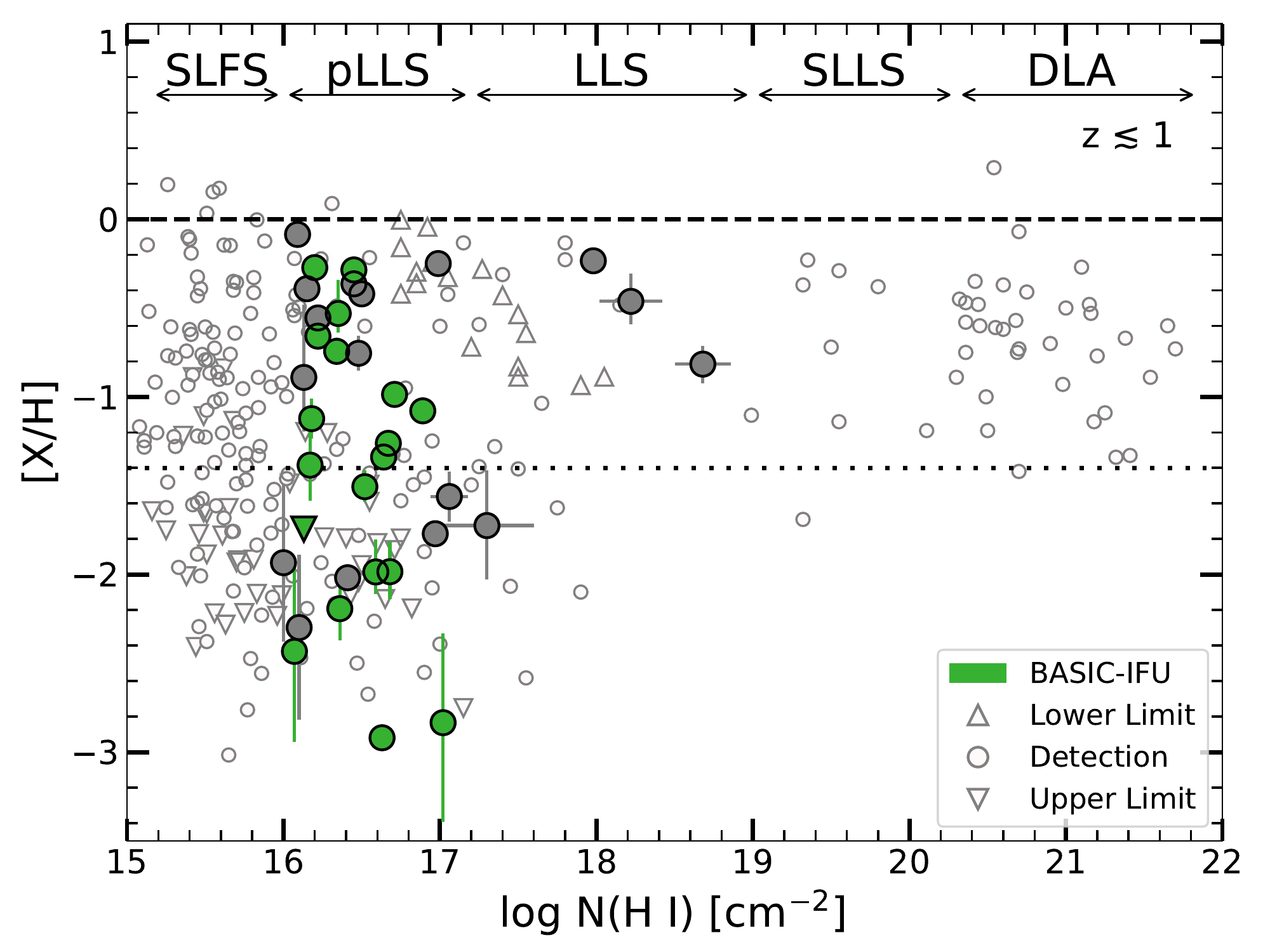}
    \caption{Absorber metallicity versus \logNHI\ for the CCC sample at \z\ $\lesssim 1$. The open circles display the median values of the metallicity posterior PDFs, while the upward-pointing triangles and downward-pointing triangles (lower and upper limits) represent the 10th and 90th percentiles, respectively. (Error bars are not shown for clarity; all absorber metallicities are from \citealt{lehner2019}). The BASIC absorbers are highlighted in grey (long-slit observations) and green (IFU observations and the focus of this paper) symbols. The dashed and dotted lines show solar and 4\% solar metallicity, respectively. The vertical error bars represent the 68\% confidence interval. The absorber labels are strong Ly$\alpha$ forest systems (SLFSs), partial Lyman limit systems (pLLSs), Lyman limit systems (LLSs), super Lyman limit systems (SLLSs), and damped Ly$\alpha$ absorbers (DLAs).} 
\label{fig:sample}
\end{figure}

\begin{deluxetable*}{lccc}
\tablecaption{Absorber Information \label{tab:absinfo}}
\tablehead{\colhead{QSO Field} & \colhead{$z_{\rm abs}$} & \colhead{\logNHI} & \colhead{[X/H]\tablenotemark{a}}}
\startdata
HE0153$-$4520 & 0.225958 & $16.71 \pm 0.07$ & $-1.06, -0.99, -0.91$ \\
PHL1377 & 0.322464 & $16.07 \pm 0.01$ & $-2.94, -2.43, -1.91$ \\
PHL1377 & 0.738890 & $16.67 \pm 0.01$ & $-1.30, -1.26, -1.22$ \\
PKS0405$-$123 & 0.167160 & $16.45 \pm 0.05$ & $-0.32, -0.29, -0.25$ \\
HE0439$-$5254 & 0.614962 & $16.20 \pm 0.01$ & $-0.29, -0.27, -0.25$ \\
PKS0552$-$640 & 0.345149 & $17.02 \pm 0.03$ & $-3.39, -2.83, -2.33$ \\
HE1003+0149 & 0.418522 & $16.89 \pm 0.04$ & $-1.14, -1.08, -1.02$ \\
HE1003+0149 & 0.836989 & $16.52 \pm 0.02$ & $-1.59, -1.51, -1.41$ \\
HE1003+0149 & 0.837390 & $16.36 \pm 0.02$ & $-2.37, -2.19, -2.05$ \\
HE1003+0149 & 0.839400 & $16.13 \pm 0.01$ & $<-4.59, <-3.11, <-1.74$ \\
PG1338+416 & 0.348827 & $16.34 \pm 0.01$ & $-0.77, -0.74, -0.72$ \\
J1419+4207 & 0.288976 & $16.35 \pm 0.05$ & $-0.64, -0.53, -0.34$ \\
J1419+4207 & 0.425592 & $16.17 \pm 0.05$ & $-1.58, -1.38, -1.18$ \\
J1435+3604 & 0.372981 & $16.68 \pm 0.05$ & $-2.14, -1.98, -1.82$ \\
J1435+3604 & 0.387594 & $16.18 \pm 0.06$ & $-1.23, -1.12, -1.01$ \\
PG1522+101 & 0.518500 & $16.22 \pm 0.02$ & $-0.69, -0.66, -0.63$ \\
PG1522+101 & 0.728885 & $16.63 \pm 0.05$ & $-2.97, -2.92, -2.87$ \\
J1619+3342 & 0.269400 & $16.59 \pm 0.03$ & $-2.11, -1.99, -1.80$ \\
J1619+3342 & 0.470800 & $16.64 \pm 0.02$ & $-1.37, -1.34, -1.31$ \\
\enddata
\tablecomments{The above values are taken from CCC \citep{lehner2018,lehner2019}.}\tablenotetext{a}{The lower and upper bounds for each quantity represent the 68\% CI for detections and the 80\% CI for the upper limit.}
\end{deluxetable*}

The metallicities and \HI\ column densities of the BASIC-IFU sample are summarized in Table~\ref{tab:absinfo}. All 19 absorbers are pLLSs, and the sample includes eight low-metallicity absorbers and 11 metal-enriched absorbers. In this paper, we define an absorber of 4\% solar metallicity or below as a low-metallicity system ($\xh \le -1.4$) and above 4\% solar metallicity as a metal-enriched system. The dividing value is chosen such that 95\% of the metallicity probability distribution function (PDF) of the damped Ly$\alpha$ absorbers (DLAs, \logNHI\ $\ge 20.3$) reported in CCC \citep{lehner2018,lehner2019,wotta2019} is above this value.\footnote{Absorbers with $\xh \la -1.4$ were defined in CCC as very metal-poor absorbers.} In Figure~\ref{fig:sample}, we compare the BASIC absorbers (long-slit in grey and IFU in green) with the entirety of the CCC absorbers. In this figure, the median values (circles) of the metallicity posterior PDFs are shown with 68\% confidence errors, except for lower limits (upward-pointing triangles) and upper limits (downward-pointing triangles), where the values represent the $10{\rm th}$ and $90{\rm th}$ percentiles, respectively. The full posterior metallicity distribution function (MDF) for the combined BASIC sample is shown in Figure~\ref{fig:mdf}, and the redshift distribution of the absorbers is plotted in Figure~\ref{fig:zabs}. The lack of a clear bimodal MDF arises from the combination of 1) using the combined PDFs of the absorbers (instead of their mean or median metallicities), and 2) having a large sample of absorbers in BASIC at $z\la 0.5$ (for the BASIC-IFU sample 12/19 absorbers are at $z\la 0.5$). As shown by \citet{lehner2019}, the bimodal MDF is only strongly present at $0.5 \la z < 1$, becoming unimodal at smaller redshifts (see their Figure~12). We ran a dip test on both the BASIC and BASIC-IFU samples using the median metallicity values, but the results were inconclusive.

\begin{figure}
    \epsscale{1.17}
    \plotone{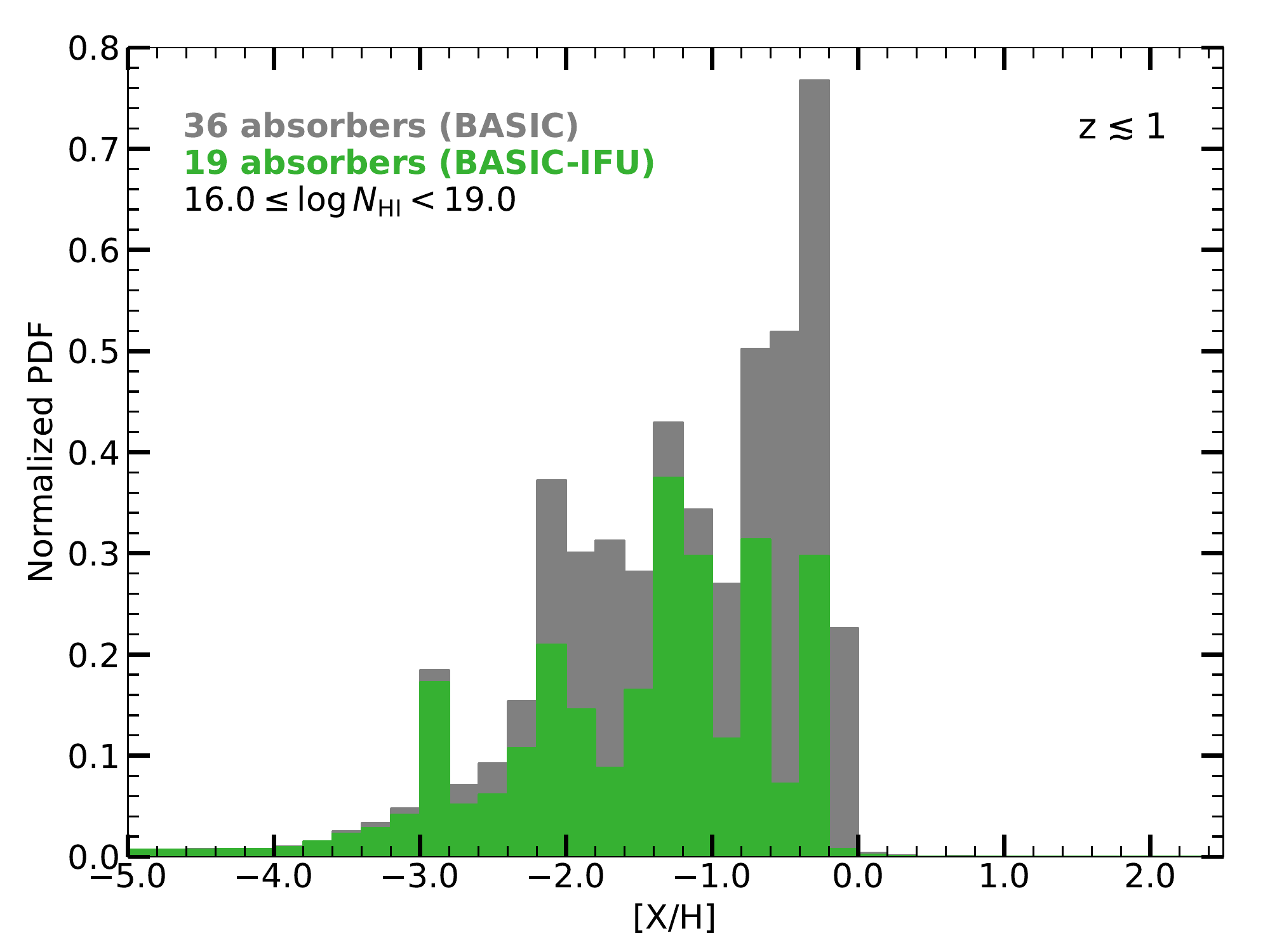}
    \caption{Posterior metallicity PDF of the full BASIC absorber sample and the BASIC-IFU subsample. The region extending to low metallicities displays the contribution of the one upper limit to the distribution (see the text for more details; the original metallicity PDFs are from \citealt{lehner2019}).} \label{fig:mdf}
\end{figure}

\begin{figure}
    \epsscale{1.17}
    \plotone{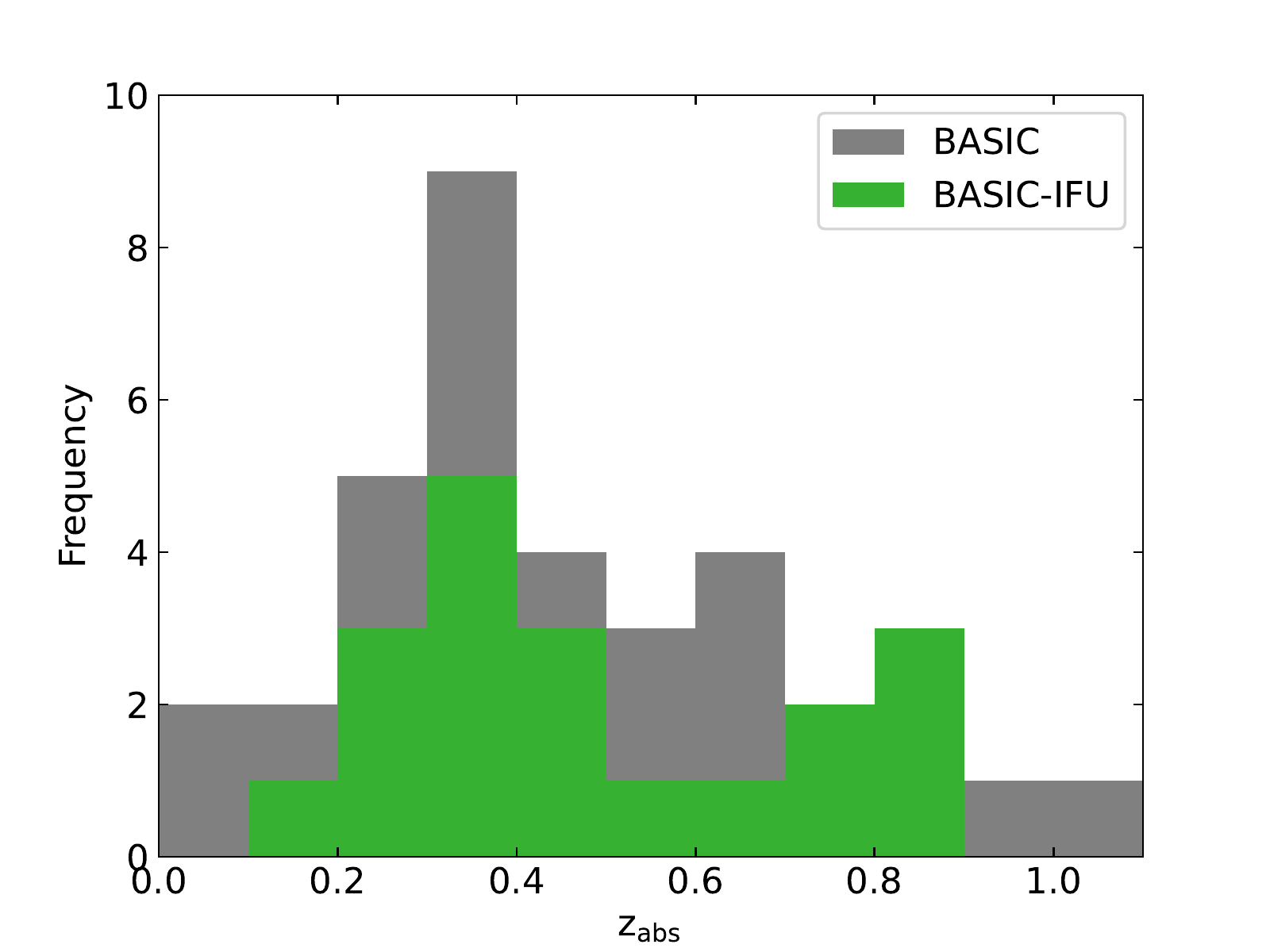}
    \caption{Distribution of absorber redshifts for the BASIC survey and the BASIC-IFU subsample (redshifts are adopted from CCC in \citealt{lehner2018}).} \label{fig:zabs}
\end{figure}

\section{Observations} \label{sec:obs}

In this section, we detail the observations taken for the BASIC-IFU fields. We also describe the data reduction, spectral extraction, and galaxy identification processes. 

\subsection{HST Imaging}\label{sec:hst}

By definition, all of the \qsosample\ BASIC fields have \hst\ imaging. Fourteen of the fields are drawn from our Cycle 23 ACS imaging program (PID 14269, PI: Lehner); the remaining nine have archival imaging (WFPC2 PIDs 5099, 5143, 5849, 6303; WFC3 PID 11598; or ACS PIDs 9418, 13024). The archival observations were retrieved from the Mikulski Archive for Space Telescope (MAST). The data were reduced using the standard \hst\ reduction pipeline ({\tt AstroDrizzle} version 2.1.3, \citealt{stsci2012}). The individual dithered images were flagged for bad pixels, corrected for geometric distortion, registered to a common origin, cleaned for cosmic rays, and combined using drizzle. All \ifuqsosample\ fields in the BASIC-IFU sample have (Cycle 23 or archival) ACS imaging with the F814W filter. The ACS images have a field of view of 202\arcsec $\times$ 202\arcsec\ and reach a 5$\sigma$ limiting magnitude of AB(F841W) = 25.6 mag for those taken in Cycle 23. In Figure~\ref{fig:HE0153} we show an example \hst\ image of one of the fields with an outline of the IFU pointing. The remaining field images are provided in Appendix~\ref{sec:appa}. All the \hst\ imaging data used in this paper can be found in MAST: \dataset[10.17909/zw9p-mg69]{http://dx.doi.org/10.17909/zw9p-mg69}.

\begin{figure}
    \epsscale{1.17}
    \plotone{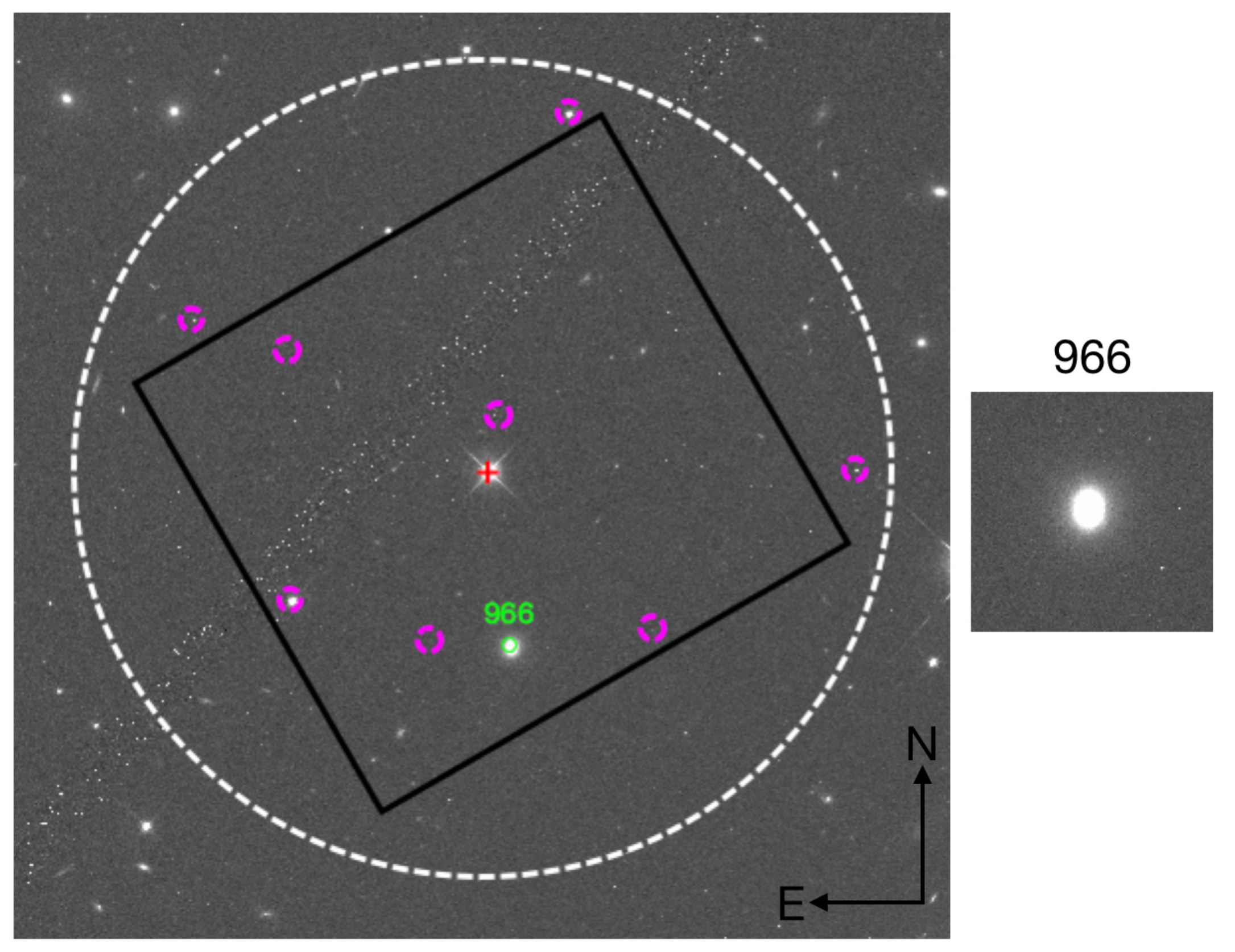}
    \caption{\hst/ACS image with outline of VLT/MUSE pointing for the field of HE0153$-$4520. North is up, and East is to the left. The QSO is marked with the red cross, stars are indicated with magenta circles, and the white dashed circle has a radius of 200 kpc at the redshift of the absorber. The black box designates the MUSE FOV. The candidate galaxy of the absorber is marked by the green circle and labelled with the SE\#.} \label{fig:HE0153}
\end{figure}

We employed {\tt Source Extractor} (hereafter SE; \citealt{ba1996}) version 2.19.5 to identify objects in the fields to compare to the IFU observations and to measure the galaxy photometry in AB magnitudes. We used a detection threshold of 3.0$\sigma$ above the background for source identification. The archival ACS fields have excess cosmic rays along the detector chip boundary that are not wholly removed. SE labels these objects as stars, so they are easily removed in the resulting catalog. 

\subsection{VLT/MUSE IFU Spectroscopy}\label{sec:muse}

Seven of the fields (see Table~\ref{tab:fieldinfo}) were observed with the Multi-Unit Spectroscopic Explorer (MUSE, \citealt{bacon2006,bacon2010}) on the Very Large Telescope (VLT). These observations were taken as part of the MUSE Quasar-field Blind Emitter Survey (MUSE-QuBES, \citealt{muzahid2020,muzahid2021}) program (PID: 094.A-0131, PI: J. Schaye). MUSE is an integral field spectrograph with wavelength coverage from 4650--9300 \AA\ and a spectral resolution of $R=2000$--4000. The observations were taken in wide field mode, resulting in a $1\arcmin \times 1\arcmin$ field of view with a pixel scale of $0\farcs2$.

Exposures were taken in increments of 900 seconds, with dithering and position angle offsets applied to smooth the tiling pattern of the combined cubes. The observations vary in depth: six of the cubes have exposures of 2--3 hours, while PKS0405$-$123 has a much longer exposure of 9.75 hours. The observations were reduced using the standard MUSE reduction pipeline version 1.6 \citep{weilbacher2020}, with additional post-processing using tools from the {\tt CubExtractor} package (CubEx version 1.5, \citealt{cantalupo2019}; Cantalupo et al., in preparation) to coadd the individual MUSE exposures. This additional step improves the flat-fielding and sky subtraction \citep{marino2018}. 

We created whitelight and narrowband images of the cubes using {\tt PyMUSE}\footnote{\url{https://github.com/ismaelpessa/PyMUSE}} \citep{pessa2020}. Due to the redshifts of the associated absorbers and the wavelength coverage of MUSE, we created narrowband images centered on restframe \OIId, H$\beta$, \OIIId, and H$\alpha$ for our absorbers (though not all fields cover the \OII\ doublet or H$\alpha$ at the redshift of the absorber). These narrowband images (created by summing the data within $\pm$ 15\AA\ of line center) were solely used for quick identification of potential galaxies at our redshifts of interest, not for spectral extraction. We note there were no occurrences of any sources only seen in the narrowband images. We used SE to identify all of the objects with a detection threshold of 6.0$\sigma$ above the background in the whitelight images to compare to the \hst\ observations by eye. (The \hst\ and MUSE observations reach similar magnitude depths, except for the much deeper PKS0405$-$123 cube.) We used this limit in an effort to mitigate erroneous edge effect detections. For HE0153$-$4520 and HE1003+0149 we lowered the threshold to 4.0$\sigma$ due to the prevalence of small, lower-magnitude objects in these fields. Since PKS0405$-$123 is a much deeper cube, we also changed the DEBLEND\_MINCONT parameter to 0.0001 following \citet{lofthouse2020}. We use and report the \hst\ object positions instead of the MUSE object positions throughout the paper because of its higher resolution.

We used {\tt PyMUSE} to extract the spectra of the galaxies identified with SE in the whitelight image. Specifically, we transformed the SE catalog into a {\tt ds9} \citep{sao2000} region file and extracted the spectra within the defined elliptical regions, using the whitelight-weighted mean method of extraction without any smoothing. The flux within the region is multiplied by the weights, summed, and normalized to match the total integrated flux \citep{pessa2020}. The error is treated in the same way. We extracted all non-stellar sources in the fields. Stars were identified in the \hst\ images as having the SE parameter CLASS\_STAR $>$ 0.2. All potential stellar object spectra were checked using {\tt QFitsView}\footnote{\url{https://www.mpe.mpg.de/~ott/dpuser/qfitsview.html}} \citep{ott2012} to make sure we did not remove any galaxies from the sample. The field of PKS0552$-$640 lies toward the Large Magellanic Cloud, so there is a large number of stellar contaminants. For this field, we extracted all objects identified with SE and examined the spectra by eye to remove stars from the sample. 

The associated galaxies of the pLLS in the PKS0405--123 field lie outside of the MUSE field of view. For these two objects we make use of the Du Pont/WFCCD \citep{weymann2001} spectrum reported in \citet{prochaska2006}\footnote{\url{http://www.ucolick.org/~xavier/WFCCDOVI/}} and the Magellan Baade/IMACS \citep{dressler2011} spectrum from \citet{johnson2013} (S. D. Johnson 2020, private communication). We found no other associated galaxy candidates within the MUSE field.

\subsection{Keck/KCWI IFU Spectroscopy}\label{sec:kcwi}

The other four fields (see Table~\ref{tab:fieldinfo}) were observed with the Keck Cosmic Web Imager (KCWI, \citealt{morrissey2018}) on the Keck II Telescope. All of the observations were taken with the large IFU slicer and the BL grating. This configuration resulted in a field of view $33\arcsec \times 20.4\arcsec$, a wavelength coverage from 3500--5600 \AA, a spectral resolution of $R\approx900$, and a pixel scale of $1\farcs35$. For the observations of J1419+4207, J1619+3342, and PG1338+416, the blue filter was removed to extend the wavelength coverage to 5750 \AA. This was done to cover the \OIId\ doublet of any galaxies associated with the higher redshift absorbers. Removing the filter causes unequal wavelength coverage for each slice at the red end; the total throughput does not change significantly when entering this extended wavelength region. The \OIId\ doublet is the primary emission line we use for galaxy identification for these observations because of the instrument wavelength coverage and redshifts of the absorbers. 

Due to the smaller field of view of KCWI, we tiled the pointings around the QSO within roughly 200 kpc at the redshift of the lowest-redshift absorber for a given sightline. In Appendix~\ref{sec:appa} we provide images detailing the KCWI pointings on the \hst\ image of each field. For PG1338+416 we were only able to acquire one pointing, so we chose a section of sky with the highest galaxy density. For J1619+3342 we prioritized sections of the sky with higher galaxy density in our pointings. We note this approach could bias our observations in these two fields because we are not evenly sampling dense and underdense regions of the sky. However, in the MUSE observations that have full field coverage the emission always corresponded to an identifiable galaxy in the \hst\ image.

The data were reduced using {\tt KDERP}\footnote{\url{https://github.com/Keck-DataReductionPipelines/KcwiDRP}} \citep{morrissey2018} version 1.1.0, except for data from the J1619+3342 field which were reduced with version 1.2.0. These versions of {\tt KDERP} do not transform the observations from the geocentric to the heliocentric frame, so we applied heliocentric-corrections to the cubes after the reduction.\footnote{There are now two python-based KCWI pipelines; {\tt KDERP} is no longer supported.}

Composite images of the field pointings were created using {\tt Montage} version 6.0 \citep{berriman2007}. We aligned the composite images to the \hst\ images using the QSOs as reference points (for PG1338+416 we used the brightest galaxy in the pointing) to aid in galaxy identification. We created whitelight and narrowband images centered on the \OIId\ doublet at the redshift of the absorbers from the composite images. We used SE to identify all of the objects with a detection threshold of 3.5$\sigma$ above the background in the whitelight and \OII\ narrowband images to compare to the \hst\ observations. Since the resolution of the KCWI whitelight images is much lower than that of the \hst\ images, we use and report the \hst\ object positions instead of the KCWI object positions throughout this paper.

We extracted the spectra of all galaxies detected by SE in the whitelight and \OII\ narrowband images. The \OII\ narrowband images help us identify star-forming galaxies at the redshift of the absorbers, while the whitelight images help us identify galaxies without strong emission. We used the Python package {\tt kcwitools}\footnote{\url{https://github.com/pypeit/kcwitools}} to perform a boxcar extraction within a rectangular region of the cube files. We identified stars in the \hst\ images as described above.

\section{Galaxy Properties}\label{sec:galprop}

This section details the different methods used for the characterization of the associated galaxies in each field. We describe the properties derived from the IFU observations in \S~\ref{sec:ifuprops}, the properties derived from the \hst\ imaging in \S~\ref{sec:hstprops}, and other additional properties in \S~\ref{sec:addprops}. We detail how we assess detection limits for each field in \S~\ref{sec:limits}. Our associated galaxy sample selection criteria are presented in \S~\ref{sec:groups}.

\subsection{Properties from the IFU Observations}\label{sec:ifuprops}

\subsubsection{Redshift}\label{sec:redshift}

Once we have extracted the spectra for all of the galaxies of interest in a field, we determine their redshifts to identify which galaxies are associated with the absorbers. We use the code {\tt REDROCK}\footnote{\url{https://github.com/desihub/redrock}} created by the DESI collaboration to determine galaxy redshifts via spectral template fitting. {\tt REDROCK} returns the three best-fit models and redshifts for a QSO, galaxy, and stellar template, ranking the nine results by the model's $\chi^2$ value when compared with the data. The redshift errors are of order $\sigma_z \approx 10^{-5}$--$10^{-4}$. See \citet{bolton2012} and \citet{ross2020} for more information. 

We review the results of each object and assign a redshift flag of [0, 1, 3, 4] indicating our confidence in the redshift results.\footnote{We purposefully skip setting a flag of 2 for the ease of separating the data.} Our flag assignments are modeled after those used in \citet{wilde2021}. A galaxy receives a flag of 0 if the spectrum has too low a S/N to be useful ($\approx 1$ per pixel without a discernible continuum). A flag of 1 indicates the spectrum has no emission or absorption lines even though the S/N is sufficient to identify any lines. If there is one emission or absorption line in the spectrum, the galaxy is assigned a flag of 3. This flag is also set if the galaxy continuum is well-matched by the template continuum without any lines (for MUSE spectra only). When there are multiple lines in the spectrum, we assign a flag of 4 indicating a high confidence in the results. In a very few instances we determine the redshifts by hand. This happens when there are blended sources, or {\tt REDROCK} does not identify a line in the spectrum. Due to the low S/N of some of our spectra, {\tt REDROCK} occasionally picks up on spurious features in the spectra (e.g., poorly-subtracted sky lines) and misidentifies a line. 
 
As we are focusing on the \OIId\ doublet in the KCWI observations, it is important to understand if we are identifying this line correctly in our spectra since we do not resolve the doublet. The {\tt REDROCK} results consistently label the emission line we see in the spectra as either \OII, \HeIline, or H$\beta$. For the template models that identify the line as H$\beta$, the width of the model line is too narrow compared to the emission line in the galaxy spectrum. Additionally, the wavelength coverage of KCWI is such that either H$\beta$ would be seen along with the \OIII\ $\lambda$4958 line, or it would be seen along with the \OII\ doublet. The \OII\ doublet is the only line that would appear prominently and by itself over the redshift range of the KCWI spectrum. 

{\tt REDROCK} also identifies single emission lines as \HeIline\ in some of the best-fit galaxy results, though we consider this unlikely. It is not possible to identify \CaII\ H \& K (or the H8, H9, and H10 lines) in the spectrum unless the S/N in the region redward of \OII\ is $>5$. As a consequence of this, in some fits {\tt REDROCK} favors \HeI\ over \OII\ in the absence of these absorption lines. However, the emission we see is often strong. In the {\tt REDROCK} templates where these lines are seen in emission, the \OII\ doublet is 16 to $>100\times$ stronger than the \HeI\ line. We search the literature for galaxy spectra covering both of these lines and find no reported cases where \OIId/\HeIline\ $\lesssim10$ \citep{rose2011,florian2020}. Additionally, in MUSE spectra where both lines are covered, we do not see the \HeIline\ line in any case. Therefore, we are confident that the emission line we see in the single-line KCWI spectra is \OIId.

A final possibility for a singular emission line is Ly$\alpha$. For the KCWI observations, this places the objects at $z\approx3-3.5$. {\tt REDROCK} does not label these lines as Ly$\alpha$ because the galaxy templates are limited to $z<1.7$. When observing a Ly$\alpha$ emitter (LAE), a decrement in the flux is usually observed around the Ly$\alpha$ line and blueward. Though the S/N of the KCWI spectra are low, we do not see any such decrement in the continuum. We can also inspect the galaxy morphology. LAEs exhibit circular morphology in imaging \citep{wisotzki2016,ouchi2020}. We have several objects that are circular in appearance, and a few galaxies that are extended objects (see Appendix~\ref{sec:appa}). In MUSE observations of LAEs at $z\approx3.3$, \citet{muzahid2021} notes their Ly$\alpha$ lines exhibit an asymmetric profile with a prominent red wing. We do not see any such profiles in our spectra. Taken all together, we can rule out this singular line as Ly$\alpha$. 

In the instances where we need to determine the redshift of a galaxy by hand, we model the \OII\ emission line and use a fitter to determine the central redshift. We model the continuum around the line as a Chebyshev polynomial of degree three and use a linear least squares fitter to fit the spectrum. For the doublet, we model each line as a Gaussian using an initial redshift guess to set the mean wavelength. We fit the lines and continuum simultaneously, tying the redshifts and widths of the two lines. The redshift errors are of order $10^{-4}$ (i.e. $\approx 30\!-\!100$ \kms).

\subsubsection{Star Formation Rate}\label{sec:sfr}

We calculate the star formation rates (SFRs) for the associated galaxies in the KCWI fields using the \OIId\ doublet. We derive line fluxes and errors from Gaussian fits as described in section~\ref{sec:redshift} with redshifts fixed to the values from {\tt REDROCK}.

For the MUSE associated galaxies, we also use the \OIId\ doublet to derive SFRs because it is covered in the associated galaxy spectra in all but two cases. We use the H$\beta$ line instead for the other two associated galaxies. The stellar continuum is well-defined for the MUSE associated galaxies, so we fit and remove it using {\tt pPXF} \citep{cappellari2004,vazdekis2010,cappellari2017} prior to measuring the line fluxes and errors as described above.\footnote{We do not remove the stellar continuum for the KCWI spectra because the \OII\ line is not subject to stellar absorption, and the wavelength coverage does not extend to other lines where we would need to take this into account.} We have a larger spectral range for the MUSE associated galaxies, so we simultaneously fit all emission lines in the spectra with tied redshifts and ratios (when linked by simple atomic physics). We calculate the line luminosities from fluxes with luminosity distances derived from the {\tt REDROCK} redshifts for our chosen cosmology.

We treat the associated galaxies in the PKS0405$-$123 field differently. For the WFCCD spectrum, we proceed as with the MUSE associated galaxies, but we also add a correction to the line fluxes because the entire galaxy is not covered by the slit. We determine the magnitude of the associated galaxy in the rest-frame band nearest to the emission line (using Bessel $V$, $R$, or $I$) and compare it to the magnitude we derive in that filter from the spectrum. Using this magnitude ratio, we determine the percentage of flux we are missing to bring the integrated spectral fluxes in alignment with the associated galaxy apparent magnitudes. The IMACS spectrum is not flux-calibrated. Therefore, we measure line equivalent widths and convert to line luminosities using the absolute magnitude of the associated galaxy in the rest-frame band closest to the emission line (using Bessel $B$, $V$, or $R$). To account for any gas emission we might be missing in the equivalent width measurement, we apply a correction to the line luminosities by comparing the ratio of the line fluxes measured in the stellar continuum-subtracted spectrum with those in the original spectrum. The line luminosity errors are determined from carrying the equivalent width errors through these calculations. 

To calculate the SFRs, we use the calibrations between line luminosity and SFR given in \citet{moustakas2006} for H$\beta$ and \OII\ (their Tables 1 and 2), adopting the median percentile of the relations ($P_{50}$). The \citet{moustakas2006} calibrations depend on the galaxy's $B$-band luminosity (Vega). We infer this luminosity from the estimated Vega $B$-band absolute magnitude derived from the {\tt kcorrect} code (v4\_3)\footnote{Available through \url{http://kcorrect.org} or \url{https://github.com/blanton144/kcorrect}.} of \citet{blanton2007}. This code calculates spectral energy distribution fits using the \citet{bruzual2003} spectral templates assuming the initial mass function of \citet{chabrier2003} to assess the k-corrections and mass-to-light ratios of galaxies. As input, we use the \hst\ F814W apparent magnitudes calculated from SE (i.e., MAG\_AUTO and MAGERR\_AUTO), corrected for Milky Way extinction. The extinction corrections are calculated using {\tt gdpyc}\footnote{\url{https://github.com/ruizca/gdpyc}} with the high resolution \citet{schlegel1998} dust map and the \citet{schlafly2011} filter conversion factors. We do not correct for internal reddening because the \citet{moustakas2006} relations are not corrected either, which partly leads to their luminosity dependence. We run {\tt kcorrect} with the F814W apparent magnitudes and then project the derived spectra into the Vega $B$ bandpass. We choose the SFR calibration by matching the closest $B$-band magnitude listed in the tables to those we derive. 

\subsubsection{Metallicity}\label{sec:gmet}

We can derive galaxy metallicities for the MUSE associated emission line galaxies. All of these spectra have H$\beta$ and \OIII\ $\lambda$5006 coverage, so we use the O3 metallicity relation of \citet{curti2017} with a slight correction reported in \citet{howk2018} Table 3. Due to the proximity of these lines in wavelength, a dust correction is unnecessary. We utilize the {\tt pyMCZ}\footnote{\url{https://github.com/nyusngroup/pyMCZ}} code of \citet{bianco2016} to calculate metallicities and errors. {\tt pyMCZ} uses a Monte Carlo sampling of the input emission line intensity distributions to calculate the posterior distribution functions of the oxygen abundances ($\epsilon$(O) $\equiv$ 12 + log(O/H)) and their uncertainties. This code is open-source, so we add the \citet{curti2017} abundance scales. We sample the input line intensities 4000 times assuming a normal distribution with central values and dispersions given by the measurements. As noted in \citet{howk2018}, the \citet{curti2017} O3 scale has a dispersion of 0.178 dex between the strong emission line technique and the direct technique. In light of this, we add the statistical errors reported from {\tt pyMCZ} and this dispersion value in quadrature as an estimate of the total metallicity error. In this paper, we report the galaxy metallicities relative to the solar abundance as [O/H] using the \citet{asplund2009} solar oxygen abundance value of $\log(\rm O/\rm H)_{\odot} = 8.69$.

\subsection{Properties from the HST/ACS Imaging}\label{sec:hstprops}

We calculate the impact parameter (the proper distance between the galaxy and absorber, $\rho$) using the associated galaxy coordinates determined with SE from the \hst\ imaging, and the QSO coordinates taken from the Hubble Legacy Archive. Using the associated galaxy redshift determined with {\tt REDROCK}, we then convert this value to physical kpc with our chosen cosmology. 

Although we have \hst/ACS imaging for all of the BASIC-IFU fields, we are unable to determine the physical morphology for most of the associated galaxies due to their small size or high redshifts. Instead, we use the spectral morphology to classify our associated galaxies as absorption-line-dominated or emission-line-dominated and utilize the imaging to constrain the galaxy orientation and inclination.

To investigate the azimuthal angle dependence on absorber metallicity, we use {\tt GALFIT} version 3 \citep{peng2002,peng2010} to characterize the components of the associated galaxies. For the input point spread function we use a 2D Gaussian.\footnote{A more in depth decomposition should use a function closer to the \hst\ point spread function, but this is adequate for our measurements.} We fit a bulge and disk component to the associated galaxy using S\'ersic profiles. In a few instances only a bulge component is fit to the associated galaxy. The residual of the fit is inspected by eye. Only one associated galaxy and one cut candidate galaxy give poor fits; we do not include these in the orientation analysis. One of these galaxies is very small, spanning only four pixels, and the other appears to be a face-on spiral for which the orientation calculation is ambiguous (no clear axis can be identified, see Figure~\ref{fig:HE1003}). 

We use the position angle determined from the fits to calculate the azimuthal angle between the associated galaxy disk and QSO. For nearly circular galaxies, the position angle error can be greater than ten degrees. We remove three associated galaxies and two cut candidate galaxies from the orientation analysis due to their large uncertainty. The typical uncertainties are 1--2$^{\circ}$. With the galaxy position angle and the coordinates of the associated galaxy and QSO, we calculate the azimuthal angle where $\Phi = 0^{\circ}$ is along the galaxy minor axis (pole) and $\Phi = 90^{\circ}$ is along the major axis (disk).

The galaxy inclination is also determined using {\tt GALFIT}, as described above. We do not remove any other associated or cut candidate galaxies from this analysis except for the two that give poor fits. Using the computed axis ratio, we simply use the standard equation $\cos i = 1 - \varepsilon$, where $i$ is the inclination and $\varepsilon = 1- \frac{b}{a}$ is the ellipticity ($b$ and $a$ are the best-fit semi-minor and semi-major axes, respectively). An inclination of 0$^{\circ}$ is a face-on galaxy, and an inclination of 90$^{\circ}$ is an edge-on galaxy. Errors for the inclination are five to ten degrees.

\subsection{Additional Properties}\label{sec:addprops}

To derive stellar masses for the associated galaxies, we supplement the \hst\ F814W magnitudes with additional photometry from surveys hosted on NOIRLab's Astro Data Lab.\footnote{\url{https://datalab.noirlab.edu}} Specifically, we use these catalogs: Legacy Survey data release 8 ($g$, $r$, $z$, W1, W2 filters; \citealt{dey2019}), NOIRLab Source Catalog data release 2 ($g$, $r$, $i$, $z$ filters; \citealt{nidever2018,nidever2021}), 2MASS point source catalog ($J$, $H$, $K_{s}$ filters; \citealt{skrutskie2006}), and unWISE (W1, W2 filters; \citealt{schlafly2019}). The NOIRLab Source Catalog also includes photometry in the $Y$ and $VR$ bands, but we do not use those filters for this analysis. All but three of the associated galaxies and three of the cut candidate galaxies are covered in these surveys. We also have two associated galaxies that are blended with other objects, so we do not use the photometry from these catalogs. Of those that are detected, each associated galaxy is covered in at least four bands. 

We primarily use the Legacy Survey photometry, deferring to the NOIRLab Source Catalog for fields not covered by the Legacy Survey. We correct the unWISE and 2MASS magnitudes to AB mags, and note that a Milky Way extinction correction is already accounted for in the Legacy Survey and NOIRLab Source Catalog. This correction is unnecessary for the unWISE and 2MASS values. We use {\tt kcorrect} (described in \S~\ref{sec:sfr}) with at least four AB magnitudes as input to determine the stellar mass ($M_{\star}$) for each associated galaxy. The masses are output in units of ${\rm M}_{\sun}\,h^{-1}$, which we correct to our chosen cosmology. The errors on these masses are at least $\pm$0.3 dex \citep{conroy2013}.

For four of the other six galaxies not covered by the above surveys, we determine the galaxy magnitudes using $U_{\rm spec}$, $g$, $r$, and $i$ imaging from the Large Binocular Cameras (LBC) on the Large Binocular Telescope (LBT). We use SE to extract the magnitudes with the published zeropoints\footnote{\url{https://sites.google.com/a/lbto.org/lbc/phase-ii-guidelines/sensitivities}} and correct for the airmass of the observations, color offsets, and Milky Way extinction (see \S~\ref{sec:sfr}). We use {\tt kcorrect}, as described above, to calculate the stellar masses using at least four bands. 

The other two galaxies were undetected in more than one band of the LBC imaging. To derive more accurate stellar masses, we use {\tt synphot} \citep{stsci2018} to calculate synthetic magnitudes in the $r$ and $i$ bands. We also apply this analysis to one of the blended galaxies whose spectrum is not contaminated by neighboring objects. The synthetic magnitudes are corrected for Milky Way extinction, and we assign a conservative error of 0.5 mags. We use {\tt kcorrect} to calculate the stellar masses and note only three bands are used for these estimates.

The final galaxy spectrum is blended with a lower redshift galaxy (though both can be seen in the \hst\ image), so we cannot determine accurate synthetic magnitudes. We instead use a mass-to-light ratio ($M/L$) to determine the stellar mass. \citet{portinari2004} constructed chemo-photometric models of spiral galaxy disks with varying initial mass functions (IMFs) to explore the $M/L$ ratio variations in the rest frame $I$ band. They report total $M/L_I$ ratios calculated with different IMFs in their Table 7, broken down by spiral morphological type. The associated galaxy has an emission-line-dominated spectrum, so we infer a spiral morphology. We cannot distinguish between the spiral morphological types for the associated galaxy, so instead we take the average of their three $M/L_I$ values for Sa/Sab, Sb, and Sbc/Sc using the \citet{chabrier2003} IMF, giving an $M/L_I$ = 1.61 M$_{\odot}$/L$_{\odot}$. We convert the \hst\ F814W magnitude to absolute $I$ band magnitude using {\tt kcorrect}, calculate the $I$ band luminosity, and apply the $M/L_I$ ratio to determine the stellar mass.

The halo masses ($M_{\rm h}$) of the associated galaxies are calculated using the stellar mass--halo mass (SMHM) relation from \citet{rodriguez-puebla2017}, as described in \citet{berg2019}. In brief, we use their Equation 66 to determine the mean halo mass given a stellar mass $\langle\logMhalo(M_{\star})\rangle$. This SMHM relation takes into account the asymmetric scatter and covers a wide range in masses, allowing for consistent treatment of dwarfs to high-mass galaxies. We note that all of our halo masses are within the minimum and maximum bounds where this relation can be trusted at these redshifts (their Table 3). Additionally, the 1$\sigma$ confidence intervals for the values of \logMstar\ over the probed range in \logMhalo\ are below 0.05 for the redshifts covered by our associated galaxies as shown in their Figure 6. When applying this equation, we assume each associated galaxy to be a central. As such, we estimate the satellite fraction of our associated galaxy sample to be $0.27$ from the satellite fraction distribution shown in Figure 4 of \citet{wt2018} for our average associated galaxy stellar mass of $\logMstar = 10.1$ \msun.

We define the virial radius of the associated galaxy as $\Rvir = R_{200}$, which is the radius where the enclosed average density is equal to 200 times the critical density of the universe at the redshift of the associated galaxy, $R_{200} \equiv (3M_{h}/(4\pi\Delta\rho_{c}))^{1/3}$, where $\Delta = 200$ and $\rho_{c}$ is the critical density of the universe. We simply refer to this scale as \Rvir\ throughout the paper.

The escape velocity, $\vesc$, of the halo is calculated as the circular velocity at \Rvir\ assuming a Navarro-Frenk-White potential \citep{nfw1996,nfw1997}. For the halo concentration parameter, we use the relations derived in \citet{dm2014}. We use the parameterizations for quantities calculated with respect to 200$\rho_{c}$ to match our calculations. Specifically, we use their Equation 7 with the redshift evolution fit parameters given in their Equations 10 and 11. 

\subsection{Detection Limits}\label{sec:limits}

Each field in our sample has been observed to a different depth (see Table~\ref{tab:fieldinfo}), so we do not have a uniform detection limit for our observations. To understand the limits of our survey, we start with determining the redshift completeness of sources in the \hst\ images. We simply determine how many of the \hst\ sources (cleaned of stars, spikes, and cosmic rays) identified with SE are also identified in the MUSE whitelight images and have confident redshifts. In a handful of cases, there are galaxies close to the QSO that are identified in the \hst\ image, but lie in the QSO glow in the MUSE whitelight images. We have extracted these objects by hand and checked that they are not at the redshifts of interest for this survey. Figure~\ref{fig:redcomp} displays the results for the MUSE fields, with the dashed vertical lines marking the 90\% F814W magnitude completeness limits. We average the results of five fields with 2-3 hour exposures, but we plot the PKS0552-640 field by itself due to the plethora of foreground stars in the field. Without spectroscopy of higher spatial resolution, we are not confident that we have been able to fully clean the SE catalog for the \hst\ image. The redshift completeness distribution for PKS0405$-$123 (9.75 hours exposure) drops quickly below $\sim$60\% because many objects in this deep MUSE observation are not detected in the \hst\ image. We do not determine the redshift completeness for the KCWI fields because the observations are of lower S/N and integration time, resulting in many of the objects in the fields without reliable redshifts. This would biases the redshift completeness results towards only bright galaxies in the fields.

\begin{figure}
    \epsscale{1.2}
    \plotone{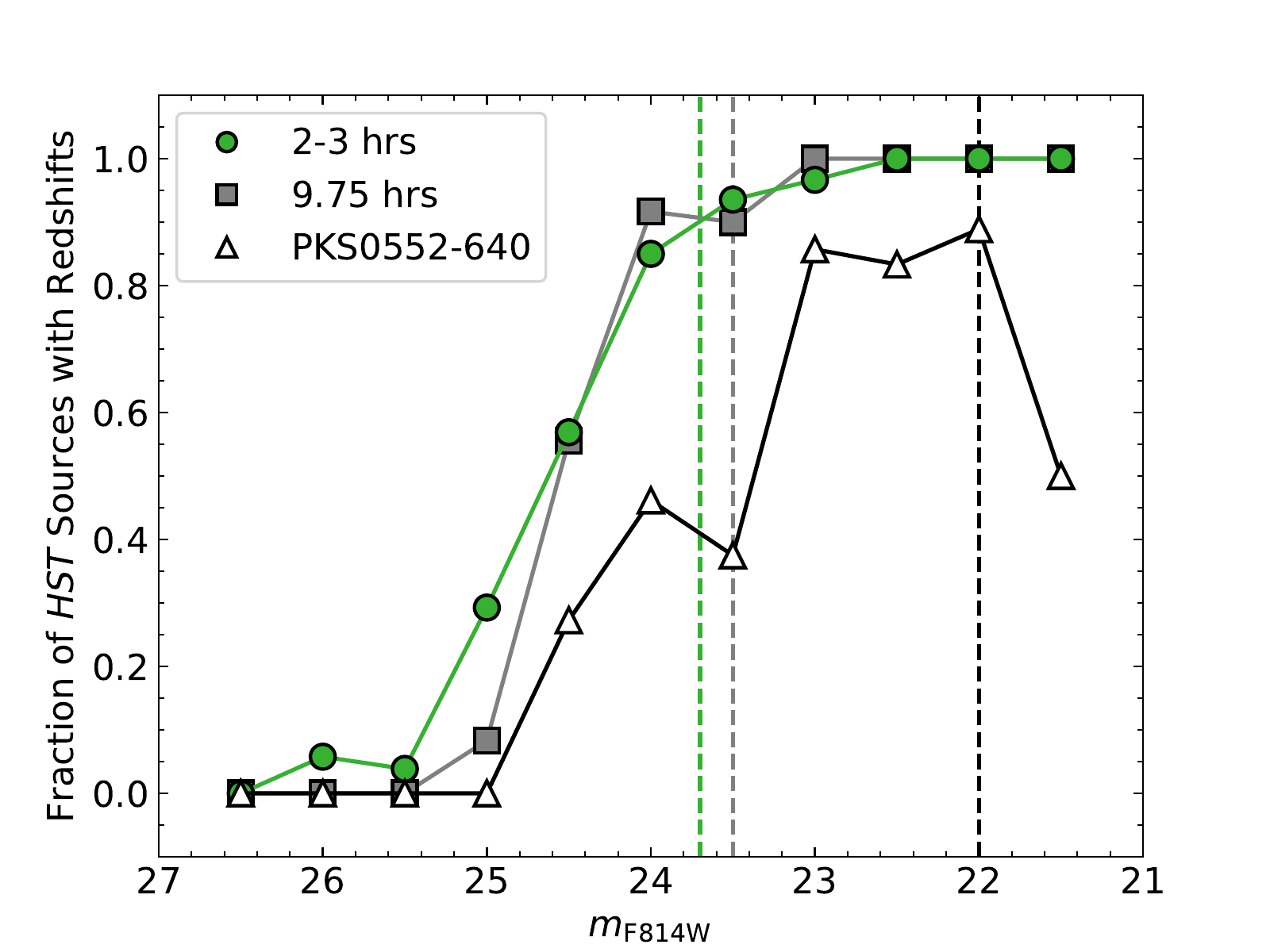}
    \caption{Fraction of extracted \hst\ sources with redshifts in the MUSE fields versus the F814W magnitude. The dashed vertical lines show the 90\% completeness limits for the distributions. We average the results of five fields with 2-3 hour exposures, leaving out PKS0552$-$640. This field is filled with foreground stars which makes pure galaxy identifications difficult in the \hst\ image.} \label{fig:redcomp}
\end{figure}

Next, we turn to the IFU observations to understand the completeness of our source identifications. For these calculations, we utilize the codebase written by M. Fumagalli available on GitHub\footnote{\url{https://github.com/mifumagalli/mypython/blob/master/ifu/muse\_mocks.py}} (see also \citealt{lofthouse2020}). We start by injecting simulated continuum sources with an exponential profile and constant flux across all wavelengths into the whitelight images and then run SE to determine the fraction of recovered injected sources. For the MUSE fields, we inject 100 sources with a FWHM of 5 pixels and scale length of 1 pixel for 5,000 iterations. Figure~\ref{fig:musecomp} displays the results of these continuum source injections. The dashed vertical lines mark the 90\% completeness limits of the distributions in AB magnitudes. We have again averaged the fields with 2-3 hour exposures. Given the smaller FOV of the KCWI observations, we inject 10 continuum sources per FOV with a FWHM of 1 pixel and scale length of 1.5 pixels. The number of iterations varies from 2,000 to 10,0000 depending on how many FOVs cover the field (we average the individual results together per field). We display the KCWI results in Figure~\ref{fig:kcwicomp}. We average the results for J1419+4207 and J1435+3604 because they have the same number of FOVs and exposure times. We also separate the J1619+3342 results for the observations at different exposure times. 

\begin{figure}
    \epsscale{1.2}
    \plotone{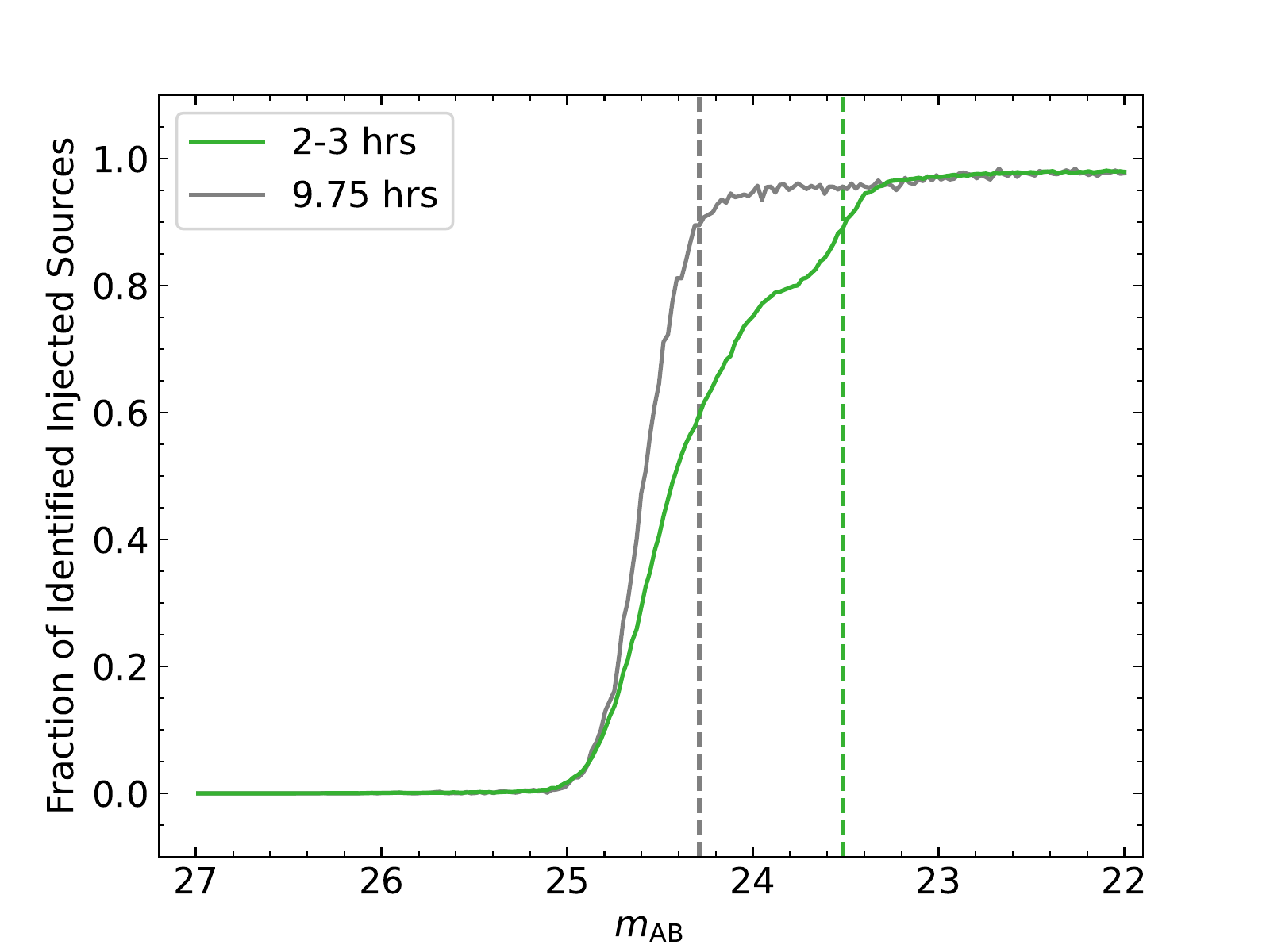}
    \caption{Fraction of simulated continuum sources detected versus magnitude for the MUSE fields. The dashed vertical lines show the 90\% completeness limits for the distributions. We average the results of all six fields with 2-3 hour exposures.} \label{fig:musecomp}
\end{figure}

\begin{figure}
    \epsscale{1.2}
    \plotone{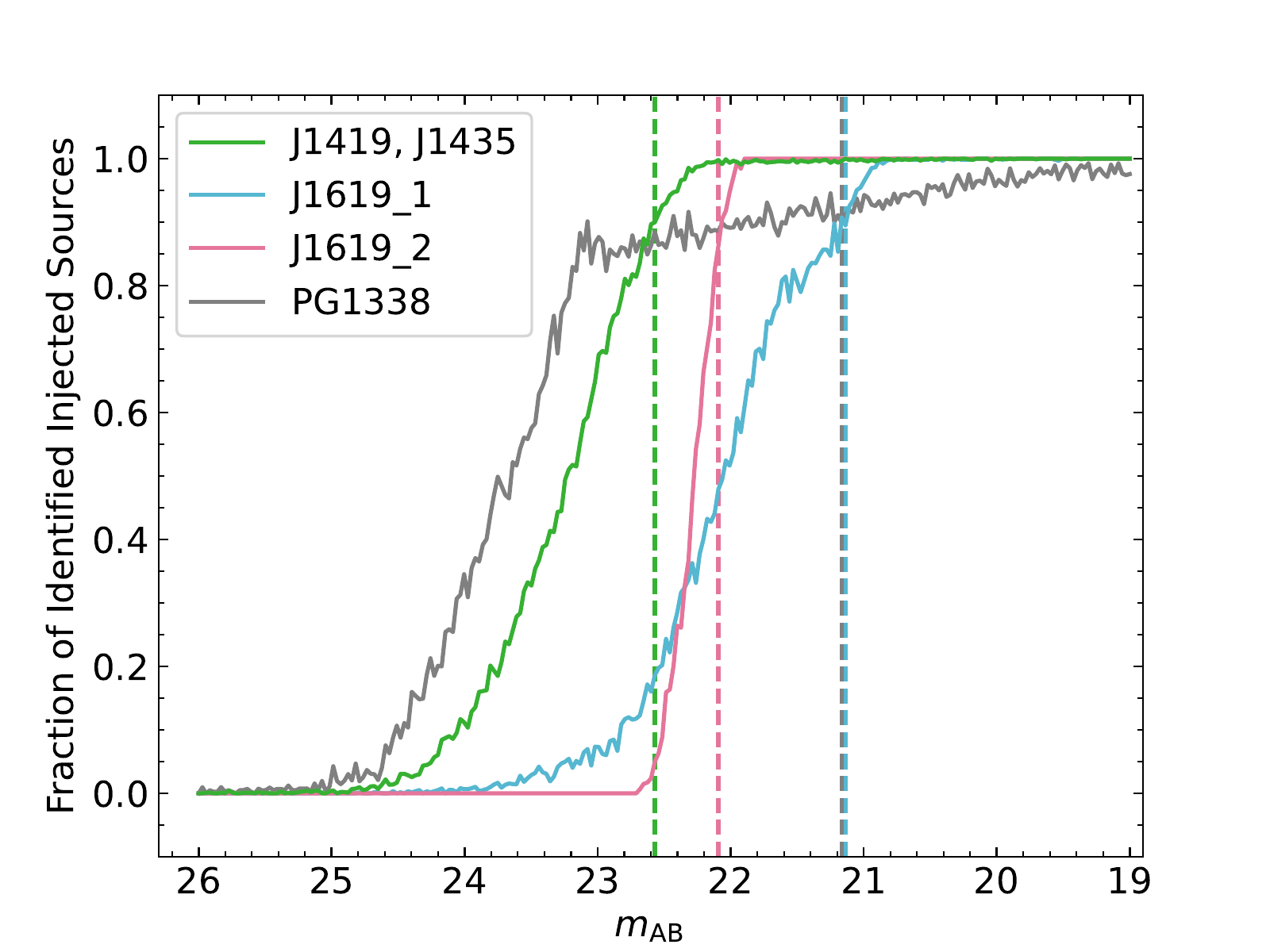}
    \caption{Fraction of simulated continuum sources detected versus magnitude for the KCWI fields. The dashed vertical lines show the 90\% completeness limits for the distributions. We average the results for J1419+4207 and J1435+3604 because they have the same number of observations and exposure times. The J1619+3342\_1 results are from the 900 second exposures, and the J1619+3342\_2 results are from the 1200 second exposures.} \label{fig:kcwicomp}
\end{figure}

Since we utilized \OII\ narrowband images for source identification in the KCWI observations, we also inject emission-line sources into the cubes, create whitelight images, and then run SE to determine the fraction of recovered injected emission-line sources in the images. We are only interested in detections at the wavelengths of interest, so we cut the cubes down to within $\pm$15\AA\ of the \OII\ line at the absorber redshift. The emission-line sources have a Gaussian profile in the spectral direction with a FWHM of 8 pixels and an exponential profile in the spatial direction. The number of sources, FWHM, scale length, and iterations are the same as with the continuum source injections. Figure~\ref{fig:kcwilinecomp} displays the results of the emission-line source injections. The dashed and dotted vertical lines mark the 90\% completeness limits of the distributions in flux. Almost all of the KCWI sightlines have two absorbers, so we plot the lower redshift absorbers in the left panel and the higher redshift absorbers in the right panel for clarity. To better understand these 90\% flux completeness limits, we calculate the corresponding 90\% SFR and stellar mass completeness values (see below) and report them in Table~\ref{tab:kcwi90fluxlimits}.

\begin{figure*}
    \epsscale{1.0}
    \plotone{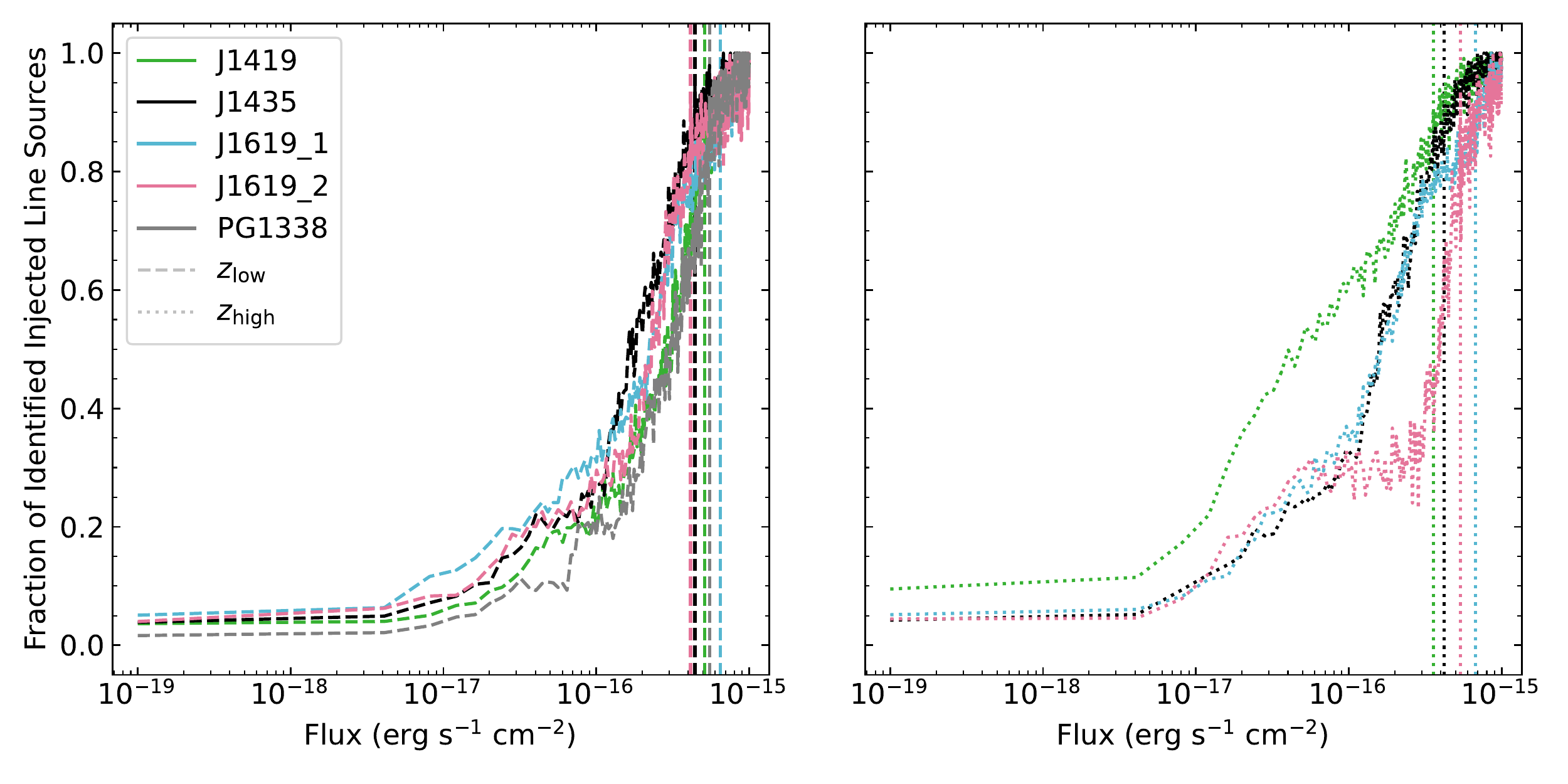}
    \caption{Fraction of simulated emission-line sources detected versus flux for the KCWI fields. The dashed and dotted vertical lines show the 90\% completeness limits for the distributions. The left panel displays the results for the lowest redshift absorber along the line of sight, and the right panel displays the results for the highest redshift absorber along the line of sight. The J1619+3342\_1 results are from the 900 second exposures, and the J1619+3342\_2 results are from the 1200 second exposures.} \label{fig:kcwilinecomp}
\end{figure*}

\begin{deluxetable*}{lcccc}
\tablecaption{KCWI 90\% Flux Limits \label{tab:kcwi90fluxlimits}}
\tablehead{\colhead{QSO Field} & \colhead{$z_{\rm abs}$} & \colhead{\OII\ Flux} & \colhead{SFR} & \colhead{$\log M_{\star}$} \\
\colhead{} & \colhead{} & \colhead{(erg s$^{-1}$ cm$^{-2}$)} & \colhead{(\msun\ yr$^{-1}$)} & \colhead{[\msun]}}
\startdata
PG1338+416 & 0.348827 & 5.50$\times 10^{-16}$ & 3.80 & 9.9 \\
J1419+4207 & 0.288976 & 5.10$\times 10^{-16}$ & 2.28 & 9.7 \\
J1419+4207 & 0.425592 & 3.57$\times 10^{-16}$ & 3.94 & 9.9 \\
J1435+3604 & 0.372981 & 4.42$\times 10^{-16}$ & 3.57 & 9.8 \\
J1435+3604 & 0.387594 & 4.22$\times 10^{-16}$ & 3.73 & 9.9 \\
J1619+3342\_1 & 0.269400 & 6.47$\times 10^{-16}$ & 2.46 & 9.7 \\
J1619+3342\_1 & 0.470800 & 6.75$\times 10^{-16}$ & 9.46 & 10.2 \\
J1619+3342\_2 & 0.269400 & 4.14$\times 10^{-16}$ & 1.58 & 9.5 \\
J1619+3342\_2 & 0.470800 & 5.38$\times 10^{-16}$ & 7.55 & 10.1 \\
\enddata
\tablecomments{We take the 90\% flux completeness levels for each sightline and treat them as an \OII\ flux to calculate a corresponding 90\% SFR and $\log M_{\star}$ completeness level at the absorber redshift. The J1619+3342\_1 results are from the 900 second exposures, and the J1619+3342\_2 results are from the 1200 second exposures.}
\end{deluxetable*}

Finally, we estimate the 3$\sigma$ SFR and stellar mass upper limits in each field that we have been able to detect down to for associated galaxy identification at the redshift of each absorber. We extract a region of background (identical to the regions used for the galaxies) and measure the dispersion in the residual flux at the appropriate wavelengths as an estimate of the 1$\sigma$ uncertainty. Our line flux limits are given at the 3$\sigma$ level, which we convert to SFRs as described above. To determine the appropriate SFR calibration, we adopt the average Vega $B$-band magnitude of $-$19.43 from the most common probable host galaxies (see below) in our sample between stellar masses of \logMstar\ = 9.0 to 10.0.

From the limiting SFRs, we calculate an equivalent stellar mass from the SFR-M$_{\star}$ parameterization derived in \citet{schreiber2015}. We use their Equation 9 which takes redshift into account. By using this conversion, we are inherently assuming the galaxy to be star-forming. If it were a passive galaxy, the stellar mass limits would be higher. However, we argue it is unlikely that a massive, quiescent galaxy within the IFU FOVs would be missed by SE, given that these galaxies tend to be quite luminous. It is possible we are missing redder, low-mass galaxies in the fields. The lowest mass absorption-line dominated associated galaxy we detect is at $\logMstar$ = 9.6. If we take this as our upper limit for the stellar mass limits, it would not change our finding of low-metallicity absorbers not associated with the halos of $\sim$$L^*$ galaxies, but the possibility of the absorbers being associated with lower-mass galaxies below our detection limits instead of overdense regions of the universe would become more likely (see the following sections for more details). The 3$\sigma$ SFR and stellar mass limits we calculate for each absorber are summarized in Table~\ref{tab:fieldlimits}. 

\begin{deluxetable*}{lcccc}
\tablecaption{Field Detection Limits \label{tab:fieldlimits}}
\tablehead{\colhead{QSO Field} & \colhead{$z_{\rm abs}$} & \colhead{SFR} & \colhead{$\log M_{\star}$}  & \colhead{FOV} \\
\colhead{} & \colhead{} & \colhead{(\msun\ yr$^{-1}$)} & \colhead{[\msun]}  & \colhead{(kpc)}}
\startdata
HE0153$-$4520 & 0.225958 & 0.01 & 7.3 & 224 $\times$ 224 \\
PHL1377 & 0.322464 & 0.02 & 7.6 & 290 $\times$ 290 \\
PHL1377 & 0.738890 & 0.08 & 8.1 & 451 $\times$ 451 \\
PKS0405$-$123 & 0.167160 & 0.01 & 7.5 & 177 $\times$ 177 \\
HE0439$-$5254 & 0.614962 & 0.02 & 7.5 & 418 $\times$ 418 \\
PKS0552$-$640 & 0.345149 & 0.29 & 8.8 & 303 $\times$ 303 \\
HE1003+0149 & 0.418522 & 0.05 & 8.0 & 341 $\times$ 341 \\
HE1003+0149 & 0.836989 & 0.14 & 8.3 & 470 $\times$ 470 \\
HE1003+0149 & 0.837390 & 0.22 & 8.5 & 470 $\times$ 470 \\
HE1003+0149 & 0.839400 & 0.13 & 8.2 & 471 $\times$ 471 \\
PG1338+416 & 0.348827 & 0.58 & 9.1 & 168 $\times$ 104 \\
J1419+4207 & 0.288976 & 0.51 & 9.0 & 148 $\times$ 91 \\
J1419+4207 & 0.425592 & 0.87 & 9.2 & 190 $\times$ 117 \\
J1435+3604 & 0.372981 & 0.38 & 8.9 & 175 $\times$ 108 \\
J1435+3604 & 0.387594 & 0.38 & 8.9 & 179 $\times$ 111 \\
PG1522+101 & 0.518500 & 0.01 & 6.9 & 385 $\times$ 385 \\
PG1522+101 & 0.728885 & 0.02 & 7.5 & 448 $\times$ 448 \\
J1619+3342 & 0.269400 & 0.12 & 8.4 & 141 $\times$ 87 \\
J1619+3342 & 0.470800 & 0.36 & 8.8 & 201 $\times$ 124 \\
\enddata
\tablecomments{The SFR and $\log M_{\star}$ values are the 3$\sigma$ limits calculated from the background residual flux. We assume the galaxy to be star-forming when determining the stellar mass limits. The final column displays the dimensions of one instrument field of view at the redshift of the absorber.}
\end{deluxetable*}

\subsection{Galaxy Selection Criteria and Group Environments}\label{sec:groups}

When determining if a galaxy in the field could be associated with an absorber, we take into account the velocity offset ($|\Delta v|$) between the absorber and the galaxy redshifts, the normalized impact parameter ($\rho/\Rvir$), and the normalized velocity offset ($|\Delta v|/\vesc$). The details of these parameters are described below. The velocity offset between a galaxy and absorber is calculated using the standard ${\Delta}v=c(z_{\rm gal}-\zabs)/(1+\zabs)$. The absorber redshifts come from CCC \citep{lehner2018}. We make an initial list of candidates within $|{\Delta}v| <$ 1,000\,\kms\ of the absorber redshift and characterize them as described above. We use this conservative velocity cut to ensure any potentially associated galaxies are retained in case the absorbers are part of an outflow or housed within a high-mass halo. We then instate a second cut based on the proximity of the absorber in distance and velocity space to the candidate galaxy. Recent results have characterized the extent of the CGM around M31 \citep{lehner2020} and an ensemble of galaxies covering a similar mass range to our sample \citep{wilde2021}. Both of these surveys show the CGM can extend farther than the virial radius of a galaxy, \Rvir. Additionally, the majority of the absorbers identified in the COS-Halos survey \citep{tumlinson2013} fall within \vesc\ for their given halo (see their Figure 10). With these results in mind, we keep candidate galaxies with $\rho/\Rvir \le$ 1.5 and $|\Delta v|/\vesc \le$ 1.5 relative to the absorber for our final sample. We refer to these objects as associated galaxies throughout the paper. We are again applying a conservative cut so that we have an accurate picture of the available galaxies in the fields that could potentially be associated with the absorbers. This second cut removes \ifucutgalsample\ candidate galaxies from the total sample (see Appendix~\ref{sec:appa} for the list of cut galaxies). 

As other galaxy-absorber surveys have shown (e.g., \citealt{burchett2016,chen2019,manuwal2019,chen2020,hamanowicz2020}), it is possible for multiple galaxies to be associated with an absorber. For the proximate absorber along J1619+3342, we have clear evidence that the associated galaxies are within the QSO galaxy group, but the distinction of a galaxy group is much harder to determine when only two or three associated galaxies are detected. We note that a galaxy survey out to 5 Mpc in the field of HE1003+0149 for the absorbers at $z \approx 0.837$ has recently been conducted by \citet{narayanan2021}. They find 21 other galaxies in the field within $\pm$1,000 \kms\ of the absorbers. The CASBaH program \citep[][and references therein]{haislmaier2021} includes the PHL1377, PG1338+416, and PG1522+101 fields, where they have taken spectra of galaxies within 10 comoving Mpc of the QSO down to a median $\logMstar \approx 10.1$ \citep{prochaska2019}. We search the DR1 database\footnote{Available through \url{https://specdb.readthedocs.io/en/latest/casbah.html}} for galaxy matches within $\pm$1,000 \kms\ of our absorber redshifts, but do not find any close matches. No galaxies are identified for the PG1338+416 absorber or the PG1522+101 \zabs\ = 0.728885 absorber, two galaxies with $\rho >$ 2,400 kpc are identified for the PHL1377 \zabs\ = 0.322464 absorber, one galaxy with $\rho >$ 3,000 kpc is identified for the PHL1377 \zabs\ = 0.738890 absorber, and 13 galaxies with $\rho > 700$ kpc are identified for the PG1522+101 \zabs\ = 0.518500 absorber. Given the number of galaxies located around the PG1522+101 sightline and near to the absorber redshift, this absorber could potentially be in a galaxy group as well. We also search within the Hectospec catalog (J. Burchett 2018 private communication) for galaxy matches within $\pm$1,000 \kms\ for the absorbers along the J1619+3342 sightline. Four galaxies with $\rho >$ 400 kpc are identified for the \zabs\ = 0.470800 absorber, and 26 galaxies with $\rho >$ 200 kpc are identified for the \zabs\ = 0.269400 absorber. The closest galaxy ($\approx$0.6$L^*$) lies just outside of the 200 kpc region marked in Figure~\ref{fig:J1619} on the left side. Using this value, we determine it would not meet the selection criteria listed above. Instead this absorber seems to be located on the edge of a galaxy group. For the remaining fields, it is possible they show a similar excess of galaxies at the redshifts of the absorbers, though we have not undertaken observations to confirm this. 

In the following sections, we report all galaxies that are associated with an absorber (i.e., those that meet the two requirements stated above) and designate the most probable host galaxy (MPG) for an absorber. This is done by determining which associated galaxy has the value of ${\rm MPG}={\rm min}\left(\sqrt{(\rho/\Rvir)^2+(|\Delta v|/\vesc)^2}\right)$ out of all of the associated galaxies for that absorber. In the case of the QSO galaxy group, we do not label the QSO host galaxy as the MPG because it is unclear if this absorber is actually associated with the QSO host galaxy. 

\section{The Galaxy Properties of the \HI-selected Absorbers}\label{sec:results}

The \HI\ selection of our absorbers and their well-characterized metallicities allow us to investigate the origins of these absorbers. The metal-enriched absorbers may have a higher likelihood of being associated with galaxies. The low-metallicity systems could be associated with the IGM or the CGM. (Systems with [X/H] $< -2$ imply no redshift evolution in their metal content since $z \approx 2$--3, since a majority of the MDF for pLLSs/LLSs lies below this value at these redshifts; \citealt{lehner2016,lehner2019,lehner2022}.) Here we present the galaxy associations we find for each absorber and an overview of the associated galaxy sample characteristics. There are no apparent correlations between the absorber column densities and host galaxy properties discussed in this section.

\subsection{Galaxy Association}\label{sec:galassoc}

Absorbers in the pLLS and LLS column density regimes represent baryon overdensities around $\delta \equiv \rho/(\Omega_b\rho_c) \sim 10^2$--$10^3$ at $z < 1$ (see \citealt{schaye2001,wotta2016}). These overlap the typical overdensities of gaseous galaxy halos seen in simulations (e.g., \citealt{wiersma2009,wiersma2011}). Additionally, there are strong correlations between these absorbers and gaseous galaxy halos (e.g., \citealt{lanzetta1995,prochaska2011,tejos2014}). A principle goal of our work is to investigate whether there are differences in the galaxies that are associated with low-metallicity and metal-enriched absorbers. 

After instating our two cuts to the full galaxy sample in all \ifuqsosample\ fields ($|{\Delta}v| <$ 1,000 \kms\ and keeping galaxies with $\rho/\Rvir$ and $|\Delta v|/\vesc$ $\le$ 1.5 with respect to the absorber; see \S~\ref{sec:groups}), we are left with a total of \ifugalsample\ unique galaxies associated with our \ifuabssample\ absorbers (all pLLSs). The derived properties for the associated galaxies of all the absorbers are reported in Table~\ref{tab:galinfo}, and the galaxy location information is in Table~\ref{tab:galinfoapp}. Additionally, we report the derived properties for the \ifucutgalsample\ cut candidate galaxies in Table~\ref{tab:galinfo_other} for the interested reader. Due to the small offsets in the redshifts of the absorbers along the HE1003+0149 sightline, the same galaxy is associated with the absorbers at $\zabs = 0.836989$ and $\zabs = 0.837390$. Additionally, the host of the QSO J1619+3342 is included as an associated galaxy for the proximate absorber found in the QSO galaxy group. This brings the full sample to 25 associated galaxies. We note that the properties of the host of the QSO J1619+3342 have large uncertainties (including its precise velocity); for this reason we do not include the QSO host as a candidate for association with this absorber.

In Figure~\ref{fig:hosts} we display the frequency distribution of the number of associated galaxies per absorber. We do not identify any associated galaxies for \ifunonsamplet\ of the \ifuabssample\ absorbers. When galaxies meeting our criteria are detected, we find on average \hostsperabs\ galaxy that can be associated with an absorber. The absorber with seven associated galaxies is the proximate absorber toward J1619+3342 located in the QSO galaxy group, with the QSO host included in this count.

\begin{figure}
    \epsscale{1.17}
    \plotone{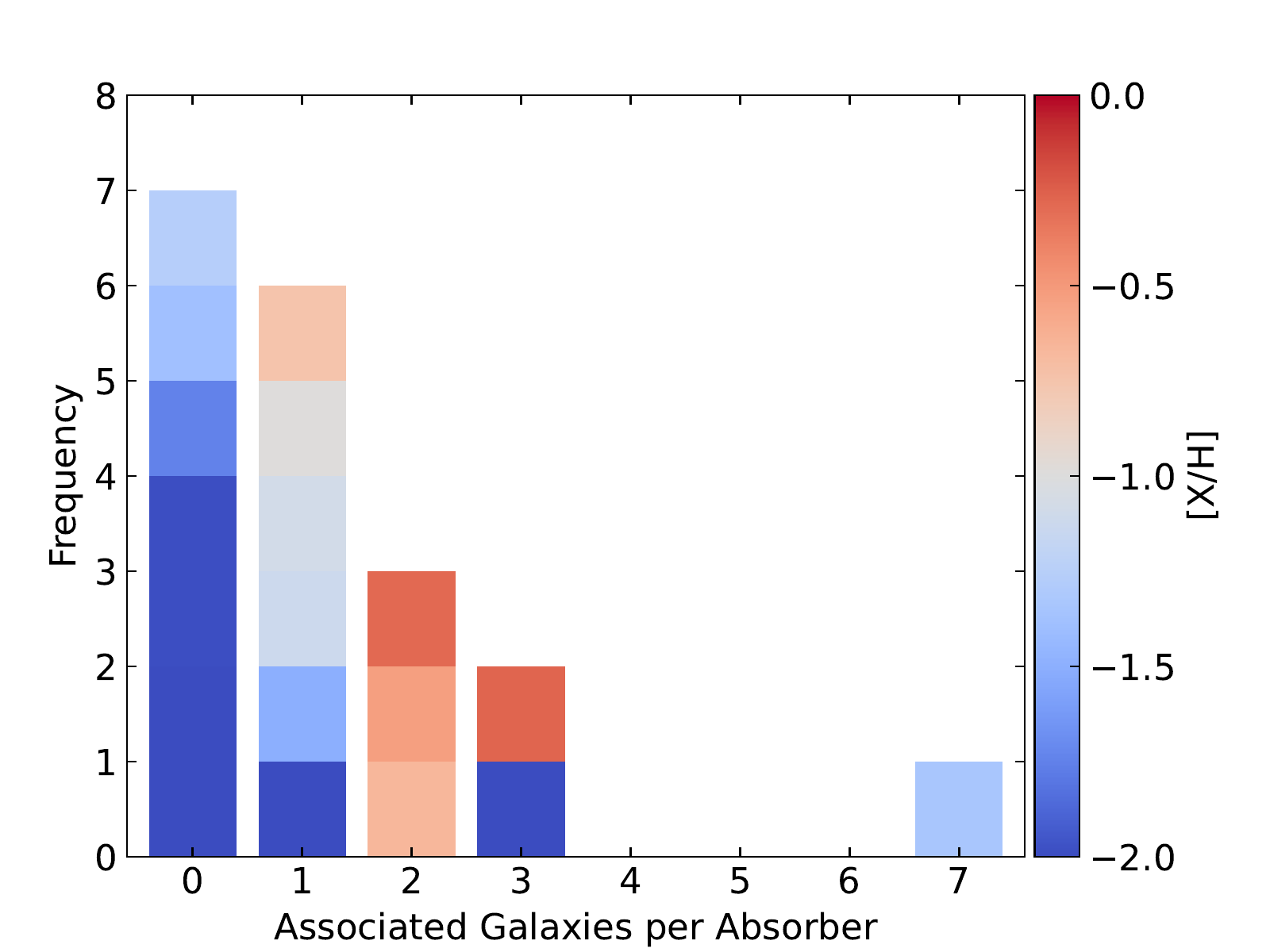}
    \caption{Number of associated galaxies identified per absorber in the IFU observations. The bars are colored by the absorber metallicity. The color bar extends only to $-$2.0 to keep the bar symmetric around $-1.0$ dex, but we include all absorbers below this value in the dark blue coloring. The absorber with seven associated galaxies is a proximate absorber toward J1619+3342 found in the QSO galaxy group; all the others are intervening absorbers. The absorbers with no associated galaxy detections all have [X/H] $< -1$.} \label{fig:hosts}
\end{figure}

There is no clear correlation between the absorber metallicity and the number of associated galaxies. However, there is a metallicity distinction: all of the absorbers with no associated galaxy detections have [X/H] $< -1$. The absorbers without associated galaxies are along the sightlines J1419+4207 at $\zabs = 0.425592$, J1435+3604 at $\zabs = 0.372981$, J1619+3342 at $\zabs = 0.269400$, HE1003+0149 at $\zabs = 0.839400$, PHL1377 at $\zabs = 0.322464$ and $\zabs = 0.738890$, and PG1522+101 at $\zabs = 0.728885$. The first three of these systems are in KCWI fields, while the last four are in MUSE fields. With the IFU observations, we attempted to reach field coverage to a radius of $\rho \le$ 200 kpc centered on the QSO at the redshifts of the absorbers. In some fields (and depending on the redshift of the absorber) we were successful, while in others we did not uniformly sample this range. In Table~\ref{tab:fieldlimits} we list the dimensions for one instrument field of view at the redshift of the absorber, and in Appendix~\ref{sec:appa} we display the IFU pointings on the \hst/ACS images of the fields with circles of radius 200 kpc at the absorber redshifts. For the HE1003+0149 absorber, the PG1522+101 absorber, and the higher redshift PHL1377 absorber, our observations cover this area completely. With the pointings of J1419+4207 and J1435+3604, we estimate our coverage to be 90\% and 85\%, respectively. There are a few galaxies that fall outside of our KCWI pointings, but the majority of the remaining area is empty space in \hst\ imagery. The low redshift absorber along the PHL1377 sightline has 65\% coverage of this area because this radius extends farther than the MUSE field of view. However, if we look at the sections of the sky that are not covered, only the northern arc has a substantial number of galaxies that were not observed. For J1619+3342, we have $<$50\% coverage of this area at the redshift of the absorber. Though there are portions of this field where we do not see any galaxies, there are still several in the field that have not been ruled out as an associated galaxy for the absorber. 

In addition to the distances covered by the observations, we also need to take into account the depth of the observations. All of the MUSE fields were observed for at least two hours, if not longer. The HE1003+0149 and PG1522+101 fields were observed for two hours, while the PHL1377 field was observed for three hours. It is then quite surprising that with the field coverage and depth of these observations no associated galaxies were detected for the four absorbers found along these sightlines. (See Appendix~\ref{sec:appa} for a full description of the pointings and Table~\ref{tab:fieldinfo} for the observing times.)

To quantify the limits of our observations, we determine SFR and stellar mass upper limits for all of the fields as discussed in \S~\ref{sec:limits} and summarized in Table~\ref{tab:fieldlimits}. These field limits are very constraining for the absorbers without associated galaxies, reaching stellar mass limits of $\logMstar < 9.2$ for star-forming galaxies. Outside of the absorber at $\zabs = 0.269400$ toward J1619+3342, it seems unlikely more observations within 200 kpc would yield detectable associated galaxies with $\logMstar \gtrsim 9.0$ due to the field limits and field coverage already achieved by our current observations. However, there are certainly dwarf galaxies below our detection limits that could be associated with the absorbers \citep{rs2014}, or there could be large-scale structures beyond 200 kpc (see \S~\ref{sec:groups}). Additionally, there may be galaxies associated with the absorbers that lie outside of the observed area, as in the case of the associated galaxies for the PKS0405$-$123 absorber. If an absorber is associated with a higher-mass halo, the virial radius of the galaxy would extend to much larger distances than 200 kpc, and even an $L^*$ galaxy could be missed with our observations at lower redshift where the absorber would be associated with the outer halo. 

\startlongtable
\begin{deluxetable*}{lccccccccccccc}
\tabletypesize{\footnotesize}
\tabcolsep=2pt
\tablecaption{Associated Galaxy Information \label{tab:galinfo}}
\tablehead{\colhead{SE\#} & \colhead{$\rho$} & \colhead{$|\Delta v|$} & \colhead{$\log M_{\star}$} & \colhead{$\log M_{\rm h}$} & \colhead{$R_{\rm vir}$} & \colhead{$v_{\rm esc}$} & \colhead{$M_{F814W}$} & \colhead{SFR} & \colhead{$\log {\rm sSFR}$} & \colhead{[O/H]\tablenotemark{a}} & \colhead{$\Phi$} & \colhead{$i$\tablenotemark{b}} & \colhead{MPG\tablenotemark{c}} \\
\colhead{} &\colhead{(kpc)} &\colhead{(\kms)} &\colhead{[\msun]} &\colhead{[\msun]} &\colhead{(kpc)} &\colhead{(\kms)} &\colhead{(mag)} &\colhead{(\msun\ yr$^{-1}$)} &\colhead{[yr$^{-1}$]} &\colhead{} &\colhead{(deg)} &\colhead{(deg)} &\colhead{}}
\startdata
\hline
\multicolumn{14}{c}{HE0439$-$5254, \hspace{0.5mm} $z_{\rm abs}$=0.614962, \hspace{0.5mm} \logNHI=$16.20 \pm 0.01$, \hspace{0.5mm} [X/H]=$-0.27 \pm 0.02$} \\
\hline
1009 & 131 & $22 \pm 7$ & 9.8 & 11.6 & 126 & 224 & $-$19.8 & $0.68 \pm 0.03$ & $-$9.98 & $0.00 \pm 0.18$ & $6.6 \pm 3.9$ & 44 & 0 \\
1083 & 48 & $27 \pm 1$ & 9.4 & 11.4 & 108 & 192 & $-$20.0 & $4.68 \pm 0.03$ & $-$8.75 & $-0.07 \pm 0.18$ & $48.5 \pm 1.4$ & 68 & 1 \\
1309 & 108 & $49 \pm 1$ & 9.5 & 11.5 & 112 & 199 & $-$20.0 & $6.95 \pm 0.05$ & $-$8.67 & $-0.14 \pm 0.18$ & $67.5 \pm 0.3$ & 81 & 0 \\
\hline
\multicolumn{14}{c}{PKS0405$-$123, \hspace{0.5mm} $z_{\rm abs}$=0.167160, \hspace{0.5mm} \logNHI=$16.45 \pm 0.05$, \hspace{0.5mm} [X/H]=$-0.29 \pm 0.04$} \\
\hline
1822 & 117 & $51 \pm 10$ & 10.3 & 11.9 & 183 & 248 & $-$22.2 & $1.69 \pm 0.23$ & $-$10.10 & $0.01 \pm 0.18$ & $15.8 \pm 1.6$ & 46 & 1 \\
3207 & 101 & $51 \pm 3$ & 8.5 & 10.9 & 87 & 115 & $-$18.7 & $0.73 \pm 0.06$ & $-$8.63 & $-0.13 \pm 0.18$ & $73.1 \pm 0.6$ & 66 & 0 \\
\hline
\multicolumn{14}{c}{J1419+4207, \hspace{0.5mm} $z_{\rm abs}$=0.288976, \hspace{0.5mm} \logNHI=$16.35 \pm 0.05$, \hspace{0.5mm} [X/H]=$-0.53^{+0.19}_{-0.11}$} \\
\hline
1003 & 54 & $99 \pm 23$ & 9.8 & 11.6 & 135 & 195 & $-$19.8 & $0.37 \pm 0.04$ & $-$10.19 & \nodata & $77.6 \pm 0.3$ & 80 & 1 \\
1163 & 110 & $145 \pm 23$ & 8.7 & 11.0 & 90 & 128 & $-$16.1 & $0.03 \pm 0.01$ & $-$10.22 & \nodata & \nodata & 44 & 0 \\
\hline
\multicolumn{14}{c}{PG1522+101, \hspace{0.5mm} $z_{\rm abs}$=0.518500, \hspace{0.5mm} \logNHI=$16.22 \pm 0.02$, \hspace{0.5mm} [X/H]=$-0.66 \pm 0.03$} \\
\hline
1077\tablenotemark{d} & 171 & $39 \pm 39$ & 10.0 & 11.7 & 139 & 232 & $-$19.1 & $< 0.03$ & $-$11.47 & \nodata & $6.7 \pm 2.4$ & 53 & 1 \\
1151 & 182 & $310 \pm 2$ & 9.9 & 11.7 & 136 & 227 & $-$19.9 & $1.53 \pm 0.03$ & $-$9.72 & $-0.01 \pm 0.18$ & $83.5 \pm 1.9$ & 64 & 0 \\
\hline
\multicolumn{14}{c}{PG1338+416, \hspace{0.5mm} $z_{\rm abs}$=0.348827, \hspace{0.5mm} \logNHI=$16.34 \pm 0.01$, \hspace{0.5mm} [X/H]=$-0.74 \pm 0.02$} \\
\hline
1118 & 95 & $96 \pm 7$ & 9.9 & 11.6 & 138 & 207 & $-$20.1 & $3.04 \pm 0.07$ & $-$9.37 & \nodata & $10.4 \pm 1.2$ & 61 & 1 \\
\hline
\multicolumn{14}{c}{HE0153$-$4520, \hspace{0.5mm} $z_{\rm abs}$=0.225958, \hspace{0.5mm} \logNHI=$16.71 \pm 0.07$, \hspace{0.5mm} [X/H]=$-0.99 \pm 0.07$} \\
\hline
966 & 88 & $100 \pm 1$ & 10.3 & 11.9 & 183 & 256 & $-$22.2 & $0.04 \pm 0.03$ & $-$11.75 & \nodata & $83.1 \pm 0.4$ & 47 & 1 \\
\hline
\multicolumn{14}{c}{HE1003+0149, \hspace{0.5mm} $z_{\rm abs}$= 0.418522, \hspace{0.5mm} \logNHI=$16.89 \pm 0.04$, \hspace{0.5mm} [X/H]=$-1.08 \pm 0.06$} \\
\hline
974 & 125 & $55.4 \pm 0.4$ & 9.0 & 11.2 & 100 & 156 & $-$18.7 & $0.65 \pm 0.01$ & $-$9.20 & $-0.31 \pm 0.18$ & $76.0 \pm 2.4$ & 65 & 1 \\
\hline
\multicolumn{14}{c}{J1435+3604, \hspace{0.5mm} $z_{\rm abs}$=0.387594, \hspace{0.5mm} \logNHI=$16.18 \pm 0.06$, \hspace{0.5mm} [X/H]=$-1.12 \pm 0.11$} \\
\hline
1060\tablenotemark{e} & 116 & $20 \pm 65$ & 9.7 & 11.6 & 133 & 204 & $-$20.0 & $0.60 \pm 0.12$ & $-$9.95 & \nodata & $71.2 \pm 0.3$ & 80 & 1 \\
\hline
\multicolumn{14}{c}{PHL1377, \hspace{0.5mm} $z_{\rm abs}$=0.738890, \hspace{0.5mm} \logNHI=$16.67 \pm 0.01$, \hspace{0.5mm} [X/H]=$-1.26 \pm 0.04$} \\
\hline
\nodata & \nodata & \nodata & $< 8.1$ & \nodata & \nodata & \nodata & \nodata & $< 0.08$ & \nodata & \nodata & \nodata & \nodata & \nodata \\
\hline
\multicolumn{14}{c}{J1619+3342, \hspace{0.5mm} $z_{\rm abs}$=0.470800, \hspace{0.5mm} \logNHI=$16.64 \pm 0.02$, \hspace{0.5mm} [X/H]=$-1.34 \pm 0.03$} \\
\hline
903\tablenotemark{f} & 0 & $163 \pm 71$ & 11.7 & 13.8 & 723 & 1237 & $-$24.9 & $73.83 \pm 1.01$ & $-$9.82 & \nodata & \nodata & 28 & 0 \\
946 & 74 & $57 \pm 14$ & 10.9 & 12.5 & 260 & 430 & $-$21.2 & $12.16 \pm 0.48$ & $-$9.78 & \nodata & $14.5 \pm 0.1$ & 81 & 1 \\
1006 & 49 & $204 \pm 20$ & 9.4 & 11.4 & 112 & 180 & $-$20.0 & $1.13 \pm 0.16$ & $-$9.30 & \nodata & \nodata & 22 & 0 \\
1028 & 51 & $41 \pm 20$ & 10.2 & 11.8 & 157 & 256 & $-$19.5 & $0.46 \pm 0.05$ & $-$10.53 & \nodata & $47.3 \pm 7.6$ & 38 & 0 \\
1137 & 140 & $82 \pm 122$ & 9.5 & 11.5 & 119 & 192 & $-$19.4 & $0.13 \pm 0.05$ & $-$10.41 & \nodata & $61.8 \pm 2.3$ & 61 & 0 \\
1179 & 92 & $122 \pm 6$ & 9.8 & 11.6 & 133 & 216 & $-$19.6 & $6.13 \pm 0.14$ & $-$9.02 & \nodata & $73.1 \pm 7.3$ & 40 & 0 \\
1180 & 62 & $285 \pm 18$ & 10.7 & 12.3 & 217 & 356 & $-$20.8 & $1.73 \pm 0.23$ & $-$10.44 & \nodata & $8.3 \pm 1.1$ & 58 & 0 \\
\hline
\multicolumn{14}{c}{J1419+4207, \hspace{0.5mm} $z_{\rm abs}$=0.425592,  \hspace{0.5mm} \logNHI=$16.17 \pm 0.05$, \hspace{0.5mm} [X/H]=$-1.38 \pm 0.20$} \\
\hline
\nodata & \nodata & \nodata & $< 9.2$ & \nodata & \nodata & \nodata & \nodata & $< 0.87$ & \nodata & \nodata & \nodata & \nodata & \nodata \\
\hline
\multicolumn{14}{c}{HE1003+0149, \hspace{0.5mm} $z_{\rm abs}$=0.836989, \hspace{0.5mm} \logNHI=$16.52 \pm 0.02$, \hspace{0.5mm} [X/H]=$-1.51 \pm 0.09$} \\
\hline
1001\tablenotemark{g} & 157 & $171 \pm 2$ & 10.4 & 12.0 & 155 & 321 & $-$18.8 & $1.03 \pm 0.03$ & $-$10.39 & $-0.16 \pm 0.19$ & $56.9 \pm 4.1$ & 59 & 1 \\
\hline
\multicolumn{14}{c}{HE1003+0149, \hspace{0.5mm} $z_{\rm abs}$=0.839400, \hspace{0.5mm} \logNHI=$16.13 \pm 0.01$, \hspace{0.5mm} [X/H] $<-1.74$} \\
\hline
\nodata & \nodata & \nodata & $< 8.2$ & \nodata & \nodata & \nodata & \nodata & $< 0.13$ & \nodata & \nodata & \nodata & \nodata & \nodata \\
\hline
\multicolumn{14}{c}{J1435+3604, \hspace{0.5mm} $z_{\rm abs}$=0.372981, \hspace{0.5mm} \logNHI=$16.68 \pm 0.05$, \hspace{0.5mm} [X/H]=$-1.98^{+0.17}_{-0.15}$} \\
\hline
\nodata & \nodata & \nodata & $< 8.9$ & \nodata & \nodata & \nodata & \nodata & $< 0.38$ & \nodata & \nodata & \nodata & \nodata & \nodata \\
\hline
\multicolumn{14}{c}{J1619+3342, \hspace{0.5mm} $z_{\rm abs}$=0.269400, \hspace{0.5mm} \logNHI=$16.59 \pm 0.03$, \hspace{0.5mm} [X/H]=$-1.99^{+0.18}_{-0.13}$} \\
\hline
\nodata & \nodata & \nodata & $< 8.4$ & \nodata & \nodata & \nodata & \nodata & $< 0.12$ & \nodata & \nodata & \nodata & \nodata & \nodata \\
\hline
\multicolumn{14}{c}{HE1003+0149, \hspace{0.5mm} $z_{\rm abs}$=0.837390, \hspace{0.5mm} \logNHI=$16.36 \pm 0.02$, \hspace{0.5mm} [X/H]=$-2.19^{+0.15}_{-0.18}$} \\
\hline
1001\tablenotemark{g} & 157 & $237 \pm 2$ & 10.4 & 12.0 & 155 & 321 & $-$18.8 & $1.03 \pm 0.03$ & $-$10.39 & $-0.16 \pm 0.19$ & $56.9 \pm 4.1$ & 59 & 1 \\
\hline
\multicolumn{14}{c}{PHL1377, \hspace{0.5mm} $z_{\rm abs}$=0.322464, \hspace{0.5mm} \logNHI=$16.07 \pm 0.01$, \hspace{0.5mm} [X/H]=$-2.43^{+0.52}_{-0.51}$} \\
\hline
\nodata & \nodata & \nodata & $< 7.6$ & \nodata & \nodata & \nodata & \nodata & $< 0.02$ & \nodata & \nodata & \nodata & \nodata & \nodata \\
\hline
\multicolumn{14}{c}{PKS0552$-$640, \hspace{0.5mm} $z_{\rm abs}$=0.345149, \hspace{0.5mm} \logNHI=$17.02 \pm 0.03$, \hspace{0.5mm} [X/H]=$-2.83^{+0.50}_{-0.56}$} \\
\hline
5939 & 138 & $310 \pm 4$ & 10.3 & 11.9 & 167 & 251 & $-$20.5 & $0.07 \pm 0.01$ & $-$11.40 & \nodata & $58.2 \pm 0.2$ & 65 & 0 \\
6264 & 84 & $276 \pm 7$ & 9.6 & 11.5 & 126 & 188 & $-$19.4 & $< 0.01$ & $-$11.61 & \nodata & \nodata & 24 & 0 \\
78334\tablenotemark{d} & 78 & $76 \pm 2$ & 9.2 & 11.3 & 108 & 161 & $-$18.3 & $0.031 \pm 0.003$ & $-$10.72 & $-0.21^{+0.18}_{-0.19}$ & \nodata & \nodata & 1 \\
\hline
\multicolumn{14}{c}{PG1522+101, \hspace{0.5mm} $z_{\rm abs}$=0.728885, \hspace{0.5mm} \logNHI=$16.63 \pm 0.05$, \hspace{0.5mm} [X/H]=$-2.92 \pm 0.05$} \\
\hline
\nodata & \nodata & \nodata & $< 7.5$ & \nodata & \nodata & \nodata & \nodata & $< 0.02$ & \nodata & \nodata & \nodata & \nodata & \nodata \\
\enddata
\tablecomments{The above absorber column density and metallicity values are taken from the COS CGM Compendium \citep{lehner2018,lehner2019}. The error bars on the median absorber and galaxy metallicity represent the 68\% confidence interval. The galaxy associated with the HE1003+0149 absorbers at $z_{\rm abs}$=0.836989 and $z_{\rm abs}$=0.837390 is the same. We list the stellar mass and SFR 3$\sigma$ limits for absorbers without associated galaxies. Two other associated galaxies at $\rho >$ 200 kpc have been identified for the HE1003+0149 absorber at $z_{\rm abs}$=0.418522 (S. D. Johnson 2022, private communication).}\tablenotetext{a}{The galaxy metallicity reported here is [O/H] = $\epsilon$(O) $-$ 8.69.}\tablenotetext{b}{The inclination values have errors around 5 to 10 degrees.}\tablenotetext{c}{The most probable host galaxy is designated by the lowest value of $\sqrt{(\rho/R_{\rm vir})^2 + (|\Delta v|/v_{\rm esc})^2}$.}\tablenotetext{d}{The stellar mass of this galaxy was calculated using synthetic magnitudes.}\tablenotetext{e}{The stellar mass of this galaxy was calculated using an M/L ratio.}\tablenotetext{f}{The redshift for this QSO is reported in \citet{hw2010}. The characteristics for this galaxy are uncertain.}\tablenotetext{g}{This mass estimate is in conflict with the value reported in \citet{narayanan2021} and is likely due to the differences in the assumed star formation history. We independently ran {\tt kcorrect} and {\tt eazy-py} (\citealt{brammer2008}, \url{https://github.com/gbrammer/eazy-py}) with photometry from the Hyper Suprime-Cam Subaru Strategic Program Data Release 3 \citep{aihara2022} and obtained a mass estimate of \logMstar\ = 9.8 and \logMstar\ = 9.9, respectively. Our measurement is consistent within the expected errors using this method.}
\end{deluxetable*}

In Figures~\ref{fig:nhinorm} and \ref{fig:metmorphology} we show the full associated galaxy sample in distance and velocity space with respect to the absorber column density and metallicity. We connect the galaxy data points that are associated with the same absorber, and we mark the most probable host galaxy for the absorber as defined in \S~\ref{sec:groups}. On these plots we include the candidate galaxies that were removed from the sample in our second cut ($\times$ symbols). The majority of the most probable host galaxies are well within $\Rvir$ and $\vesc$ from the absorber. We find no significant correlation between the absorber metallicity and the galaxy normalized impact parameter.

\begin{figure*}
    \epsscale{1.0}
    \plotone{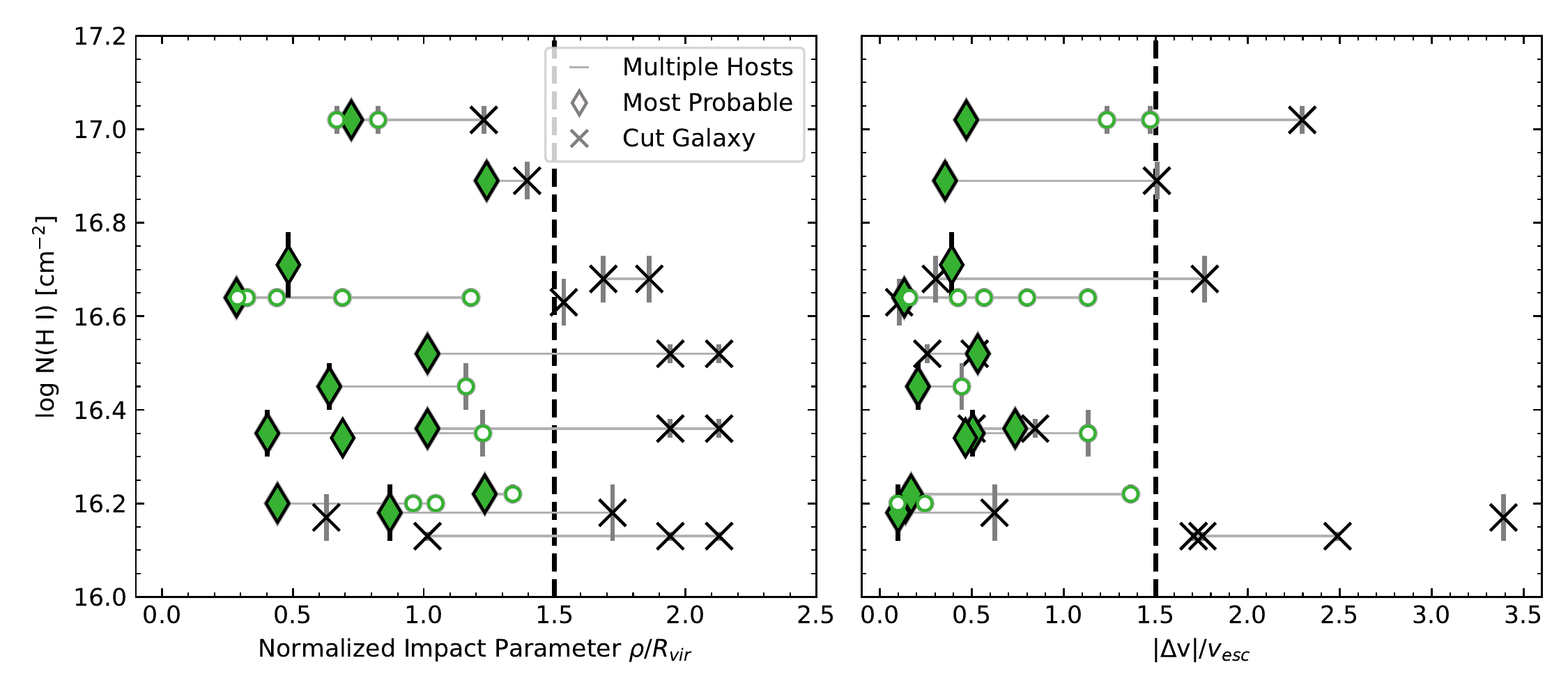}
    \caption{Absorber \HI\ column density versus associated galaxy normalized impact parameter and normalized velocity offset. The velocity offset is normalized by the escape velocity of each associated galaxy halo calculated at \Rvir. Multiple associated galaxies for an absorber are connected with grey lines, and the most probable host (the associated galaxy with the lowest value of $\sqrt{(\rho/R_{\rm vir})^2 + (|\Delta v|/\vesc)^2}$) is designated with a filled diamond. We include the candidate galaxies ($\times$ symbols) that were cut from the final associated galaxy sample for having $\rho/\Rvir$ or $|\Delta v|/\vesc >$ 1.5, so the reader can see where they lie in relation to the associated galaxies. The vertical dashed line marks this value on both plots.} \label{fig:nhinorm}
\end{figure*}

\begin{figure*}
    \epsscale{1.0}
    \plotone{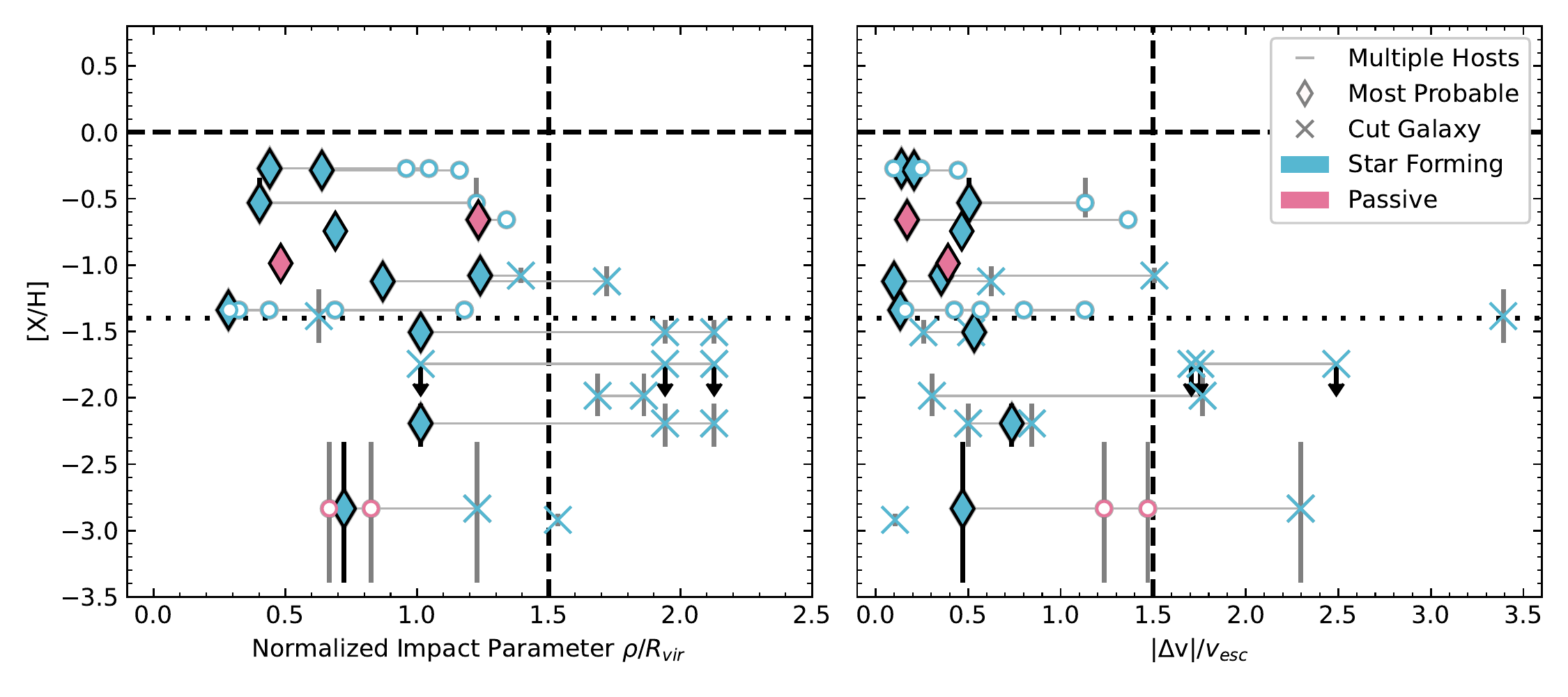}
    \caption{Absorber metallicity versus associated galaxy normalized impact parameter and normalized velocity offset with associated galaxy spectral morphology. The velocity offset is normalized by the escape velocity of each associated galaxy halo calculated at \Rvir. Multiple associated galaxies for an absorber are connected with grey lines, and the most probable host (the associated galaxy with the lowest value of $\sqrt{(\rho/R_{\rm vir})^2 + (|\Delta v|/\vesc)^2}$) is designated with a filled diamond. The dashed and dotted horizontal lines display solar and 4\% solar metallicity, respectively. An associated galaxy is denoted as star forming if the spectrum is emission-line dominated, and it is denoted as passive if the spectrum is absorption-line dominated. We include the candidate galaxies ($\times$ symbols) that were cut from the final associated galaxy sample for having $\rho/\Rvir$ or $|\Delta v|/\vesc >$ 1.5, so the reader can see where they lie in relation to the associated galaxies. The vertical dashed line marks this value on both plots.} \label{fig:metmorphology}
\end{figure*}

We color-code the data points in Figure~\ref{fig:metmorphology} based on each galaxy's spectral classification. If the spectrum is emission-line dominated we label it as star forming, and if the spectrum is absorption-line dominated we label it as passive. The majority of the associated galaxy sample is star-forming, but we have a few detections of passive galaxies. We find that when passive galaxies are the most probable host, they are associated with a metal-enriched absorber.

\subsection{Absorber Properties and Galaxy Stellar Mass}\label{sec:galstmass}

In Figure~\ref{fig:sfrstmass}, we show the SFR versus stellar mass of our associated galaxy sample compared to the galaxies in SDSS DR7 with $z < 0.7$ whose properties are derived and reported in the Max Planck for Astrophysics--Johns Hopkins University (MPA--JHU) catalogs \citep{kauffmann2003,brinchmann2004,salim2007}. We include the absorbers with no associated galaxy detections on this plot using the field limits from Table~\ref{tab:fieldlimits}. No high-mass galaxies ($\logMstar > 11$ \msun) are associated with our absorbers.

\begin{figure}
    \epsscale{1.17}
    \plotone{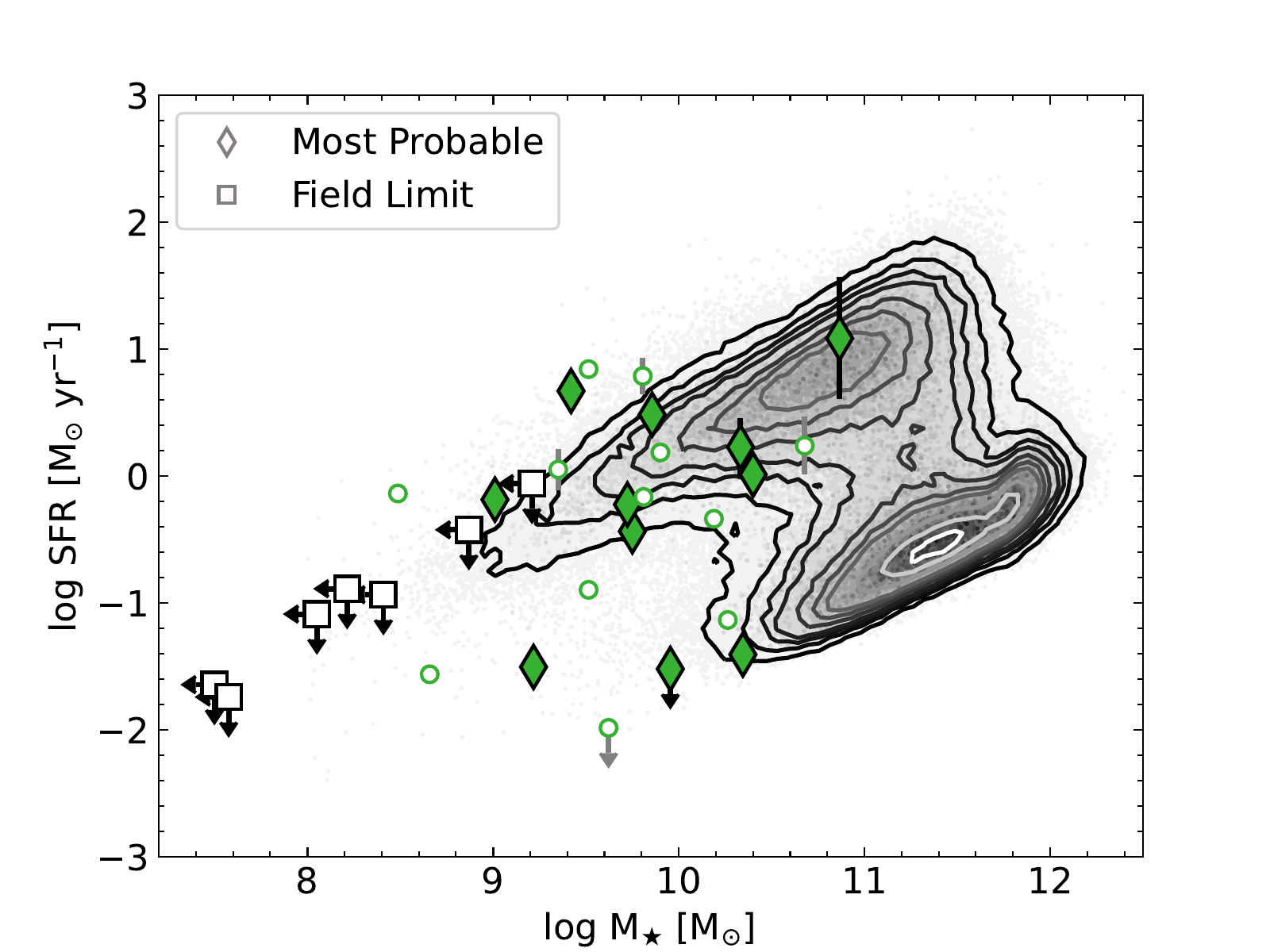}
    \caption{Distribution of SFR versus stellar mass for our associated galaxy sample compared with the general SDSS sample. We plot the SFR and stellar mass limits for the absorbers without associated galaxies from Table~\ref{tab:fieldlimits}. The contours are made using the MPA--JHU catalogs of the galaxies in SDSS DR7 with $z < 0.7$ \citep{kauffmann2003,brinchmann2004,salim2007}. The majority of the absorber-associated galaxies have $\logMstar \lesssim 10.5$ \msun.} \label{fig:sfrstmass}
\end{figure}

Figure~\ref{fig:metstmass} displays the absorber metallicity versus the host galaxy stellar mass with the points color-coded by normalized impact parameter, $\rho/\Rvir$. Here, and in the following sections, we only consider the most probable host galaxies in the figure. (In Appendix~\ref{sec:appb}, we display the same figures as in the main text with the entire candidate galaxy sample.) We see a difference in the stellar mass values of the hosts between the low-metallicity and metal-enriched absorbers, with the low-metallicity absorbers generally showing no host galaxies. In contrast, the metal-enriched system hosts span a broad range of stellar masses with \logMstar\ $\ge 9.0$. This wide range in stellar masses may be reflecting the diversity in possible origins for the metal-enriched absorbers. 

We investigate the possibility of a correlation by running an Anderson-Darling 2-sample test and a Kendall tau test with censored data (using {\tt pymccorrelation},\footnote{\url{https://github.com/privong/pymccorrelation}} \citealt{isobe1986,privon2020}) on the host galaxy and limit stellar mass values. Both test results are statistically significant with $p$-values of 0.01. Using a binomial distribution with the posterior distribution described by a beta function \citep{cameron2011}, we estimate the likelihood of identifying a host galaxy with $\logMstar \ge 9.0$ for the metal-enriched absorbers to be $0.78^{+0.10}_{-0.13}$ (9/11 absorbers), while for the low-metallicity absorbers it is $0.39^{+0.16}_{-0.15}$ (3/8 absorbers). The error bars represent the 68\% confidence interval. The low-metallicity systems are rarely found to be associated with $L^*$ galaxies (a conclusion also advanced by \citealt{prochaska2017} based on the lack of low-metallicity absorbers found around COS-Halos $\sim$$L^*$ galaxies at $\rho \la 160$ kpc). We find no significant correlation between the absorber metallicity and galaxy stellar mass.

In addition to the stellar mass discrepancies, there is also a difference in the distribution of impact parameters. The low-metallicity systems are rarely found within the halo of an associated galaxy. Only 1/8 of the low-metallicity absorbers is found within $\rho \le \Rvir$ of an associated galaxy, whereas 7/11 of the metal-enriched systems are found within $\rho \le \Rvir$. (Similar fractions are obtained when considering absorbers found within $\rho \le 150$ kpc.) We calculate the likelihood of identifying an absorber within \Rvir\ using the binomial distribution as above to be $0.18^{+0.14}_{-0.10}$ for the low-metallicity absorbers and $0.62^{+0.13}_{-0.14}$ for the metal-enriched absorbers. This result strongly suggests these low-metallicity absorbers are associated with overdense regions of the universe ($\rho/\bar{\rho} \sim 10-10^2$) instead of with galaxy halos ($\rho/\bar{\rho} \sim 10^2-10^3$). (See Figure 6 of \citealt{oppenheimer2006} for more information.) This result also suggests outflows have enriched the halos of \logMstar\ $\ge 9.0$ galaxies and/or there has been efficient mixing of incoming low-metallicity gas with the ambient halo gas. However, this sample is small, and the ``cold accretion" absorber identified in \citet{ribaudo2011b} provides a counter example to efficient mixing.

\begin{figure}
    \epsscale{1.17}
    \plotone{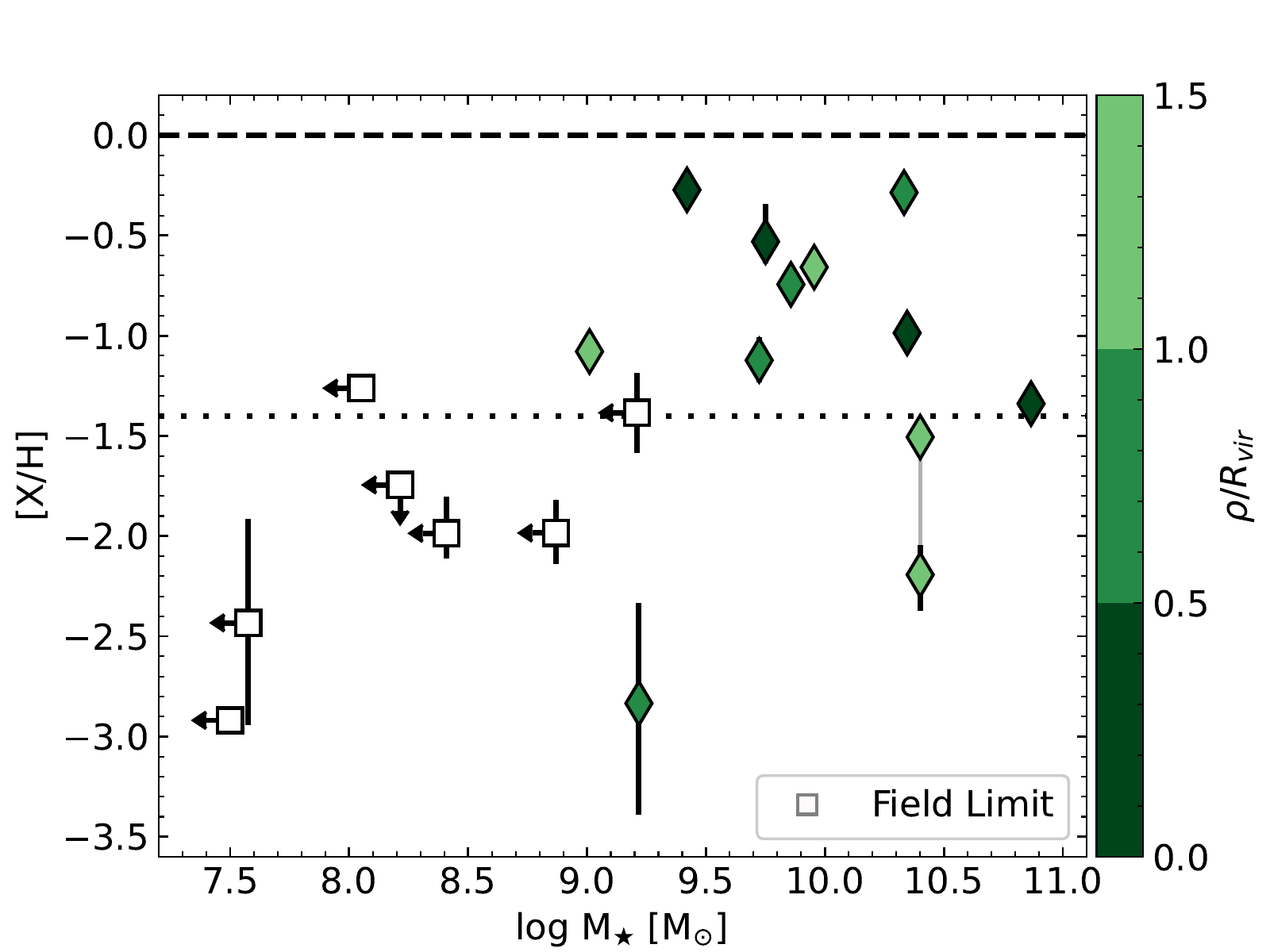}
    \caption{Absorber metallicity versus host galaxy stellar mass. The dashed and dotted horizontal lines display solar and 4\% solar metallicity, respectively. Only the most probable host galaxies are plotted for clarity. The points are colored by the galaxy normalized impact parameter. We plot the stellar mass limits for the absorbers without associated galaxies from Table~\ref{tab:fieldlimits}. The two points connected with a vertical bar are the same galaxy that is associated with two absorbers along the HE1003+0149 sightline. The metal-enriched absorber hosts span a broad range in stellar mass, while we have few host galaxies identified for the low-metallicity absorbers.} \label{fig:metstmass}
\end{figure}

\subsection{Absorber Properties and Galaxy Orientation and Inclination}\label{sec:galorient}

We show the azimuthal angle (orientation) and inclination of the host galaxies with respect to the absorber metallicity in Figures~\ref{fig:metaz} and \ref{fig:metincl}. Only one host galaxy is not included on these plots due to poor resultant fits. For the azimuthal angle, we see an over-density of points close to both the major and minor axes of the host galaxies for the metal-enriched systems. The major axis is aligned with the disk of a galaxy (when present), while the minor axis is aligned with the pole of the galaxy. Given the metallicities of these systems, the absorbers found along the disk could represent new material or recycling material being re-accreted onto the disk. For the low-metallicity systems, it is hard to interpret these results due to the large distances between the absorbers and host galaxies, though they are aligned closer to the disks than the poles of the galaxies. Using a two-sample Kolmogorov-Smirnov test comparing the distributions of the metal-enriched absorbers above and below 45$^{\circ}$, we reject the null hypothesis that they are drawn from the same distribution due to a $p$-value of 0.02. We also compare the absorbers around the poles ($< 22.5^{\circ}$) and disks ($> 67.5^{\circ}$) to those at intermediate angles and cannot reject the null hypothesis to greater than 56\%. When the full sample is compared to a thousand random draws from a Gaussian distribution repeated 100 times, we cannot reject the null hypothesis to greater than 74\% (the median $p$-value of the 100 comparisons). Taken all together, we see a difference in the metal-enriched absorbers orientation distributions, but the full sample is similar to a random distribution. 

Other surveys have also identified clusterings of absorbers along the disks or poles of galaxies. Several surveys have used \MgII\ equivalent widths (e.g., \citealt{bordoloi2011,bouche2012,kacprzak2012}) and concluded the gas is partaking in inflows around the disk and outflows at the poles. \citet{ho2017} and \citet{martin2019} have determined that \MgII\ absorbers at low orientation angles are co-rotating with the galaxy disks, a clear signature of inflows. However, we note that the above surveys include sightlines at smaller impact parameters ($\rho <$ 50 kpc) than observed in our survey. \citet{pointon2019} classified their absorbers with respect to metallicity and found similar results to our distribution, with no trend in galaxy orientation with absorber metallicity when considering the pLLS/LLS column density range and an excess of absorbers around the pole and disk of the galaxies. This survey covers an impact parameter distribution closer to our survey, though they still have several sightlines within $\rho <$ 50 kpc. Additionally, \citet{peroux2020} investigated this azimuthal dependence using the IllustrisTNG50 \citep{nelson2019,pillepich2019} and EAGLE \citep{schaye2015} simulations and found a clear signal of inflows and outflows along separate axes for galaxies of $8.5 < \logMstar < 10.5$ at $z < 1$.

\begin{figure}
    \epsscale{1.17}
    \plotone{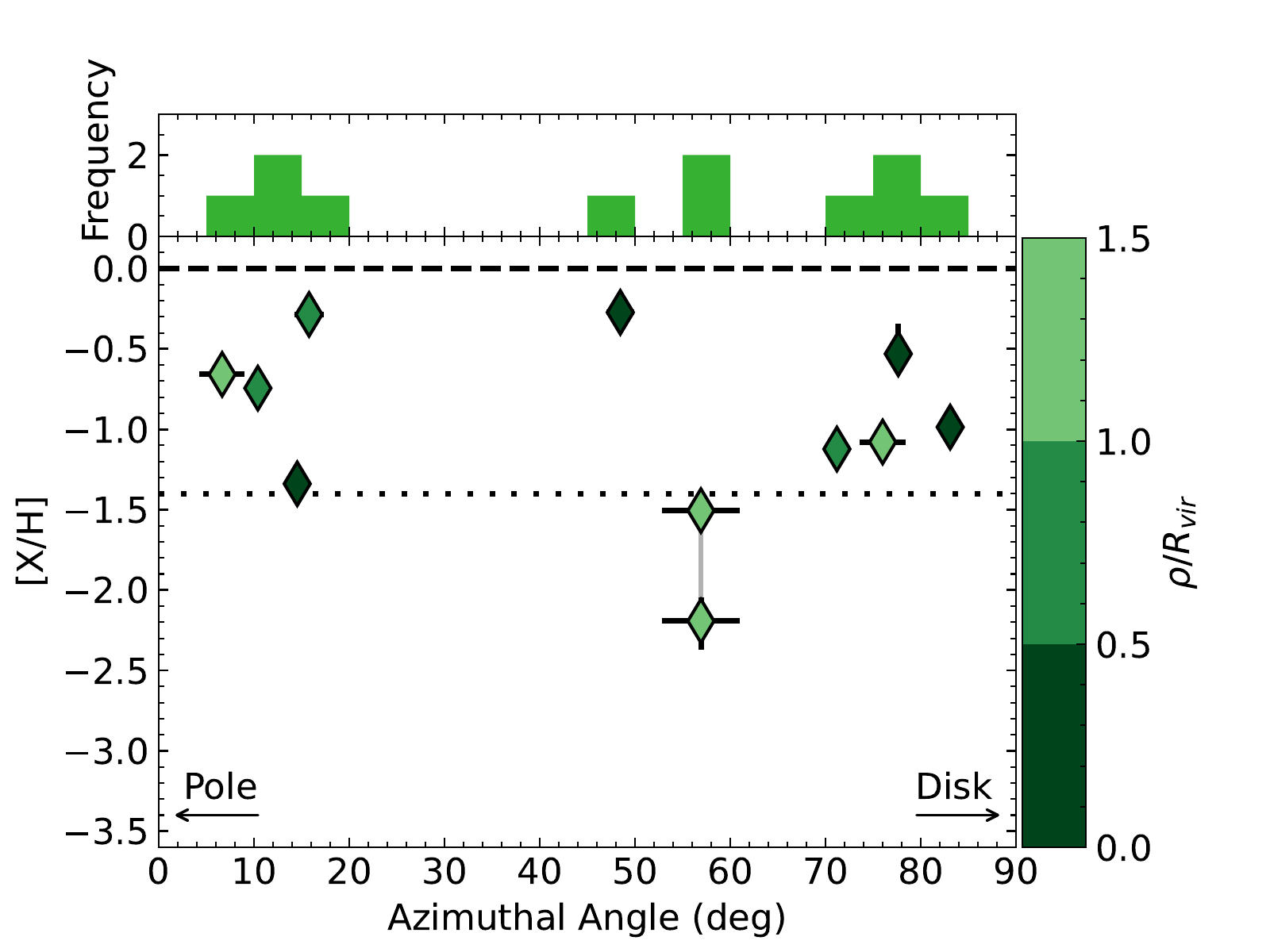}
    \caption{Absorber metallicity versus host galaxy azimuthal angle. The azimuthal angle is measured between the QSO and host galaxy relative to the galaxy major axis. Lower azimuthal angle values are located around the galaxy pole where outflows are expected, while higher azimuthal angles are located around the disk where inflows are expected. The dashed and dotted horizontal lines display solar and 4\% solar metallicity, respectively. Only the most probable host galaxies are plotted for clarity. The points are colored by the host galaxy normalized impact parameter. The two points connected with a vertical bar are the same galaxy that is associated with two absorbers along the HE1003+0149 sightline. We include a histogram of the azimuthal angle in the top panel. One host galaxy is not included on this plot due to poor fitting. The metal-enriched systems are found around the pole and disk.} \label{fig:metaz}
\end{figure}

We find no evidence for a correlation between the host galaxy inclination angle and the absorber metallicity in our sample. We do not have host galaxies with low inclination angles (face-on galaxies), and the majority of our host galaxies have $i > 45^{\circ}$.\footnote{This result is expected due to the smaller amount of solid angle subtended when the galaxy is face-on.} We note these angles are uncertain and have errors of $\pm$5--10$^{\circ}$.

\begin{figure}
    \epsscale{1.17}
    \plotone{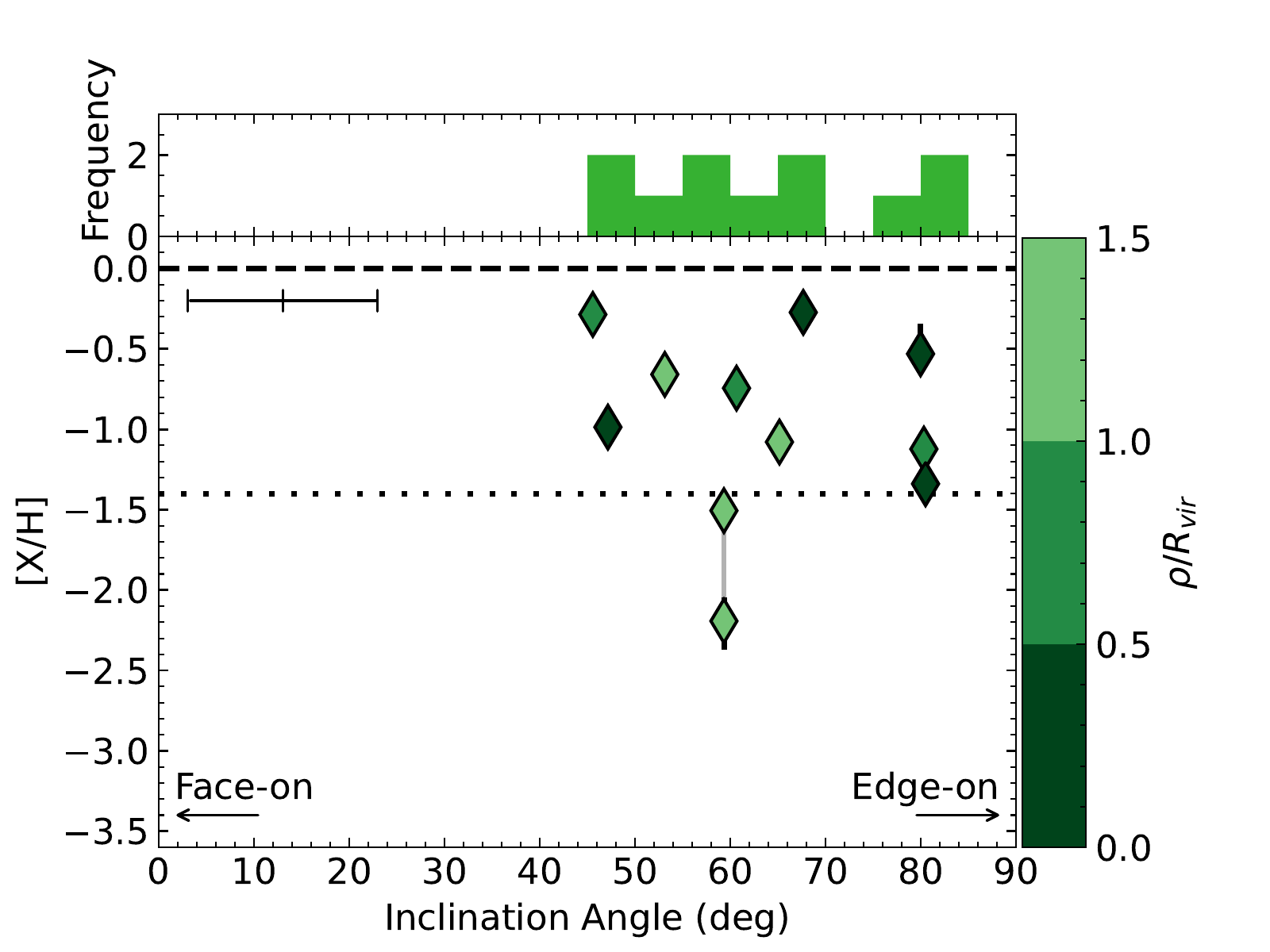}
    \caption{Absorber metallicity versus host galaxy inclination angle. Lower inclination angle values indicate a face-on galaxy, while higher inclination angles indicate an edge-on galaxy. The dashed and dotted horizontal lines display solar and 4\% solar metallicity, respectively. Only the most probable host galaxies are plotted for clarity. The points are colored by the galaxy normalized impact parameter. The two points connected with a vertical bar are the same galaxy that is associated with two absorbers along the HE1003+0149 sightline. We include a histogram of the inclination angle in the top panel. The typical inclination error bar is shown in the upper left of the plot. One host galaxy is not included on this plot due to poor fitting. The majority of the host galaxies tend to be more edge-on.} \label{fig:metincl}
\end{figure}

\subsection{Absorber Properties and Galaxy Metallicity}\label{sec:galmet}

We are only able to calculate a metallicity for half of the host galaxies due to wavelength coverage or the absence of emission lines in the spectrum. We display the absorber metallicity versus the host galaxy metallicity in Figure~\ref{fig:metgalmet}. It is immediately apparent that we see no correlation in host galaxy metallicity with the absorber metallicity. However, the absence of a correlation could be due to the paucity of galaxy metallicities. All of the host galaxies are metal-rich and close to the solar value. Even though the absorber metallicities span over three dex, the host galaxy metallicities are all within 0.4 dex of each other. Nearly all of the absorber metallicities are $\gtrsim$1.0 dex lower than their host galaxy metallicities, indicating a trend for metallicity to decrease steeply when moving from the stellar disk into the halo. This clearly signifies that the absorbers are not part of any active, unmixed outflows in these galaxies, though they could have formed from an ancient outflow.

\begin{figure}
    \epsscale{1.17}
    \plotone{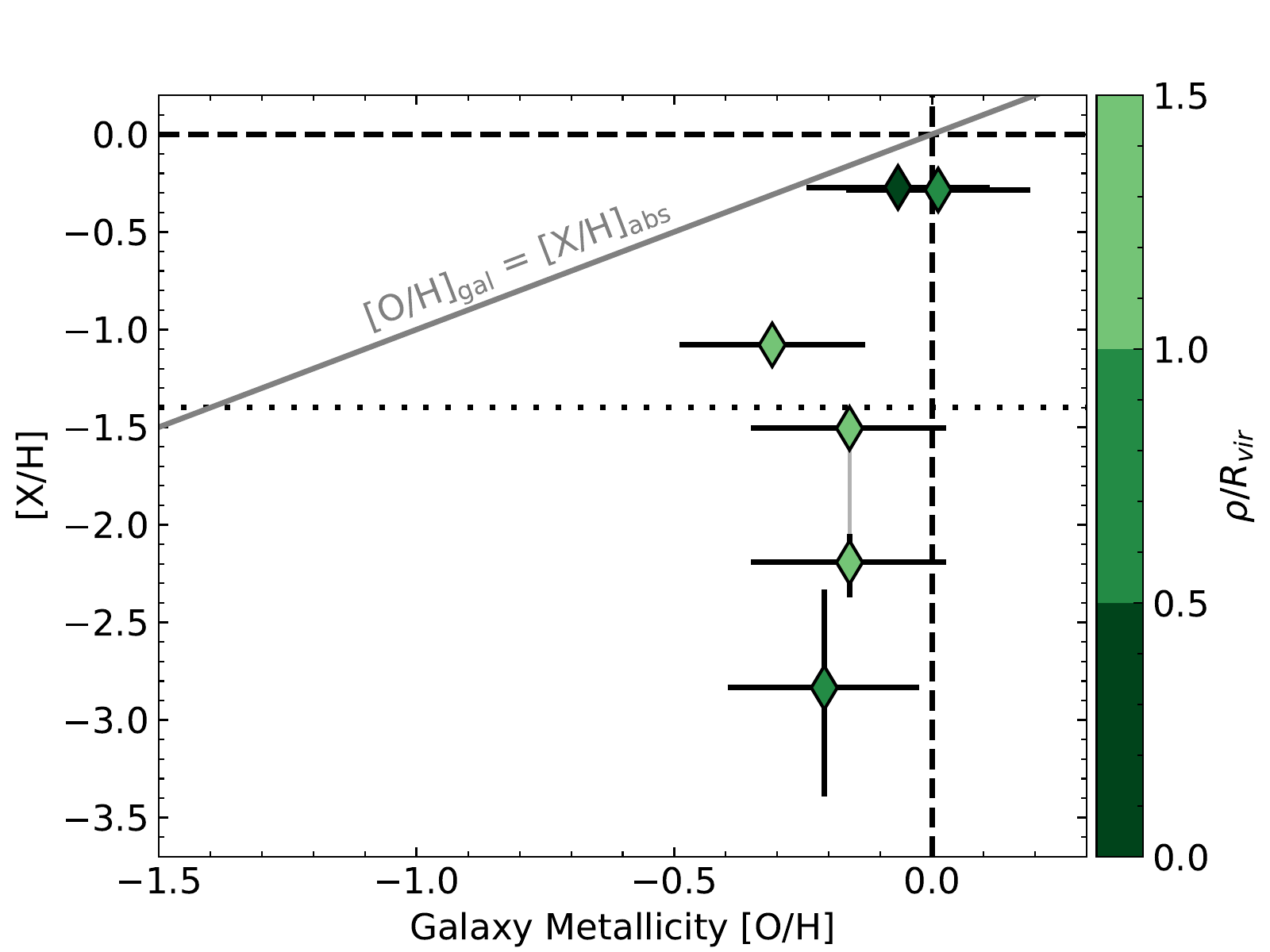}
    \caption{Absorber metallicity versus host galaxy metallicity. The dashed and dotted horizontal lines display solar and 4\% solar metallicity, respectively. The grey line shows the one-to-one correspondence between the two metallicities. Only the most probable host galaxies are plotted for clarity. The points are colored by the galaxy normalized impact parameter. The two points connected with a vertical bar are the same galaxy that is associated with two absorbers along the HE1003+0149 sightline. The host galaxy metallicity is calculated using the O3 relation from \citet{curti2017}. The oxygen abundances are given relative to the solar abundance from \citet{asplund2009}. We are unable to calculate the galaxy metallicity for six of the host galaxies due to the wavelength range of the spectra. The metal-rich absorbers do not trace recent outflows because they are located below the grey line.} \label{fig:metgalmet}
\end{figure}

\section{Discussion}\label{sec:discussion}

\subsection{Preamble}\label{sec:disc-pre}

With BASIC, we use a sample of pLLSs and LLSs at low redshift largely drawn from the initial survey by \citetalias{lehner2013}, which showed for the first time that low-metallicity ($\xh \le -1.4$) absorbers are, surprisingly, an important population at $z<1$. As part of this initial survey, \citetalias{lehner2013} gathered galaxy information from the literature. They found six metal-enriched absorbers, all with associated galaxies within 34--127 kpc and $|\Delta v| \le 50$ \kms, and four out of five low-metallicity absorbers with galaxies within 37--104 kpc and $|\Delta v| = 0$--354 \kms; for the remaining low-metallicity absorber, no $>0.3 L^*$ galaxy was found. \citetalias{lehner2013} noted larger $|\Delta v|$ values in the low-metallicity absorber sample, but they did not appreciate that some of the low-metallicity absorbers may be more loosely connected to galaxies. Among these five low-metallicity absorbers, only three would fit our criteria of $\rho/\Rvir$ and $|\Delta v|/\vesc \le 1.5$, i.e., 60 $\pm$ 20\% of the \citetalias{lehner2013} low-metallicity absorbers are found in the CGM of a galaxy, while 100\% (6/6) of the metal-enriched absorbers are within the CGM of a galaxy. Obviously the literature galaxy information available to \citetalias{lehner2013} represented an inhomogeneous sample with different depth and completeness, but there was a hint already that low-metallicity absorbers may have different origins. 

Combining IFU observations and \hst/ACS imaging, we have now increased the \citetalias{lehner2013} sample by a factor $\approx$2 and, importantly, the galaxy survey is more homogeneous with a much better understanding of the galaxy completeness level. We have identified and characterized a sample of \ifugalsample\ galaxies associated with \ifuabssample\ pLLSs at $z < 1$ in \ifuqsosample\ QSO fields, spanning a wide range in absorber metallicities. For \ifunonsamplet\ absorbers with $\xh <-1$ we are unable to identify any associated galaxies with \logMstar\ $\gtrsim 9.0$ within our criteria. We find a distinction in the stellar mass and impact parameter distributions between the low-metallicity and metal-enriched absorber counterpart galaxies: the low-metallicity absorbers have a probability of $0.39^{+0.16}_{-0.15}$ to be found at $\rho/\Rvir \le 1.5$ of galaxies with \logMstar\ $\ge 9.0$, while the metal-enriched absorbers have a probability of $0.78^{+0.10}_{-0.13}$.

\begin{figure}
    \epsscale{1.2}
    \plotone{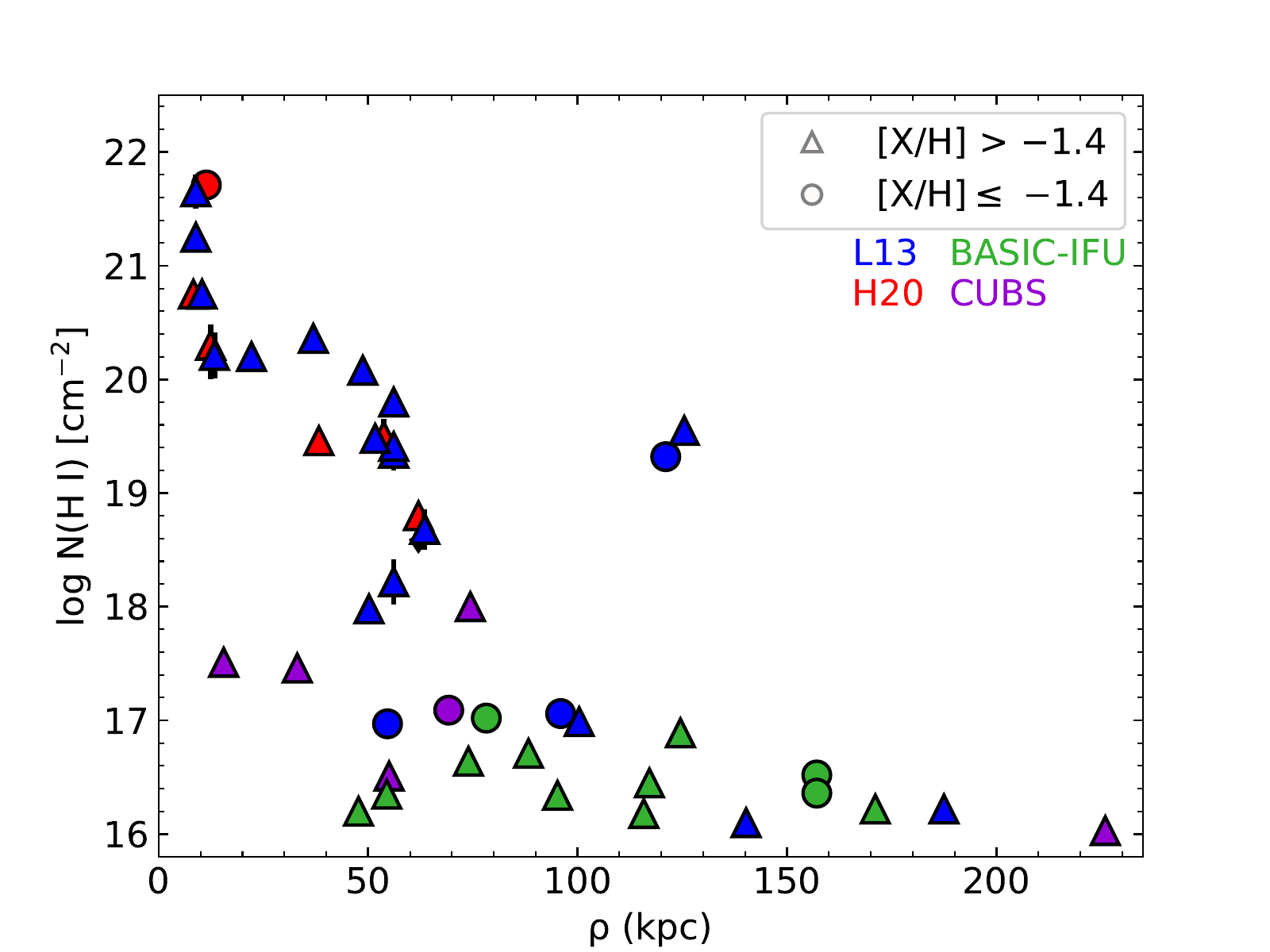}
    \caption{Absorber \HI\ column density versus galaxy impact parameter for the closest host galaxy identified in the field or the most probable host galaxy for the BASIC-IFU sample. We include the \HI\ absorbers studied in \citet{hamanowicz2020}, the CUBS survey \citep{chen2020,cooper2021}, and those compiled in \citetalias{lehner2013} (see references therein). We have recalculated the impact parameters using our cosmology and differentiated the absorbers based on their metallicity. The galaxies associated with pLLSs and LLSs tend to be located farther away from the absorber than galaxies associated with denser absorbers. Few low-metallicity absorbers have had an associated galaxy identified.} \label{fig:nhirhosam}
\end{figure}

In Figure~\ref{fig:nhirhosam}, we update Figure~9 of \citetalias{lehner2013}, showing the projected distances between absorbers and their counterpart galaxies over a range of \HI\ column densities from pLLSs to DLAs. We plot the closest galaxy in $\rho$ identified in the field for the absorbers in \citet{hamanowicz2020}, the CUBS survey \citep{chen2020,cooper2021}, and those compiled in \citetalias{lehner2013} (see references therein) along with the most probable host galaxies identified for our BASIC-IFU sample.\footnote{We do not include the absorbers identified in \citealt{kulkarni2022} on this plot due to the $\rho$-based survey design.} As noted by \citetalias{lehner2013} (but now with a factor $\approx$2 larger number of galaxy-absorber pairs), there is a clear decrease of the impact parameter with increasing \NHI. All the galaxies associated with DLAs are found within 15 kpc, while most of the SLLSs are projected within 50 kpc of galaxies. With distances of 50 kpc or less, the higher column density absorbers are typically located well within \Rvir\ of their host galaxy. On the other hand, galaxies associated with pLLSs span a much broader range in impact parameter. A similar trend is found in the COS-Halos survey \citep[see Figure 4 in][]{prochaska2017}, albeit with a sample that extends to somewhat lower \HI\ column densities.

In Figure~\ref{fig:nhirhosam}, we also differentiate absorbers based on our dividing metallicity at $\xh = -1.4$ (the 95\% confidence limit of the metallicity distribution function of the DLAs identified in CCC, see \citealt{lehner2019,wotta2019}, with lower metallicities being rare ($\lesssim$ 5\%) for these column densities). While for DLAs (by definition) and SLLSs it is not surprising we do not see more low-metallicity absorbers, it is striking that for pLLSs and LLSs this figure is dominated by metal-enriched absorbers. Only 20\% of the pLLSs and LLSs in Figure~\ref{fig:nhirhosam} have $\xh \le -1.4$ despite the fact that about 50\% of this absorber population has metallicities $\xh \le -1.4$ \citep{wotta2019,lehner2019}. This is consistent with the result presented here: metal-enriched pLLSs and LLSs are nearly systematically found in the CGM of galaxies, while only about half of the low-metallicity pLLSs and LLSs population is directly associated with galaxies. Below, we discuss the possible origins of these absorbers, the implications of our results, and compare our results to other CGM surveys.

\subsection{Origins of the Metal-Enriched Absorbers}\label{sec:interpmet}

In the majority of cases, we are able to identify a most probable host galaxy for the metal-enriched absorbers. We see a broad range in the host galaxy stellar masses of $9.0 \lesssim \logMstar < 11.0$. We also find the metal-enriched absorbers are preferentially located along the pole or disk of the associated galaxy. The expected origins of these absorbers are stellar or AGN outflows, tidally-stripped material from satellite galaxies, or recycling inflowing material from an ancient outflow. The absorbers located along the disk of their host galaxy may be participating in a recycling inflow, while those along the pole of the galaxy may be indicative of outflowing material. However, we emphasize that due to the large offsets between the absorbers and host galaxy metallicities, these absorbers cannot be part of any active, unmixed outflows. Tidal stripping could also be a plausible origin for some of these absorbers, especially given that some absorbers have multiple associated galaxies and the possibility of lower-mass dwarf galaxies below our detection limits. The efficient mixing of incoming low-metallicity gas with the ambient halo gas could also be occurring in these halos to form these absorbers.

The host galaxy characteristics of the metal-enriched absorbers are similar to those found in \citet{hafen2017} who studied the CGM of pLLS and LLS host galaxies at $z < 1$ using the cosmological hydrodynamical zoom-in FIRE simulations \citep{hopkins2014}. In these simulations of halos with \logMhalo\ $\approx$ 9--13, the majority of the metal-enriched absorbers originate in outflowing or recycling material within halos of $10 \lesssim \logMhalo \lesssim 12$. The metal-enriched absorbers are mostly found within the CGM of a galaxy.

\subsection{Origins of the Low-Metallicity Absorbers}\label{sec:interplow}

Host galaxies for the low-metallicity absorbers are not as frequently identified as for the metal-enriched absorbers: only $40 \pm 15\%$ of the low-metallicity absorbers in our sample have a host galaxy within $\rho/\Rvir \le 1.5$ (in these cases, their stellar masses are $\logMstar \approx 9.2$ or $\logMstar \approx 10.4$), while for 60\% of the absorbers no galaxy was found. We note there are also a few examples in the literature of galaxies with detections of low-metallicity pLLSs and LLSs in the CGM. In the first metallicity survey of pLLSs and LLSs at $z<1$, \citetalias{lehner2013} reported four out five low-metallicity absorbers with galaxies within \Rvir\ (none of these are part of the IFU sample presented here, see also \citealt{prochaska2011,ribaudo2011b,kacprzak2012,thom2012}). The only low-metallicity absorber at that time with no associated galaxy was reported by \citet{cooksey2008}. More recently, other studies have also identified low-metallicity absorbers associated with galaxies, such as, e.g., \citet{norris2021} for a galaxy with $\logMstar \approx 10.3$, and \citet{zahedy2021} for a galaxy with $\logMstar \approx 10.4$ (see \S~\ref{sec:lowz} for more details). Considering the pLLSs and LLSs with accurately measured \NHI, \citet{kacprzak2019} have identified two low-metallicity pLLSs within $\la 0.5 \Rvir$ of galaxies with $\logMstar \approx 9.8$--10.0. There are also two detections in the much higher-mass halos of luminous red galaxies \citep{berg2019,chen2019}, but these galaxies have an extremely large CGM (i.e., \Rvir) relative to the BASIC-IFU sample. These absorbers identified within the CGM have too low metallicities for them to originate from ancient outflows, recent outflows, or tidally-stripped material. These low-metallicity absorbers must instead be material newly accreted into the CGM of the galaxies, possibly from IGM filaments (see also, e.g., \citealt{ribaudo2011b,norris2021}). However, it is possible some of these detections are not within the CGM given that a gas velocity cut can include material far from galaxy halos along the line of sight (see \citealt{ho2021} for more details).

On the other hand, the other 60\% of pLLSs with $\xh \le -1.4$ cannot be located directly in the CGM since no galaxy is found within $\le1.5 \Rvir$ of these absorbers. However, there could be associated galaxies that fall outside of the observing area or dwarf galaxies below our detection limits, though we have placed stringent limits for the stellar mass of any potential dwarf host galaxies. These absorbers are instead likely found in overdense regions of the universe ($\rho/\bar{\rho} \sim 10-10^2$). Although these absorbers could be stripped gas from extremely-metal-poor dwarf galaxies, these galaxies are most likely too rare \citep{izotov2018} and hence unlikely to make an important contribution to the origin of the low-metallicity absorbers. These absorbers are, however, reminiscent of the strong \HI, extremely metal-poor LLSs found in high resolution cosmological simulations of the IGM by \citet{mandelker2019,mandelker2021} at $z \approx 3$--5. These high resolution simulations show that the IGM can be clustered in small-scale, dense-cloud structures that are not observed in lower resolution simulations \citep{mandelker2021}. In these simulations, the metallicity of this denser gas is $\xh<-3$ at $z \simeq 3$, but according to the metallicity evolution of these absorbers \citep{lehner2022}, the metallicity of such structures is expected to increase at lower redshift. Future high resolution simulations should shed light on the possibility of such dense IGM structures being common at $z<1$. 

When investigating the metallicity distribution for pLLSs and LLSs in the CGM of galaxy halos at $z < 1$ using the FIRE simulations, \citet{hafen2017} were unable to reproduce the amount of low-metallicity gas seen in CCC within their halos. With our new results, the reason could be that these low-metallicity systems are typically located outside of the CGM of $\logMstar \ge 9.0$ galaxies. We note that \citet{rahmati2018}, using the EAGLE simulation, were able to reproduce the fractions of low-metallicity and metal-enriched pLLSs within galaxy halos at $z < 1$, but the absorbers in these simulations exhibit a much stronger redshift evolution in metallicity than observed \citep{lehner2019}.

\subsection{Implications for the Metallicity of the IGM}\label{sec:igm}

 The low-metallicity absorbers, especially those without associated galaxies, provide a constraint on the metallicity of the IGM at low redshift, since we expect metallicity to decrease towards lower overdensities \citep{schaye2003}. A tentative observational range of the IGM metallicity at $z < 0.4$ is $-1.7 \lesssim\ \xh \lesssim -1.4$ \citep{shull2014}, where the lower bound is estimated from the Ly$\alpha$ forest and the upper bound is estimated from the IGM warm-hot ionized medium using \hst/COS observations.\footnote{\citet{manuwal2021} offers another IGM metallicity value of $\xh \approx -2.7$ at $z < 0.16$ using \CIV\ absorbers. This value is closer to our metallicity estimate.} While this range is narrow, there are large uncertainties on both the lower and upper bounds because these absorbers were selected based on their metal content and it is unclear how they can be truly differentiated as IGM or CGM absorbers. Taken at face value, this metallicity range is in the higher range of the low-metallicity absorbers in our sample and the CCC survey \citep{wotta2019,lehner2019}. 

Using the low-metallicity absorbers without associated galaxies in our sample, we estimate the unweighted geometric mean metallicity of these dense regions of the universe to be $\xh \approx -2.1$ for $z < 1$. We note that these absorbers span a broad range in redshift, but little evolution in the absorber metallicity at these values is expected over this redshift range \citep{lehner2019}. Our metallicity estimate for overdense regions is lower than the range estimated in \citet{shull2014}, but it is similar to the SLFSs in CCC detected down to $\logNHI \approx 15.2$ with the lowest metallicities in the range of $-3\la \xh \la -2.1$ (see also Figure~\ref{fig:sample}). The difference between the metallicities derived in CCC and those in \citet{shull2014} likely arises from several factors: 1) there is an inherent bias when estimating the metallicity when absorbers are selected based on their metal content as was done in \citet{shull2014}; 2) as discussed above and in \citeauthor{shull2014}, there is a large uncertainty in separating the metal contributions of the CGM and IGM; 3) the limited S/N available for the COS spectra is another bias against fully sampling the low metallicity range, especially for IGM absorbers that have by definition $\logNHI \la 15.2$ (see \citealt{lehner2018,lehner2019}). The latter is also confirmed with simulated spectra of IGM absorbers \citep{dave2010,oppenheimer2012}.

Turning to cosmological simulations, \citet{oppenheimer2006} show that the enrichment of the IGM from galaxy winds at $z= 1.5$ is in the range $-3 < \xh \lesssim -1.5$, which is found mostly within the large-scale filamentary structure (see their Figure~6). \citet{wiersma2011} also investigated the cosmic metal distribution in the IGM using the OWLS suite of simulations \citep{schaye2010} and found for the diffuse IGM $\xh \approx -2$ at $z=0$, which is consistent with our metallicity estimate. \citet{rahmati2016} calculated the column density distribution functions and cosmic densities for several metal ions using the EAGLE simulation from $z$ = 0--6. Their cosmic metal ion densities are generally a factor 2--3 lower than those derived in \citet{shull2014}. Future low-redshift simulations, especially high resolution simulations of the IGM like those presented in \citet{mandelker2019,mandelker2021}, will provide additional insight about the enrichment of the CGM and IGM and the presence or absence of the pLLSs and LLSs in the spectra of QSOs at $z<1$.

\subsection{Comparison to Other Low-$z$ Surveys}\label{sec:lowz}

\subsubsection{The COS-Halos Survey}

We now compare our results to three other low-\z\ CGM absorber surveys.\footnote{We do not compare our results to the COS-Dwarfs survey \citep{bordoloi2014} because they are unable to estimate the metallicity of the detected absorbers.} The COS-Halos survey \citep{tumlinson2013} is a galaxy-selected survey comprising 44 star-forming and quiescent galaxies ($\logMstar = 9.5$--11.5) at $z \approx 0.2$ with sightlines probing to 160 kpc (0.75\Rvir\ for their mean halo). These $L^*$ galaxies have well-constrained properties \citep{werk2012}. A wide range of \HI\ absorbers are associated with these galaxies (including SLFSs, SLLSs, and DLAs), and they identified 19 pLLSs and LLSs. Though they find low-metallicity gas around galaxies in both of their subsamples, it is rare compared to the frequency of low-metallicity absorbers found in CCC \citep{prochaska2017,lehner2019}. In \citet{prochaska2017}, they conjectured that such low-metallicity gas (with $\xh \la -1.4$) is most likely not associated with $L^*$ galaxies, though redshift evolution could also play a role in the lack of low-metallicity absorbers. \citet{lehner2019} compared the CCC and COS-Halos results over a similar redshift interval and found the metallicity distributions are more comparable, but there was still a lack of very low-metallicity absorbers in COS-Halos. They interpreted this result as pointing to a different origin for these absorbers. 

In Figure~\ref{fig:metstmassch} we plot the absorber metallicity versus galaxy stellar mass for our absorber host galaxies and those from the COS-Halos survey that are associated with pLLSs and LLSs. All of the COS-Halos points on this figure are within $0.5\Rvir$, and the absorber metallicities are taken from CCC. When comparing the two samples, it becomes clear why we see this discrepancy: the COS-Halos survey did not probe galaxies with $\logMstar < 9.5$ \msun\ nor larger impact parameters, thus missing the lowest-metallicity gas present in our sample. Therefore, rather than a large redshift evolution in the metallicities, Figure~\ref{fig:metstmassch} strongly points to a different environment (lower mass galaxies and overdense regions of the universe) that COS-Halos did not probe owing to the galaxy selection for their sample. We note there are a few detections in the COS-Halos survey of low-metallicity gas close to the galaxy that provide other examples of inefficient mixing in galaxy halos.

\begin{figure}
    \epsscale{1.17}
    \plotone{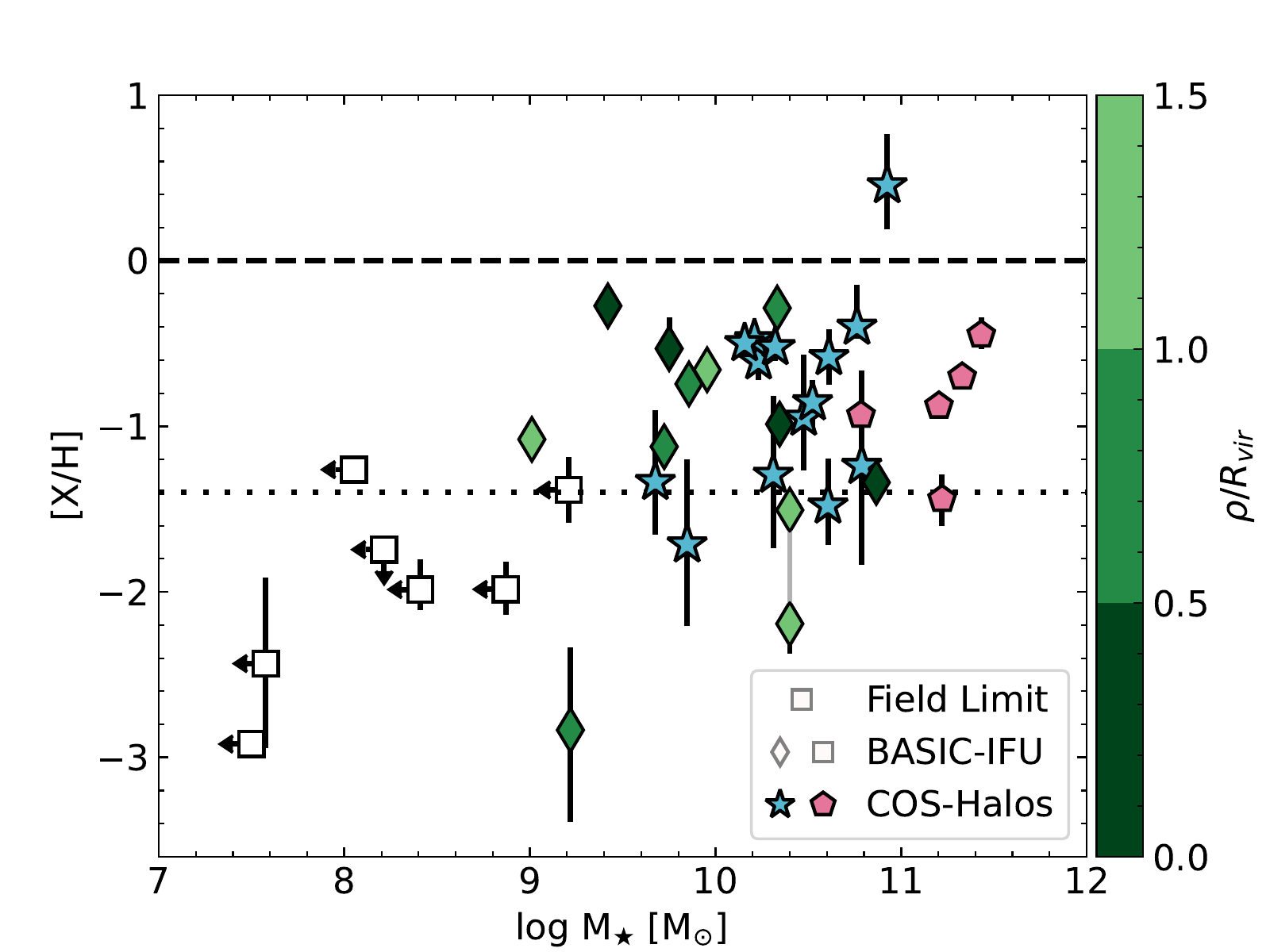}
    \caption{Absorber metallicity versus host galaxy stellar mass for the BASIC-IFU and COS-Halos samples \citep{werk2012,prochaska2017}. The dashed and dotted horizontal lines display solar and 4\% solar metallicity, respectively. Only the most probable host galaxies are plotted for clarity. The points are colored by the galaxy normalized impact parameter. We plot the stellar mass limits for the absorbers without associated galaxies from Table~\ref{tab:fieldlimits}. The two points connected with a vertical bar are the same galaxy that is associated with two absorbers along the HE1003+0149 sightline. We include only the COS-Halos galaxies associated with pLLSs/LLSs. The COS-Halos sample is split by sSFR, where $\log$ sSFR $> -11$ is the star-forming sample represented with stars and the quiescent sample is represented with pentagons. All of the COS-Halos data points are within $\rho/\Rvir \le 0.5$, and the metallicities are taken from CCC. The COS-Halos survey did not detect the lowest-metallicity gas found in CCC because those systems are either associated with lower-mass galaxies or located at larger distances in galaxy halos than the survey probed.} \label{fig:metstmassch}
\end{figure}

\subsubsection{``Multiphase Galaxy Halos" Survey}

The ``Multiphase Galaxy Halos" survey from \citet{pointon2019} and \citet{kacprzak2019} is another galaxy-selected survey comprising 47 $L^*$ galaxies ($10.8 < \logMhalo < 12.5$) at $z \approx 0.3$ with sightlines probing to 203 kpc (1.2\Rvir\ for their mean halo). These galaxies are selected to be isolated with no other neighboring galaxies within 100 kpc or $|\Delta v| < 500$ \kms\ of their redshift. They also detect a wide range of \HI\ absorbers around their galaxies, but the \NHI\ values for the absorbers are often quite uncertain (leading to uncertain metallicities). Several of the \HI\ column densities can only be confined to a range of values (all equally likely) and about half of the \HI\ column densities exhibit errors $>$0.1\,dex. 

Within the pLLS/LLS column density range, they identified eight low-metallicity absorbers, with four of these systems spanning a \logNHI\ range of $\gtrsim$1.5 dex. (In this count, we only include the absorbers with a \logNHI\ range spanning between the pLLS/LLS column density ranges and those absorbers with metallicity error bars that clearly establish that they have low-metallicity.) These absorbers are within 0.2--$0.7\Rvir$ of their host galaxies. This sample also probes a larger redshift range than the COS-Halos survey, though it is still smaller than our range of $z < 1$. Since their sample appears to be galaxy-selected, by definition they find detections of low-metallicity absorbers only within the CGM of galaxies. 

\subsubsection{The CUBS Survey}

The CUBS survey presented in \citet{chen2020} is a field-selected survey comprising 15 QSO fields with deep ground-based galaxy data to study the CGM and IGM at $z < 1$. They have identified four pLLSs/LLSs in the QSO spectra and characterized the absorber metallicities in \citet{zahedy2021} and the associated galaxies in \citet{chen2020}. 

Only one of the four absorbers exhibits low-metallicity gas in the optically-thick component ($\logNHI > 17.0$), and it is found in the halo of an $\sim$$L^*$ galaxy at $0.4\Rvir$ and $z \approx 0.4$. The other three systems are associated with a pair of interacting dwarf galaxies, a galaxy group, and a massive star-forming galaxy. Two more metal-enriched pLLSs are characterized in \citet{cooper2021} and are associated with an $\sim$$L^*$ galaxy and an overdensity of galaxies, respectively. The metal-enriched absorbers in our survey also show diverse galaxy environments and origins, such as in the halos of $\sim$$L^*$ star-forming and passive galaxies, within a QSO galaxy group, and near an overdensity of galaxies (see Table~\ref{tab:galinfo} and \S~\ref{sec:groups}). 

\subsection{Comparison to High-z Surveys}\label{sec:highz}

At higher redshifts, the KODIAQ-Z survey characterizes \HI\ absorbers with $ \logNHI \ga 14.5$ at $z \gtrsim 2$ \citep{lehner2016,lehner2022}. This survey covers a similar \HI\ column density regime as CCC, but thanks to the higher S/N of the Keck/HIRES spectra, they are able to explore even lower \NHI\ absorbers. They find that from $2.2\le z\le 3.6 $ to $z<1$ there is an overall increase of the metallicity of the gas probed by a factor $\approx$8, which applies to all \HI\ column density ranges. They show that low-metallicity absorbers (i.e., those with metallicities that are below the 95\% confidence limit of the metallicity distribution function of the DLAs) have a similar fraction at low and high redshifts, and that the high-$z$ low-metallicity absorbers have a similar metallicity range as observed in the IGM at those redshifts. The latter provides support to our argument in \S~\ref{sec:igm} that the low-metallicity pLLSs have a metallicity distribution that is representative of the IGM metallicity. 

At $z \approx 2$--3, \citet{rudie2012} identified numerous absorbers ($12.0 \le \logNHI \le 21.0$) around Lyman break galaxies, with about half of the absorbers at $\logNHI > 15.5$ found in the CGM of galaxies (defined within $\rho < 300$ kpc and $|\Delta v| \le 300$ \kms\ of the galaxy redshift). The galaxy environment of pLLSs and LLSs is, however, still very poorly known at high $z$, but several surveys are underway using in particular VLT/MUSE and Keck/KCWI. The few cases of galaxy counterpart searches for LLSs at $z > 2$ show some interesting trends. \citet{fumagalli2016} searched for the associated galaxies of two pristine LLSs with $\xh < -3.3$ using MUSE observations. No associated galaxies were detected for one system, while five Ly$\alpha$ emitters were found at a similar redshift to the other absorber. Their interpretation of these results is that the first absorber is most likely located in the IGM, while the other absorber is located in a gaseous filament feeding one or more of the detected galaxies. In the first results for the MUSE analysis of gas around galaxies (MAGG) survey designed specifically to characterize the galaxy environments of $\approx$50 LLSs at $z \approx 3$--4 using MUSE, \citet{lofthouse2020} found three Ly$\alpha$ emitters near the redshift of an LLS with $\xh = -3.41 \pm 0.26$ \citep{crighton2016}. The authors also interpret the absorber as located within a gaseous filament possibly accreting onto one of the nearby galaxies. The latest MAGG results (the characterized environments of 61 absorbers) in \citet{lofthouse2023} similarly show that LLSs at these high redshifts are either located in regions close to galaxies (3-4 \Rvir) or likely in filaments of the IGM. Though it is outside of the LLS column density range, we also note the detection of an SLLS ($19.0 \le \logNHI < 20.3$) with $\xh = -1.89 \pm 0.11$ in the halo of a star-forming galaxy at $z = 2.44$ \citep{crighton2013}. Although these results are still developing and larger samples will be needed, the wide range of environments (IGM, CGM, intragroup) found for these low-metallicity absorbers at high redshift is striking, especially in light of the BASIC-IFU findings presented here. It will be intriguing to see more results from the MAGG survey and other future surveys characterizing the environments of strong \HI\ absorbers at $z > 2$. 

\section{Summary and Concluding Remarks} \label{sec:conclusion}

In this paper we introduce the bimodal absorption system imaging campaign (BASIC) where we are investigating the galaxy environments of low-metallicity and metal-enriched gas with a sample of \abssample\ pLLSs and LLSs in \qsosample\ QSO fields at $z < 1$. This first paper focuses on \ifuabssample\ pLLSs in \ifuqsosample\ QSO fields that have been observed with the Keck/KCWI and VLT/MUSE IFUs. Combined with \hst/ACS imaging, we have identified and characterized a sample of \ifugalsample\ unique galaxies associated with this absorber subset. Our main results are as follows:

\begin{enumerate}

    \item We detect \hostsperabs\ associated galaxy per absorber on average that meets our selection criteria of being within $|\Delta v| <$ 1,000 \kms\ of the absorber redshift and within $\rho/\Rvir$ and $|\Delta v|/\vesc \le 1.5$. We are unable to detect any associated galaxies with $\logMstar \gtrsim 9.0$ for \ifunonsamplet\ absorbers ($\xh < -1$) using these criteria. Six of these absorbers have excellent field coverage to $\approx$200 kpc at the absorber redshift, while the final absorber has $<$50\% coverage of this area.

    \item We find a strong dependence of host galaxy stellar mass on absorber metallicity. The metal-enriched absorbers ($\xh > -1.4$) are associated with galaxies spanning a range of stellar masses with $9.0 \lesssim \logMstar < 11.0$, while the low-metallicity absorbers ($\xh \le -1.4$) often have no associated galaxy. The metal-enriched absorbers have a probability of $0.78^{+0.10}_{-0.13}$ of being associated with a $\logMstar \ge 9.0$ galaxy, while the low-metallicity absorbers have a probability of $0.39^{+0.16}_{-0.15}$. We do not find any strong correlations between the absorber metallicity and the host galaxy properties of spectral morphology, orientation, inclination, or metallicity.

    \item The majority of the metal-enriched absorbers ($\xh > -1.4$) are found preferentially along the major and minor axes of their host galaxy. This may be indicative of outflowing and recycling material for the absorber origins, respectively. The large discrepancy between the absorber metallicity and the galaxy metallicity suggests that these absorbers are not from recent outflows. The absorbers could also originate as tidally-stripped material given some have several associated galaxies and the possibility of dwarf galaxies below our detection limits. The metal-enriched absorbers are associated with galaxy halos. 

    \item The low-metallicity absorbers ($\xh \le -1.4$) show two populations: those associated with galaxy halos and those located in overdense regions of the universe ($\rho/\bar{\rho} \sim 10-10^2$), not within the CGM. The absorbers within the CGM of a galaxy must be newly accreted material because of their low metallicity, while the other absorbers could be gas originating from the IGM. The low-metallicity absorbers frequently observed in our \HI-selected survey were likely not regularly detected in previous galaxy-selected absorption-line surveys because of their mass and impact parameter range selections. 

    \item We estimate the unweighted geometric mean IGM metallicity at $z < 1$ to be $\xh \lesssim -2.1$ using the low-metallicity absorbers without associated galaxies. This is substantially lower than previously estimated, but consistent with some cosmological simulations of the IGM at similar redshift. 

\end{enumerate}

Though a large amount of low-metallicity, high-density gas has been detected at $z < 1$, it appears that roughly half of it lies in overdense regions of the universe rather than within the CGM. However, new questions about the enrichment level of the low-redshift IGM and the formation of such low-metallicity systems must be explored. To improve our statistical assessment of the breakdown of these two populations, more observational studies seeking the galaxy hosts of low-metallicity pLLSs and LLSs are necessary to complete our understanding of how often they are associated with galaxy halos versus overdense regions of the universe. Additional high-resolution simulation work on characterizing the structure of the IGM is also needed to understand how these low-metallicity absorbers form and move towards galaxy halos. In the next paper in this series, we will continue working toward understanding the metal enrichment of overdensities and galaxy halos at low redshift by identifying and characterizing the galaxies associated with the other half of the BASIC absorbers that have been observed with long-slit spectroscopy. 

\acknowledgments

We thank the referee and statistician for providing useful comments that helped improve the overall content of our manuscript.
MAB thanks Matthew C. Wilde and Joseph N. Burchett for their help with the DESI {\tt REDROCK} redshift analysis, Hsiao-Wen Chen for her help obtaining the PKS0405$-$123 galaxy spectrum, and Emma K. Lofthouse, Matteo Fossati, and Michele Fumagalli for their help with the completeness tests. Support for this research was provided by NASA through grants HST-GO-14269 and HST-AR-15634 from the Space Telescope Science Institute, which is operated by the Association of Universities for Research in Astronomy (AURA), Incorporated, under NASA contract NAS5-26555. 
MAB acknowledges support from the Future Investigators in NASA Earth and Space Science and Technology (FINESST) grant 80NSSC19K1408.
JS acknowledges support from Vici grant 639.043.409 from the Dutch Research Council (NWO).
KLC acknowledges partial support from NSF AST-1615296.
JKW acknowledges support from NSF AST-1812521 and from the Research Corporation for Science Advancement, Cottrell Scholar grant ID number 26842.
Some of the data used for this project were obtained from the KODIAQ project, which was funded through NASA ADAP grants NNX10AE84G an NNX16AF52G.
Some of the data presented herein were obtained at the W. M. Keck Observatory, which is operated as a scientific partnership among the California Institute of Technology, the University of California and the NASA. The Observatory was made possible by the generous financial support of the W. M. Keck Foundation. The authors wish to recognize and acknowledge the very significant cultural role and reverence that the summit of Maunakea has always had within the indigenous Hawaiian community. We are most fortunate to have the opportunity to conduct observations from this mountain.
Based on observations collected at the European Southern Observatory under ESO programs 094.A-0131(A), 095.A-0200(A), and 096.A-0222(A).
This research made use of Montage. It is funded by the NSF under Grant Number ACI-1440620, and was previously funded by the NASA's Earth Science Technology Office, Computation Technologies Project, under Cooperative Agreement Number NCC5-626 between NASA and the California Institute of Technology.
This research uses services or data provided by the Astro Data Lab at NSF's National Optical-Infrared Astronomy Research Laboratory. NOIRLab is operated by AURA, Inc. under a cooperative agreement with the NSF.
This publication makes use of data products from the Two Micron All Sky Survey, which is a joint project of the University of Massachusetts and the Infrared Processing and Analysis Center/California Institute of Technology, funded by the NASA and the NSF.
The Legacy Surveys consist of three individual and complementary projects: the Dark Energy Camera Legacy Survey (DECaLS, the Beijing-Arizona Sky Survey (BASS), and the Mayall z-band Legacy Survey (MzLS). DECaLS, BASS and MzLS together include data obtained, respectively, at the Blanco telescope, Cerro Tololo Inter-American Observatory, NSF's NOIRLab; the Bok telescope, Steward Observatory, University of Arizona; and the Mayall telescope, Kitt Peak National Observatory, NOIRLab. The Legacy Surveys project is honored to be permitted to conduct astronomical research on Iolkam Du'ag (Kitt Peak), a mountain with particular significance to the Tohono O'odham Nation.
This project used data obtained with the Dark Energy Camera (DECam), which was constructed by the Dark Energy Survey (DES) collaboration. 
BASS is a key project of the Telescope Access Program (TAP), which has been funded by the National Astronomical Observatories of China, the Chinese Academy of Sciences (the Strategic Priority Research Program ``The Emergence of Cosmological Structures" Grant \# XDB09000000), and the Special Fund for Astronomy from the Ministry of Finance. 
The Legacy Surveys imaging of the DESI footprint is supported by the Director, Office of Science, Office of High Energy Physics of the U.S. Department of Energy under Contract No. DE-AC02-05CH1123, by the National Energy Research Scientific Computing Center, a DOE Office of Science User Facility under the same contract; and by the U.S. NSF, Division of Astronomical Sciences under Contract No. AST-0950945 to NOAO.
The Legacy Survey team makes use of data products from the Near-Earth Object Wide-field Infrared Survey Explorer (NEOWISE), which is a project of the Jet Propulsion Laboratory/California Institute of Technology.  The unWISE coadded images and catalog are based on data products from the Wide-field Infrared Survey Explorer, which is a joint project of the University of California, Los Angeles, and the Jet Propulsion Laboratory/California Institute of Technology, and NEOWISE, which is a project of the Jet Propulsion Laboratory/California Institute of Technology. WISE and NEOWISE are funded by the NASA.
Database access and other data services are hosted by the Astro Data Lab at the Community Science and Data Center (CSDC) of the NSF's National Optical Infrared Astronomy Research Laboratory, operated by AURA under a cooperative agreement with the NSF. This work has been supported in part by NASA ADAP grant NNH17AE75I. 
This work made use of data from the Large Binocular Telescope. The LBT is an international collaboration among institutions in the United States, Italy and Germany. LBT Corporation partners are: The University of Arizona on behalf of the Arizona Board of Regents; Istituto Nazionale di Astrofisica, Italy; LBT Beteiligungsgesellschaft, Germany, representing the Max-Planck Society, The Leibniz Institute for Astrophysics Potsdam, and Heidelberg University; The Ohio State University, representing OSU, University of Notre Dame, University of Minnesota and University of Virginia.


\facilities{Du Pont (WFCCD), \hst\ (ACS), Keck:II (KCWI), LBT (LBC), Magellan Baade (IMACS), VLT (MUSE)}

\software{AstroDrizzle \citep{stsci2012}, astropy \citep{robitaille2013,astropy2018}, CubExtractor \citep{cantalupo2019}, DESI REDROCK (\url{https://github.com/desihub/redrock}), ds9 \citep{sao2000}, eazy-py \citep{brammer2008}, GALFIT \citep{peng2002,peng2010}, gdpyc (\url{https://github.com/ruizca/gdpyc}), kcorrect \citep{blanton2007}, kcwitools (\url{https://github.com/pypeit/kcwitools}), KDERP \citep{morrissey2018}, Matplotlib \citep{hunter2007}, Montage \citep{berriman2007}, NOIRLab Astro Data Lab (\url{https://datalab.noirlab.edu}), Numpy \citep{vanderwalt2011}, pPXF \citep{cappellari2004,vazdekis2010,cappellari2017}, pymccorrelation ({\url{https://github.com/privong/pymccorrelation}}), pyMCZ \citep{bianco2016}, PyMUSE \citep{pessa2020}, QFitsView \citep{ott2012}, SciPy \citep{scipy2020}, SExtractor \citep{ba1996}, synphot \citep{stsci2018}}


\bibliography{BASICrefs}

\begin{thebibliography}{}
\expandafter\ifx\csname natexlab\endcsname\relax\def\natexlab#1{#1}\fi
\providecommand{\url}[1]{\href{#1}{#1}}

\bibitem[{{Aihara} {et~al.}(2022){Aihara}, {AlSayyad}, {Ando}, {Armstrong},
  {Bosch}, {Egami}, {Furusawa}, {Furusawa}, {Harasawa}, {Harikane}, {Hsieh},
  {Ikeda}, {Ito}, {Iwata}, {Kodama}, {Koike}, {Kokubo}, {Komiyama}, {Li},
  {Liang}, {Lin}, {Lupton}, {Lust}, {MacArthur}, {Mawatari}, {Mineo},
  {Miyatake}, {Miyazaki}, {More}, {Morishima}, {Murayama}, {Nakajima},
  {Nakata}, {Nishizawa}, {Oguri}, {Okabe}, {Okura}, {Ono}, {Osato}, {Ouchi},
  {Pan}, {Plazas Malag{\'o}n}, {Price}, {Reed}, {Rykoff}, {Shibuya},
  {Simunovic}, {Strauss}, {Sugimori}, {Suto}, {Suzuki}, {Takada}, {Takagi},
  {Takata}, {Takita}, {Tanaka}, {Tang}, {Taranu}, {Terai}, {Toba}, {Turner},
  {Uchiyama}, {Vijarnwannaluk}, {Waters}, {Yamada}, {Yamamoto}, \&
  {Yamashita}}]{aihara2022}
{Aihara}, H., {AlSayyad}, Y., {Ando}, M., {et~al.} 2022, \pasj, 74, 247

\bibitem[{{Asplund} {et~al.}(2009){Asplund}, {Grevesse}, {Sauval}, \&
  {Scott}}]{asplund2009}
{Asplund}, M., {Grevesse}, N., {Sauval}, A.~J., \& {Scott}, P. 2009, \araa, 47,
  481

\bibitem[{{Astropy Collaboration} {et~al.}(2013){Astropy Collaboration},
  {Robitaille}, {Tollerud}, {Greenfield}, {Droettboom}, {Bray}, {Aldcroft},
  {Davis}, {Ginsburg}, {Price-Whelan}, {Kerzendorf}, {Conley}, {Crighton},
  {Barbary}, {Muna}, {Ferguson}, {Grollier}, {Parikh}, {Nair}, {Unther},
  {Deil}, {Woillez}, {Conseil}, {Kramer}, {Turner}, {Singer}, {Fox}, {Weaver},
  {Zabalza}, {Edwards}, {Azalee Bostroem}, {Burke}, {Casey}, {Crawford},
  {Dencheva}, {Ely}, {Jenness}, {Labrie}, {Lim}, {Pierfederici}, {Pontzen},
  {Ptak}, {Refsdal}, {Servillat}, \& {Streicher}}]{robitaille2013}
{Astropy Collaboration}, {Robitaille}, T.~P., {Tollerud}, E.~J., {et~al.} 2013,
  \aap, 558, A33

\bibitem[{{Bacon} {et~al.}(2006){Bacon}, {Bauer}, {Boehm}, {Boudon},
  {Brau-Nogu{\'e}}, {Caillier}, {Capoani}, {Carollo}, {Champavert}, {Contini},
  {Daguis{\'e}}, {Dall{\'e}}, {Delabre}, {Devriendt}, {Dreizler}, {Dubois},
  {Dupieux}, {Dupin}, {Emsellem}, {Ferruit}, {Franx}, {Gallou}, {Gerssen},
  {Guiderdoni}, {Hahn}, {Hofmann}, {Jarno}, {Kelz}, {Koehler}, {Kollatschny},
  {Kosmalski}, {Laurent}, {Lilly}, {Lizon}, {Loupias}, {Lynn}, {Manescau},
  {McDermid}, {Monstein}, {Nicklas}, {Par{\`e}s}, {Pasquini},
  {P{\'e}contal-Rousset}, {P{\'e}contal}, {Pello}, {Petit}, {Picat}, {Popow},
  {Quirrenbach}, {Reiss}, {Renault}, {Roth}, {Schaye}, {Soucail}, {Steinmetz},
  {Stroebele}, {Stuik}, {Weilbacher}, {Wozniak}, \& {de Zeeuw}}]{bacon2006}
{Bacon}, R., {Bauer}, S., {Boehm}, P., {et~al.} 2006, in Society of
  Photo-Optical Instrumentation Engineers (SPIE) Conference Series, Vol. 6269,
  Society of Photo-Optical Instrumentation Engineers (SPIE) Conference Series,
  ed. I.~S. {McLean} \& M.~{Iye}, 62690J

\bibitem[{{Bacon} {et~al.}(2010){Bacon}, {Accardo}, {Adjali}, {Anwand},
  {Bauer}, {Biswas}, {Blaizot}, {Boudon}, {Brau-Nogue}, {Brinchmann},
  {Caillier}, {Capoani}, {Carollo}, {Contini}, {Couderc}, {Daguis{\'e}},
  {Deiries}, {Delabre}, {Dreizler}, {Dubois}, {Dupieux}, {Dupuy}, {Emsellem},
  {Fechner}, {Fleischmann}, {Fran{\c{c}}ois}, {Gallou}, {Gharsa}, {Glindemann},
  {Gojak}, {Guiderdoni}, {Hansali}, {Hahn}, {Jarno}, {Kelz}, {Koehler},
  {Kosmalski}, {Laurent}, {Le Floch}, {Lilly}, {Lizon}, {Loupias}, {Manescau},
  {Monstein}, {Nicklas}, {Olaya}, {Pares}, {Pasquini}, {P{\'e}contal-Rousset},
  {Pell{\'o}}, {Petit}, {Popow}, {Reiss}, {Remillieux}, {Renault}, {Roth},
  {Rupprecht}, {Serre}, {Schaye}, {Soucail}, {Steinmetz}, {Streicher}, {Stuik},
  {Valentin}, {Vernet}, {Weilbacher}, {Wisotzki}, \& {Yerle}}]{bacon2010}
{Bacon}, R., {Accardo}, M., {Adjali}, L., {et~al.} 2010, in Society of
  Photo-Optical Instrumentation Engineers (SPIE) Conference Series, Vol. 7735,
  Ground-based and Airborne Instrumentation for Astronomy III, ed. I.~S.
  {McLean}, S.~K. {Ramsay}, \& H.~{Takami}, 773508

\bibitem[{{Battisti} {et~al.}(2012){Battisti}, {Meiring}, {Tripp}, {Prochaska},
  {Werk}, {Jenkins}, {Lehner}, {Tumlinson}, \& {Thom}}]{battisti2012}
{Battisti}, A.~J., {Meiring}, J.~D., {Tripp}, T.~M., {et~al.} 2012, \apj, 744,
  93

\bibitem[{{Berg} {et~al.}(2019){Berg}, {Howk}, {Lehner}, {Wotta}, {O'Meara},
  {Bowen}, {Burchett}, {Peeples}, \& {Tejos}}]{berg2019}
{Berg}, M.~A., {Howk}, J.~C., {Lehner}, N., {et~al.} 2019, \apj, 883, 5

\bibitem[{{Berg} {et~al.}(2018){Berg}, {Ellison}, {Tumlinson}, {Oppenheimer},
  {Horton}, {Bordoloi}, \& {Schaye}}]{berg2018}
{Berg}, T. A.~M., {Ellison}, S.~L., {Tumlinson}, J., {et~al.} 2018, \mnras,
  478, 3890

\bibitem[{{Berriman} {et~al.}(2007){Berriman}, {Good}, {Laity}, {Jacob},
  {Katz}, {Deelman}, {Singh}, \& {Su}}]{berriman2007}
{Berriman}, G.~B., {Good}, J.~C., {Laity}, A.~C., {et~al.} 2007, in
  Astronomical Society of the Pacific Conference Series, Vol. 382, The National
  Virtual Observatory: Tools and Techniques for Astronomical Research, ed.
  M.~J. {Graham}, M.~J. {Fitzpatrick}, \& T.~A. {McGlynn}, 179

\bibitem[{{Bertin} \& {Arnouts}(1996)}]{ba1996}
{Bertin}, E., \& {Arnouts}, S. 1996, \aaps, 117, 393

\bibitem[{{Bianco} {et~al.}(2016){Bianco}, {Modjaz}, {Oh}, {Fierroz}, {Liu},
  {Kewley}, \& {Graur}}]{bianco2016}
{Bianco}, F.~B., {Modjaz}, M., {Oh}, S.~M., {et~al.} 2016, Astronomy and
  Computing, 16, 54

\bibitem[{{Bielby} {et~al.}(2019){Bielby}, {Stott}, {Cullen}, {Tripp},
  {Burchett}, {Fumagalli}, {Morris}, {Tejos}, {Crain}, {Bower}, \&
  {Prochaska}}]{bielby2019}
{Bielby}, R.~M., {Stott}, J.~P., {Cullen}, F., {et~al.} 2019, \mnras, 486, 21

\bibitem[{{Blanton} \& {Roweis}(2007)}]{blanton2007}
{Blanton}, M.~R., \& {Roweis}, S. 2007, \aj, 133, 734

\bibitem[{{Bolton} {et~al.}(2012){Bolton}, {Schlegel}, {Aubourg}, {Bailey},
  {Bhardwaj}, {Brownstein}, {Burles}, {Chen}, {Dawson}, {Eisenstein}, {Gunn},
  {Knapp}, {Loomis}, {Lupton}, {Maraston}, {Muna}, {Myers}, {Olmstead},
  {Padmanabhan}, {P{\^a}ris}, {Percival}, {Petitjean}, {Rockosi}, {Ross},
  {Schneider}, {Shu}, {Strauss}, {Thomas}, {Tremonti}, {Wake}, {Weaver}, \&
  {Wood-Vasey}}]{bolton2012}
{Bolton}, A.~S., {Schlegel}, D.~J., {Aubourg}, {\'E}., {et~al.} 2012, \aj, 144,
  144

\bibitem[{{Bordoloi} {et~al.}(2011){Bordoloi}, {Lilly}, {Knobel}, {Bolzonella},
  {Kampczyk}, {Carollo}, {Iovino}, {Zucca}, {Contini}, {Kneib}, {Le Fevre},
  {Mainieri}, {Renzini}, {Scodeggio}, {Zamorani}, {Balestra}, {Bardelli},
  {Bongiorno}, {Caputi}, {Cucciati}, {de la Torre}, {de Ravel}, {Garilli},
  {Kova{\v{c}}}, {Lamareille}, {Le Borgne}, {Le Brun}, {Maier}, {Mignoli},
  {Pello}, {Peng}, {Perez Montero}, {Presotto}, {Scarlata}, {Silverman},
  {Tanaka}, {Tasca}, {Tresse}, {Vergani}, {Barnes}, {Cappi}, {Cimatti},
  {Coppa}, {Diener}, {Franzetti}, {Koekemoer}, {L{\'o}pez-Sanjuan},
  {McCracken}, {Moresco}, {Nair}, {Oesch}, {Pozzetti}, \&
  {Welikala}}]{bordoloi2011}
{Bordoloi}, R., {Lilly}, S.~J., {Knobel}, C., {et~al.} 2011, \apj, 743, 10

\bibitem[{{Bordoloi} {et~al.}(2014){Bordoloi}, {Tumlinson}, {Werk},
  {Oppenheimer}, {Peeples}, {Prochaska}, {Tripp}, {Katz}, {Dav{\'e}}, {Fox},
  {Thom}, {Ford}, {Weinberg}, {Burchett}, \& {Kollmeier}}]{bordoloi2014}
{Bordoloi}, R., {Tumlinson}, J., {Werk}, J.~K., {et~al.} 2014, \apj, 796, 136

\bibitem[{{Borthakur} {et~al.}(2016){Borthakur}, {Heckman}, {Tumlinson},
  {Bordoloi}, {Kauffmann}, {Catinella}, {Schiminovich}, {Dav{\'e}}, {Moran}, \&
  {Saintonge}}]{borthakur2016}
{Borthakur}, S., {Heckman}, T., {Tumlinson}, J., {et~al.} 2016, \apj, 833, 259

\bibitem[{{Bouch{\'e}} {et~al.}(2012){Bouch{\'e}}, {Hohensee}, {Vargas},
  {Kacprzak}, {Martin}, {Cooke}, \& {Churchill}}]{bouche2012}
{Bouch{\'e}}, N., {Hohensee}, W., {Vargas}, R., {et~al.} 2012, \mnras, 426, 801

\bibitem[{{Bowen} {et~al.}(1995){Bowen}, {Blades}, \& {Pettini}}]{bowen1995}
{Bowen}, D.~V., {Blades}, J.~C., \& {Pettini}, M. 1995, \apj, 448, 634

\bibitem[{{Bowen} {et~al.}(2002){Bowen}, {Pettini}, \& {Blades}}]{bowen2002}
{Bowen}, D.~V., {Pettini}, M., \& {Blades}, J.~C. 2002, \apj, 580, 169

\bibitem[{{Brammer} {et~al.}(2008){Brammer}, {van Dokkum}, \&
  {Coppi}}]{brammer2008}
{Brammer}, G.~B., {van Dokkum}, P.~G., \& {Coppi}, P. 2008, \apj, 686, 1503

\bibitem[{{Brinchmann} {et~al.}(2004){Brinchmann}, {Charlot}, {White},
  {Tremonti}, {Kauffmann}, {Heckman}, \& {Brinkmann}}]{brinchmann2004}
{Brinchmann}, J., {Charlot}, S., {White}, S.~D.~M., {et~al.} 2004, \mnras, 351,
  1151

\bibitem[{{Bruzual} \& {Charlot}(2003)}]{bruzual2003}
{Bruzual}, G., \& {Charlot}, S. 2003, \mnras, 344, 1000

\bibitem[{{Burchett} {et~al.}(2016){Burchett}, {Tripp}, {Bordoloi}, {Werk},
  {Prochaska}, {Tumlinson}, {Willmer}, {O'Meara}, \& {Katz}}]{burchett2016}
{Burchett}, J.~N., {Tripp}, T.~M., {Bordoloi}, R., {et~al.} 2016, \apj, 832,
  124

\bibitem[{{Burchett} {et~al.}(2019){Burchett}, {Tripp}, {Prochaska}, {Werk},
  {Tumlinson}, {Howk}, {Willmer}, {Lehner}, {Meiring}, {Bowen}, {Bordoloi},
  {Peeples}, {Jenkins}, {O'Meara}, {Tejos}, \& {Katz}}]{burchett2019}
{Burchett}, J.~N., {Tripp}, T.~M., {Prochaska}, J.~X., {et~al.} 2019, \apjl,
  877, L20

\bibitem[{{Cameron}(2011)}]{cameron2011}
{Cameron}, E. 2011, \pasa, 28, 128

\bibitem[{{Cantalupo} {et~al.}(2019){Cantalupo}, {Pezzulli}, {Lilly}, {Marino},
  {Gallego}, {Schaye}, {Bacon}, {Feltre}, {Kollatschny}, {Nanayakkara},
  {Richard}, {Wendt}, {Wisotzki}, \& {Prochaska}}]{cantalupo2019}
{Cantalupo}, S., {Pezzulli}, G., {Lilly}, S.~J., {et~al.} 2019, \mnras, 483,
  5188

\bibitem[{{Cappellari}(2017)}]{cappellari2017}
{Cappellari}, M. 2017, \mnras, 466, 798

\bibitem[{{Cappellari} \& {Emsellem}(2004)}]{cappellari2004}
{Cappellari}, M., \& {Emsellem}, E. 2004, \pasp, 116, 138

\bibitem[{{Chabrier}(2003)}]{chabrier2003}
{Chabrier}, G. 2003, \pasp, 115, 763

\bibitem[{{Chen} {et~al.}(2010){Chen}, {Helsby}, {Gauthier}, {Shectman},
  {Thompson}, \& {Tinker}}]{chen2010}
{Chen}, H.-W., {Helsby}, J.~E., {Gauthier}, J.-R., {et~al.} 2010, \apj, 714,
  1521

\bibitem[{{Chen} \& {Mulchaey}(2009)}]{chen2009}
{Chen}, H.-W., \& {Mulchaey}, J.~S. 2009, \apj, 701, 1219

\bibitem[{{Chen} {et~al.}(2018){Chen}, {Zahedy}, {Johnson}, {Pierce}, {Huang},
  {Weiner}, \& {Gauthier}}]{chen2018}
{Chen}, H.-W., {Zahedy}, F.~S., {Johnson}, S.~D., {et~al.} 2018, \mnras, 479,
  2547

\bibitem[{{Chen} {et~al.}(2019){Chen}, {Johnson}, {Straka}, {Zahedy}, {Schaye},
  {Muzahid}, {Bouch{\'e}}, {Cantalupo}, {Marino}, \& {Wendt}}]{chen2019}
{Chen}, H.-W., {Johnson}, S.~D., {Straka}, L.~A., {et~al.} 2019, \mnras, 484,
  431

\bibitem[{{Chen} {et~al.}(2020){Chen}, {Zahedy}, {Boettcher}, {Cooper},
  {Johnson}, {Rudie}, {Chen}, {Walth}, {Cantalupo}, {Cooksey},
  {Faucher-Gigu{\`e}re}, {Greene}, {Lopez}, {Mulchaey}, {Penton}, {Petitjean},
  {Putman}, {Rafelski}, {Rauch}, {Schaye}, {Simcoe}, \& {Weiner}}]{chen2020}
{Chen}, H.-W., {Zahedy}, F.~S., {Boettcher}, E., {et~al.} 2020, \mnras, 497,
  498

\bibitem[{{Conroy}(2013)}]{conroy2013}
{Conroy}, C. 2013, \araa, 51, 393

\bibitem[{{Cooksey} {et~al.}(2008){Cooksey}, {Prochaska}, {Chen}, {Mulchaey},
  \& {Weiner}}]{cooksey2008}
{Cooksey}, K.~L., {Prochaska}, J.~X., {Chen}, H.-W., {Mulchaey}, J.~S., \&
  {Weiner}, B.~J. 2008, \apj, 676, 262

\bibitem[{{Cooper} {et~al.}(2021){Cooper}, {Rudie}, {Chen}, {Johnson},
  {Zahedy}, {Chen}, {Boettcher}, {Walth}, {Cantalupo}, {Cooksey},
  {Faucher-Gigu{\`e}re}, {Greene}, {Lopez}, {Mulchaey}, {Penton}, {Petitjean},
  {Putman}, {Rafelski}, {Rauch}, {Schaye}, \& {Simcoe}}]{cooper2021}
{Cooper}, T.~J., {Rudie}, G.~C., {Chen}, H.-W., {et~al.} 2021, \mnras, 508,
  4359

\bibitem[{{Crighton} {et~al.}(2013){Crighton}, {Hennawi}, \&
  {Prochaska}}]{crighton2013}
{Crighton}, N. H.~M., {Hennawi}, J.~F., \& {Prochaska}, J.~X. 2013, \apjl, 776,
  L18

\bibitem[{{Crighton} {et~al.}(2016){Crighton}, {O'Meara}, \&
  {Murphy}}]{crighton2016}
{Crighton}, N. H.~M., {O'Meara}, J.~M., \& {Murphy}, M.~T. 2016, \mnras, 457,
  L44

\bibitem[{{Curti} {et~al.}(2017){Curti}, {Cresci}, {Mannucci}, {Marconi},
  {Maiolino}, \& {Esposito}}]{curti2017}
{Curti}, M., {Cresci}, G., {Mannucci}, F., {et~al.} 2017, \mnras, 465, 1384

\bibitem[{{Dav{\'e}} {et~al.}(2010){Dav{\'e}}, {Oppenheimer}, {Katz},
  {Kollmeier}, \& {Weinberg}}]{dave2010}
{Dav{\'e}}, R., {Oppenheimer}, B.~D., {Katz}, N., {Kollmeier}, J.~A., \&
  {Weinberg}, D.~H. 2010, \mnras, 408, 2051

\bibitem[{{Dey} {et~al.}(2019){Dey}, {Schlegel}, {Lang}, {Blum}, {Burleigh},
  {Fan}, {Findlay}, {Finkbeiner}, {Herrera}, {Juneau}, {Landriau}, {Levi},
  {McGreer}, {Meisner}, {Myers}, {Moustakas}, {Nugent}, {Patej}, {Schlafly},
  {Walker}, {Valdes}, {Weaver}, {Y{\`e}che}, {Zou}, {Zhou}, {Abareshi},
  {Abbott}, {Abolfathi}, {Aguilera}, {Alam}, {Allen}, {Alvarez}, {Annis},
  {Ansarinejad}, {Aubert}, {Beechert}, {Bell}, {BenZvi}, {Beutler}, {Bielby},
  {Bolton}, {Brice{\~n}o}, {Buckley-Geer}, {Butler}, {Calamida}, {Carlberg},
  {Carter}, {Casas}, {Castander}, {Choi}, {Comparat}, {Cukanovaite}, {Delubac},
  {DeVries}, {Dey}, {Dhungana}, {Dickinson}, {Ding}, {Donaldson}, {Duan},
  {Duckworth}, {Eftekharzadeh}, {Eisenstein}, {Etourneau}, {Fagrelius},
  {Farihi}, {Fitzpatrick}, {Font-Ribera}, {Fulmer}, {G{\"a}nsicke},
  {Gaztanaga}, {George}, {Gerdes}, {Gontcho}, {Gorgoni}, {Green}, {Guy},
  {Harmer}, {Hernandez}, {Honscheid}, {Huang}, {James}, {Jannuzi}, {Jiang},
  {Joyce}, {Karcher}, {Karkar}, {Kehoe}, {Kneib}, {Kueter-Young}, {Lan},
  {Lauer}, {Le Guillou}, {Le Van Suu}, {Lee}, {Lesser}, {Perreault Levasseur},
  {Li}, {Mann}, {Marshall}, {Mart{\'\i}nez-V{\'a}zquez}, {Martini}, {du Mas des
  Bourboux}, {McManus}, {Meier}, {M{\'e}nard}, {Metcalfe},
  {Mu{\~n}oz-Guti{\'e}rrez}, {Najita}, {Napier}, {Narayan}, {Newman}, {Nie},
  {Nord}, {Norman}, {Olsen}, {Paat}, {Palanque-Delabrouille}, {Peng},
  {Poppett}, {Poremba}, {Prakash}, {Rabinowitz}, {Raichoor}, {Rezaie},
  {Robertson}, {Roe}, {Ross}, {Ross}, {Rudnick}, {Safonova}, {Saha},
  {S{\'a}nchez}, {Savary}, {Schweiker}, {Scott}, {Seo}, {Shan}, {Silva},
  {Slepian}, {Soto}, {Sprayberry}, {Staten}, {Stillman}, {Stupak}, {Summers},
  {Sien Tie}, {Tirado}, {Vargas-Maga{\~n}a}, {Vivas}, {Wechsler}, {Williams},
  {Yang}, {Yang}, {Yapici}, {Zaritsky}, {Zenteno}, {Zhang}, {Zhang}, {Zhou}, \&
  {Zhou}}]{dey2019}
{Dey}, A., {Schlegel}, D.~J., {Lang}, D., {et~al.} 2019, \aj, 157, 168

\bibitem[{{Dressler} {et~al.}(2011){Dressler}, {Bigelow}, {Hare}, {Sutin},
  {Thompson}, {Burley}, {Epps}, {Oemler}, {Bagish}, {Birk}, {Clardy},
  {Gunnels}, {Kelson}, {Shectman}, \& {Osip}}]{dressler2011}
{Dressler}, A., {Bigelow}, B., {Hare}, T., {et~al.} 2011, \pasp, 123, 288

\bibitem[{{Dutton} \& {Macci{\`o}}(2014)}]{dm2014}
{Dutton}, A.~A., \& {Macci{\`o}}, A.~V. 2014, \mnras, 441, 3359

\bibitem[{{Faucher-Gigu{\`e}re} {et~al.}(2016){Faucher-Gigu{\`e}re},
  {Feldmann}, {Quataert}, {Kere{\v s}}, {Hopkins}, \& {Murray}}]{fg2016}
{Faucher-Gigu{\`e}re}, C.-A., {Feldmann}, R., {Quataert}, E., {et~al.} 2016,
  \mnras, 461, L32

\bibitem[{{Ferland} {et~al.}(2013){Ferland}, {Porter}, {van Hoof}, {Williams},
  {Abel}, {Lykins}, {Shaw}, {Henney}, \& {Stancil}}]{ferland2013}
{Ferland}, G.~J., {Porter}, R.~L., {van Hoof}, P.~A.~M., {et~al.} 2013, \rmxaa,
  49, 137

\bibitem[{{Florian} {et~al.}(2021){Florian}, {Rigby}, {Acharyya}, {Sharon},
  {Gladders}, {Kewley}, {Khullar}, {Gozman}, {Brammer}, {Momcheva}, {Nicholls},
  {LaMassa}, {Dahle}, {Bayliss}, {Wuyts}, {Johnson}, \&
  {Whitaker}}]{florian2020}
{Florian}, M.~K., {Rigby}, J.~R., {Acharyya}, A., {et~al.} 2021, \apj, 916, 50

\bibitem[{{Fumagalli} {et~al.}(2016){Fumagalli}, {O'Meara}, \&
  {Prochaska}}]{fumagalli2016}
{Fumagalli}, M., {O'Meara}, J.~M., \& {Prochaska}, J.~X. 2016, \mnras, 455,
  4100

\bibitem[{{Gibson} {et~al.}(2022){Gibson}, {Lehner}, {Oppenheimer}, {Howk},
  {Cooksey}, \& {Fox}}]{gibson2022}
{Gibson}, J.~L., {Lehner}, N., {Oppenheimer}, B.~D., {et~al.} 2022, \aj, 164, 9

\bibitem[{{Haardt} \& {Madau}(1996)}]{hm1996}
{Haardt}, F., \& {Madau}, P. 1996, \apj, 461, 20

\bibitem[{{Haardt} \& {Madau}(2012)}]{hm2012}
---. 2012, \apj, 746, 125

\bibitem[{{Hafen} {et~al.}(2017){Hafen}, {Faucher-Gigu{\`e}re},
  {Angl{\'e}s-Alc{\'a}zar}, {Kere{\v s}}, {Feldmann}, {Chan}, {Quataert},
  {Murray}, \& {Hopkins}}]{hafen2017}
{Hafen}, Z., {Faucher-Gigu{\`e}re}, C.-A., {Angl{\'e}s-Alc{\'a}zar}, D.,
  {et~al.} 2017, \mnras, 469, 2292

\bibitem[{{Hafen} {et~al.}(2019){Hafen}, {Faucher-Gigu{\`e}re},
  {Angl{\'e}s-Alc{\'a}zar}, {Stern}, {Kere{\v{s}}}, {Hummels}, {Esmerian},
  {Garrison-Kimmel}, {El-Badry}, {Wetzel}, {Chan}, {Hopkins}, \&
  {Murray}}]{hafen2019}
---. 2019, \mnras, 488, 1248

\bibitem[{{Haislmaier} {et~al.}(2021){Haislmaier}, {Tripp}, {Katz},
  {Prochaska}, {Burchett}, {O'Meara}, \& {Werk}}]{haislmaier2021}
{Haislmaier}, K.~J., {Tripp}, T.~M., {Katz}, N., {et~al.} 2021, \mnras, 502,
  4993

\bibitem[{{Hamanowicz} {et~al.}(2020){Hamanowicz}, {P{\'e}roux}, {Zwaan},
  {Rahmani}, {Pettini}, {York}, {Klitsch}, {Augustin}, {Krogager}, {Kulkarni},
  {Fresco}, {Biggs}, {Milliard}, \& {Vernet}}]{hamanowicz2020}
{Hamanowicz}, A., {P{\'e}roux}, C., {Zwaan}, M.~A., {et~al.} 2020, \mnras, 492,
  2347

\bibitem[{{Heckman} {et~al.}(2017){Heckman}, {Borthakur}, {Wild},
  {Schiminovich}, \& {Bordoloi}}]{heckman2017}
{Heckman}, T., {Borthakur}, S., {Wild}, V., {Schiminovich}, D., \& {Bordoloi},
  R. 2017, \apj, 846, 151

\bibitem[{{Hewett} \& {Wild}(2010)}]{hw2010}
{Hewett}, P.~C., \& {Wild}, V. 2010, VizieR Online Data Catalog,
  J/MNRAS/405/2302

\bibitem[{{Ho} {et~al.}(2017){Ho}, {Martin}, {Kacprzak}, \&
  {Churchill}}]{ho2017}
{Ho}, S.~H., {Martin}, C.~L., {Kacprzak}, G.~G., \& {Churchill}, C.~W. 2017,
  \apj, 835, 267

\bibitem[{{Ho} {et~al.}(2021){Ho}, {Martin}, \& {Schaye}}]{ho2021}
{Ho}, S.~H., {Martin}, C.~L., \& {Schaye}, J. 2021, \apj, 923, 137

\bibitem[{{Hopkins} {et~al.}(2014){Hopkins}, {Kere{\v{s}}}, {O{\~n}orbe},
  {Faucher-Gigu{\`e}re}, {Quataert}, {Murray}, \& {Bullock}}]{hopkins2014}
{Hopkins}, P.~F., {Kere{\v{s}}}, D., {O{\~n}orbe}, J., {et~al.} 2014, \mnras,
  445, 581

\bibitem[{{Howk} {et~al.}(2009){Howk}, {Ribaudo}, {Lehner}, {Prochaska}, \&
  {Chen}}]{howk2009}
{Howk}, J.~C., {Ribaudo}, J.~S., {Lehner}, N., {Prochaska}, J.~X., \& {Chen},
  H.-W. 2009, \mnras, 396, 1875

\bibitem[{{Howk} {et~al.}(2018){Howk}, {Rueff}, {Lehner}, {Wotta}, {Croxall},
  \& {Savage}}]{howk2018}
{Howk}, J.~C., {Rueff}, K.~M., {Lehner}, N., {et~al.} 2018, \apj, 856, 166

\bibitem[{{Hunter}(2007)}]{hunter2007}
{Hunter}, J.~D. 2007, Computing in Science and Engineering, 9, 90

\bibitem[{{Isobe} {et~al.}(1986){Isobe}, {Feigelson}, \& {Nelson}}]{isobe1986}
{Isobe}, T., {Feigelson}, E.~D., \& {Nelson}, P.~I. 1986, \apj, 306, 490

\bibitem[{{Izotov} {et~al.}(2018){Izotov}, {Thuan}, {Guseva}, \&
  {Liss}}]{izotov2018}
{Izotov}, Y.~I., {Thuan}, T.~X., {Guseva}, N.~G., \& {Liss}, S.~E. 2018,
  \mnras, 473, 1956

\bibitem[{{Johnson} {et~al.}(2013){Johnson}, {Chen}, \&
  {Mulchaey}}]{johnson2013}
{Johnson}, S.~D., {Chen}, H.-W., \& {Mulchaey}, J.~S. 2013, \mnras, 434, 1765

\bibitem[{{Johnson} {et~al.}(2017){Johnson}, {Chen}, {Mulchaey}, {Schaye}, \&
  {Straka}}]{johnson2017}
{Johnson}, S.~D., {Chen}, H.-W., {Mulchaey}, J.~S., {Schaye}, J., \& {Straka},
  L.~A. 2017, \apjl, 850, L10

\bibitem[{{Kacprzak} {et~al.}(2012){Kacprzak}, {Churchill}, \&
  {Nielsen}}]{kacprzak2012}
{Kacprzak}, G.~G., {Churchill}, C.~W., \& {Nielsen}, N.~M. 2012, \apjl, 760, L7

\bibitem[{{Kacprzak} {et~al.}(2019){Kacprzak}, {Pointon}, {Nielsen},
  {Churchill}, {Muzahid}, \& {Charlton}}]{kacprzak2019}
{Kacprzak}, G.~G., {Pointon}, S.~K., {Nielsen}, N.~M., {et~al.} 2019, \apj,
  886, 91

\bibitem[{{Kauffmann} {et~al.}(2003){Kauffmann}, {Heckman}, {Tremonti},
  {Brinchmann}, {Charlot}, {White}, {Ridgway}, {Brinkmann}, {Fukugita}, {Hall},
  {Ivezi{\'c}}, {Richards}, \& {Schneider}}]{kauffmann2003}
{Kauffmann}, G., {Heckman}, T.~M., {Tremonti}, C., {et~al.} 2003, \mnras, 346,
  1055

\bibitem[{{Keeney} {et~al.}(2017){Keeney}, {Stocke}, {Danforth}, {Shull},
  {Pratt}, {Froning}, {Green}, {Penton}, \& {Savage}}]{keeney2017}
{Keeney}, B.~A., {Stocke}, J.~T., {Danforth}, C.~W., {et~al.} 2017, \apjs, 230,
  6

\bibitem[{{Kulkarni} {et~al.}(2022){Kulkarni}, {Bowen}, {Straka}, {York},
  {Gupta}, {Noterdaeme}, \& {Srianand}}]{kulkarni2022}
{Kulkarni}, V.~P., {Bowen}, D.~V., {Straka}, L.~A., {et~al.} 2022, \apj, 929,
  150

\bibitem[{{Lanzetta} {et~al.}(1995){Lanzetta}, {Bowen}, {Tytler}, \&
  {Webb}}]{lanzetta1995}
{Lanzetta}, K.~M., {Bowen}, D.~V., {Tytler}, D., \& {Webb}, J.~K. 1995, \apj,
  442, 538

\bibitem[{{Lehner} {et~al.}(2016){Lehner}, {O'Meara}, {Howk}, {Prochaska}, \&
  {Fumagalli}}]{lehner2016}
{Lehner}, N., {O'Meara}, J.~M., {Howk}, J.~C., {Prochaska}, J.~X., \&
  {Fumagalli}, M. 2016, \apj, 833, 283

\bibitem[{{Lehner} {et~al.}(2018){Lehner}, {Wotta}, {Howk}, {O'Meara},
  {Oppenheimer}, \& {Cooksey}}]{lehner2018}
{Lehner}, N., {Wotta}, C.~B., {Howk}, J.~C., {et~al.} 2018, \apj, 866, 33

\bibitem[{{Lehner} {et~al.}(2019){Lehner}, {Wotta}, {Howk}, {O'Meara},
  {Oppenheimer}, \& {Cooksey}}]{lehner2019}
---. 2019, \apj, 887, 5

\bibitem[{{Lehner} {et~al.}(2013){Lehner}, {Howk}, {Tripp}, {Tumlinson},
  {Prochaska}, {O'Meara}, {Thom}, {Werk}, {Fox}, \& {Ribaudo}}]{lehner2013}
{Lehner}, N., {Howk}, J.~C., {Tripp}, T.~M., {et~al.} 2013, \apj, 770, 138

\bibitem[{{Lehner} {et~al.}(2020){Lehner}, {Berek}, {Howk}, {Wakker},
  {Tumlinson}, {Jenkins}, {Prochaska}, {Augustin}, {Ji}, {Faucher-Gigu{\`e}re},
  {Hafen}, {Peeples}, {Barger}, {Berg}, {Bordoloi}, {Brown}, {Fox}, {Gilbert},
  {Guhathakurta}, {Kalirai}, {Lockman}, {O'Meara}, {Pisano}, {Ribaudo}, \&
  {Werk}}]{lehner2020}
{Lehner}, N., {Berek}, S.~C., {Howk}, J.~C., {et~al.} 2020, \apj, 900, 9

\bibitem[{{Lehner} {et~al.}(2022){Lehner}, {Kopenhafer}, {O'Meara}, {Howk},
  {Fumagalli}, {Prochaska}, {Acharyya}, {O'Shea}, {Peeples}, {Tumlinson}, \&
  {Hummels}}]{lehner2022}
{Lehner}, N., {Kopenhafer}, C., {O'Meara}, J.~M., {et~al.} 2022, \apj, 936, 156

\bibitem[{{Lofthouse} {et~al.}(2020){Lofthouse}, {Fumagalli}, {Fossati},
  {O'Meara}, {Murphy}, {Christensen}, {Prochaska}, {Cantalupo}, {Bielby},
  {Cooke}, {Lusso}, \& {Morris}}]{lofthouse2020}
{Lofthouse}, E.~K., {Fumagalli}, M., {Fossati}, M., {et~al.} 2020, \mnras, 491,
  2057

\bibitem[{{Lofthouse} {et~al.}(2023){Lofthouse}, {Fumagalli}, {Fossati},
  {Dutta}, {Galbiati}, {Arrigoni Battaia}, {Cantalupo}, {Christensen}, {Cooke},
  {Longobardi}, {Murphy}, \& {Prochaska}}]{lofthouse2023}
---. 2023, \mnras, 518, 305

\bibitem[{{Lundgren} {et~al.}(2021){Lundgren}, {Creech}, {Brammer}, {Kirse},
  {Peek}, {Wake}, {York}, {Chisholm}, {Erb}, {Kulkarni}, {Straka}, {Tremonti},
  \& {van Dokkum}}]{lundgren2021}
{Lundgren}, B.~F., {Creech}, S., {Brammer}, G., {et~al.} 2021, \apj, 913, 50

\bibitem[{{Maller} \& {Bullock}(2004)}]{mb2004}
{Maller}, A.~H., \& {Bullock}, J.~S. 2004, \mnras, 355, 694

\bibitem[{{Mandelker} {et~al.}(2019){Mandelker}, {van den Bosch}, {Springel},
  \& {van de Voort}}]{mandelker2019}
{Mandelker}, N., {van den Bosch}, F.~C., {Springel}, V., \& {van de Voort}, F.
  2019, \apjl, 881, L20

\bibitem[{{Mandelker} {et~al.}(2021){Mandelker}, {van den Bosch}, {Springel},
  {van de Voort}, {Burchett}, {Butsky}, {Nagai}, \& {Oh}}]{mandelker2021}
{Mandelker}, N., {van den Bosch}, F.~C., {Springel}, V., {et~al.} 2021, \apj,
  923, 115

\bibitem[{{Manuwal} {et~al.}(2019){Manuwal}, {Narayanan}, {Muzahid},
  {Charlton}, {Khaire}, \& {Chand}}]{manuwal2019}
{Manuwal}, A., {Narayanan}, A., {Muzahid}, S., {et~al.} 2019, \mnras, 485, 30

\bibitem[{{Manuwal} {et~al.}(2021){Manuwal}, {Narayanan}, {Udhwani},
  {Srianand}, {Savage}, {Charlton}, \& {Misawa}}]{manuwal2021}
{Manuwal}, A., {Narayanan}, A., {Udhwani}, P., {et~al.} 2021, \mnras, 505, 3635

\bibitem[{{Marino} {et~al.}(2018){Marino}, {Cantalupo}, {Lilly}, {Gallego},
  {Straka}, {Borisova}, {Pezzulli}, {Bacon}, {Brinchmann}, {Carollo},
  {Caruana}, {Conseil}, {Contini}, {Diener}, {Finley}, {Inami}, {Leclercq},
  {Muzahid}, {Richard}, {Schaye}, {Wendt}, \& {Wisotzki}}]{marino2018}
{Marino}, R.~A., {Cantalupo}, S., {Lilly}, S.~J., {et~al.} 2018, \apj, 859, 53

\bibitem[{{Marra} {et~al.}(2022){Marra}, {Churchill}, {Kacprzak}, {Nielsen},
  {Trujillo-Gomez}, \& {Lewis}}]{marra2022}
{Marra}, R., {Churchill}, C.~W., {Kacprzak}, G.~G., {et~al.} 2022, arXiv
  e-prints, arXiv:2202.12228

\bibitem[{{Martin} {et~al.}(2019){Martin}, {Ho}, {Kacprzak}, \&
  {Churchill}}]{martin2019}
{Martin}, C.~L., {Ho}, S.~H., {Kacprzak}, G.~G., \& {Churchill}, C.~W. 2019,
  \apj, 878, 84

\bibitem[{{Morrissey} {et~al.}(2018){Morrissey}, {Matuszewski}, {Martin},
  {Neill}, {Epps}, {Fucik}, {Weber}, {Darvish}, {Adkins}, {Allen}, {Bartos},
  {Belicki}, {Cabak}, {Callahan}, {Cowley}, {Crabill}, {Deich}, {Delecroix},
  {Doppman}, {Hilyard}, {James}, {Kaye}, {Kokorowski}, {Kwok}, {Lanclos},
  {Milner}, {Moore}, {O'Sullivan}, {Parihar}, {Park}, {Phillips}, {Rizzi},
  {Rockosi}, {Rodriguez}, {Salaun}, {Seaman}, {Sheikh}, {Weiss}, \&
  {Zarzaca}}]{morrissey2018}
{Morrissey}, P., {Matuszewski}, M., {Martin}, D.~C., {et~al.} 2018, \apj, 864,
  93

\bibitem[{{Moustakas} {et~al.}(2006){Moustakas}, {Kennicutt}, \&
  {Tremonti}}]{moustakas2006}
{Moustakas}, J., {Kennicutt}, Jr., R.~C., \& {Tremonti}, C.~A. 2006, \apj, 642,
  775

\bibitem[{{Muzahid} {et~al.}(2020){Muzahid}, {Schaye}, {Marino}, {Cantalupo},
  {Brinchmann}, {Contini}, {Wendt}, {Wisotzki}, {Zabl}, {Bouch{\'e}},
  {Akhlaghi}, {Chen}, {Claeyssens}, {Johnson}, {Leclercq}, {Maseda}, {Matthee},
  {Richard}, {Urrutia}, \& {Verhamme}}]{muzahid2020}
{Muzahid}, S., {Schaye}, J., {Marino}, R.~A., {et~al.} 2020, \mnras, 496, 1013

\bibitem[{{Muzahid} {et~al.}(2021){Muzahid}, {Schaye}, {Cantalupo}, {Marino},
  {Bouch{\'e}}, {Johnson}, {Maseda}, {Wendt}, {Wisotzki}, \&
  {Zabl}}]{muzahid2021}
{Muzahid}, S., {Schaye}, J., {Cantalupo}, S., {et~al.} 2021, \mnras, 508, 5612

\bibitem[{{Narayanan} {et~al.}(2021){Narayanan}, {Sameer}, {Muzahid},
  {Johnson}, {Udhwani}, {Charlton}, {Mauerhofer}, {Schaye}, \&
  {Yadav}}]{narayanan2021}
{Narayanan}, A., {Sameer}, {Muzahid}, S., {et~al.} 2021, \mnras, 505, 738

\bibitem[{{Navarro} {et~al.}(1996){Navarro}, {Frenk}, \& {White}}]{nfw1996}
{Navarro}, J.~F., {Frenk}, C.~S., \& {White}, S.~D.~M. 1996, \apj, 462, 563

\bibitem[{{Navarro} {et~al.}(1997){Navarro}, {Frenk}, \& {White}}]{nfw1997}
---. 1997, \apj, 490, 493

\bibitem[{{Nelson} {et~al.}(2019){Nelson}, {Pillepich}, {Springel}, {Pakmor},
  {Weinberger}, {Genel}, {Torrey}, {Vogelsberger}, {Marinacci}, \&
  {Hernquist}}]{nelson2019}
{Nelson}, D., {Pillepich}, A., {Springel}, V., {et~al.} 2019, \mnras, 490, 3234

\bibitem[{{Nidever} {et~al.}(2018){Nidever}, {Dey}, {Olsen}, {Ridgway},
  {Nikutta}, {Juneau}, {Fitzpatrick}, {Scott}, \& {Valdes}}]{nidever2018}
{Nidever}, D.~L., {Dey}, A., {Olsen}, K., {et~al.} 2018, \aj, 156, 131

\bibitem[{{Nidever} {et~al.}(2021){Nidever}, {Dey}, {Fasbender}, {Juneau},
  {Meisner}, {Wishart}, {Scott}, {Matt}, {Nikutta}, \& {Pucha}}]{nidever2021}
{Nidever}, D.~L., {Dey}, A., {Fasbender}, K., {et~al.} 2021, \aj, 161, 192

\bibitem[{{Nielsen} {et~al.}(2013){Nielsen}, {Churchill}, \&
  {Kacprzak}}]{nielsen2013}
{Nielsen}, N.~M., {Churchill}, C.~W., \& {Kacprzak}, G.~G. 2013, \apj, 776, 115

\bibitem[{{Norris} {et~al.}(2021){Norris}, {Muzahid}, {Charlton}, {Kacprzak},
  {Wakker}, \& {Churchill}}]{norris2021}
{Norris}, J.~M., {Muzahid}, S., {Charlton}, J.~C., {et~al.} 2021, \mnras, 506,
  5640

\bibitem[{{Oppenheimer} \& {Dav{\'e}}(2006)}]{oppenheimer2006}
{Oppenheimer}, B.~D., \& {Dav{\'e}}, R. 2006, \mnras, 373, 1265

\bibitem[{{Oppenheimer} {et~al.}(2012){Oppenheimer}, {Dav{\'e}}, {Katz},
  {Kollmeier}, \& {Weinberg}}]{oppenheimer2012}
{Oppenheimer}, B.~D., {Dav{\'e}}, R., {Katz}, N., {Kollmeier}, J.~A., \&
  {Weinberg}, D.~H. 2012, \mnras, 420, 829

\bibitem[{{Oppenheimer} {et~al.}(2010){Oppenheimer}, {Dav{\'e}}, {Kere{\v s}},
  {Fardal}, {Katz}, {Kollmeier}, \& {Weinberg}}]{oppenheimer2010}
{Oppenheimer}, B.~D., {Dav{\'e}}, R., {Kere{\v s}}, D., {et~al.} 2010, \mnras,
  406, 2325

\bibitem[{{Ott}(2012)}]{ott2012}
{Ott}, T. 2012, {QFitsView: FITS file viewer}, , , ascl:1210.019

\bibitem[{{Ouchi} {et~al.}(2020){Ouchi}, {Ono}, \& {Shibuya}}]{ouchi2020}
{Ouchi}, M., {Ono}, Y., \& {Shibuya}, T. 2020, \araa, 58, 617

\bibitem[{{Peng} {et~al.}(2002){Peng}, {Ho}, {Impey}, \& {Rix}}]{peng2002}
{Peng}, C.~Y., {Ho}, L.~C., {Impey}, C.~D., \& {Rix}, H.-W. 2002, \aj, 124, 266

\bibitem[{{Peng} {et~al.}(2010){Peng}, {Ho}, {Impey}, \& {Rix}}]{peng2010}
---. 2010, \aj, 139, 2097

\bibitem[{{P{\'e}roux} {et~al.}(2020){P{\'e}roux}, {Nelson}, {van de Voort},
  {Pillepich}, {Marinacci}, {Vogelsberger}, \& {Hernquist}}]{peroux2020}
{P{\'e}roux}, C., {Nelson}, D., {van de Voort}, F., {et~al.} 2020, \mnras, 499,
  2462

\bibitem[{{Pessa} {et~al.}(2020){Pessa}, {Tejos}, \& {Moya}}]{pessa2020}
{Pessa}, I., {Tejos}, N., \& {Moya}, C. 2020, in Astronomical Society of the
  Pacific Conference Series, Vol. 522, Astronomical Data Analysis Software and
  Systems XXVII, ed. P.~{Ballester}, J.~{Ibsen}, M.~{Solar}, \&
  K.~{Shortridge}, 61

\bibitem[{{Pillepich} {et~al.}(2019){Pillepich}, {Nelson}, {Springel},
  {Pakmor}, {Torrey}, {Weinberger}, {Vogelsberger}, {Marinacci}, {Genel}, {van
  der Wel}, \& {Hernquist}}]{pillepich2019}
{Pillepich}, A., {Nelson}, D., {Springel}, V., {et~al.} 2019, \mnras, 490, 3196

\bibitem[{{Planck Collaboration} {et~al.}(2016){Planck Collaboration}, {Ade},
  {Aghanim}, {Arnaud}, {Ashdown}, {Aumont}, {Baccigalupi}, {Banday},
  {Barreiro}, {Bartlett}, \& et~al.}]{planck2016}
{Planck Collaboration}, {Ade}, P.~A.~R., {Aghanim}, N., {et~al.} 2016, \aap,
  594, A13

\bibitem[{{Pointon} {et~al.}(2019){Pointon}, {Kacprzak}, {Nielsen}, {Muzahid},
  {Murphy}, {Churchill}, \& {Charlton}}]{pointon2019}
{Pointon}, S.~K., {Kacprzak}, G.~G., {Nielsen}, N.~M., {et~al.} 2019, \apj,
  883, 78

\bibitem[{{Portinari} {et~al.}(2004){Portinari}, {Sommer-Larsen}, \&
  {Tantalo}}]{portinari2004}
{Portinari}, L., {Sommer-Larsen}, J., \& {Tantalo}, R. 2004, \mnras, 347, 691

\bibitem[{{Privon} {et~al.}(2020){Privon}, {Ricci}, {Aalto}, {Viti}, {Armus},
  {D{\'\i}az-Santos}, {Gonz{\'a}lez-Alfonso}, {Iwasawa}, {Jeff}, {Treister},
  {Bauer}, {Evans}, {Garg}, {Herrero-Illana}, {Mazzarella}, {Larson}, {Blecha},
  {Barcos-Mu{\~n}oz}, {Charmandaris}, {Stierwalt}, \&
  {P{\'e}rez-Torres}}]{privon2020}
{Privon}, G.~C., {Ricci}, C., {Aalto}, S., {et~al.} 2020, \apj, 893, 149

\bibitem[{{Prochaska} {et~al.}(2011){Prochaska}, {Weiner}, {Chen}, {Mulchaey},
  \& {Cooksey}}]{prochaska2011}
{Prochaska}, J.~X., {Weiner}, B., {Chen}, H.~W., {Mulchaey}, J., \& {Cooksey},
  K. 2011, \apj, 740, 91

\bibitem[{{Prochaska} {et~al.}(2006){Prochaska}, {Weiner}, {Chen}, \&
  {Mulchaey}}]{prochaska2006}
{Prochaska}, J.~X., {Weiner}, B.~J., {Chen}, H.-W., \& {Mulchaey}, J.~S. 2006,
  \apj, 643, 680

\bibitem[{{Prochaska} {et~al.}(2017){Prochaska}, {Werk}, {Worseck}, {Tripp},
  {Tumlinson}, {Burchett}, {Fox}, {Fumagalli}, {Lehner}, {Peeples}, \&
  {Tejos}}]{prochaska2017}
{Prochaska}, J.~X., {Werk}, J.~K., {Worseck}, G., {et~al.} 2017, \apj, 837, 169

\bibitem[{{Prochaska} {et~al.}(2019){Prochaska}, {Burchett}, {Tripp}, {Werk},
  {Willmer}, {Howk}, {Lange}, {Tejos}, {Meiring}, {Tumlinson}, {Lehner},
  {Ford}, \& {Dav{\'e}}}]{prochaska2019}
{Prochaska}, J.~X., {Burchett}, J.~N., {Tripp}, T.~M., {et~al.} 2019, \apjs,
  243, 24

\bibitem[{{Quiret} {et~al.}(2016){Quiret}, {P{\'e}roux}, {Zafar}, {Kulkarni},
  {Jenkins}, {Milliard}, {Rahmani}, {Popping}, {Rao}, {Turnshek}, \&
  {Monier}}]{quiret2016}
{Quiret}, S., {P{\'e}roux}, C., {Zafar}, T., {et~al.} 2016, \mnras, 458, 4074

\bibitem[{{Rafelski} {et~al.}(2012){Rafelski}, {Wolfe}, {Prochaska},
  {Neeleman}, \& {Mendez}}]{rafelski2012}
{Rafelski}, M., {Wolfe}, A.~M., {Prochaska}, J.~X., {Neeleman}, M., \&
  {Mendez}, A.~J. 2012, \apj, 755, 89

\bibitem[{{Rahmati} \& {Oppenheimer}(2018)}]{rahmati2018}
{Rahmati}, A., \& {Oppenheimer}, B.~D. 2018, \mnras, 476, 4865

\bibitem[{{Rahmati} \& {Schaye}(2014)}]{rs2014}
{Rahmati}, A., \& {Schaye}, J. 2014, \mnras, 438, 529

\bibitem[{{Rahmati} {et~al.}(2016){Rahmati}, {Schaye}, {Crain}, {Oppenheimer},
  {Schaller}, \& {Theuns}}]{rahmati2016}
{Rahmati}, A., {Schaye}, J., {Crain}, R.~A., {et~al.} 2016, \mnras, 459, 310

\bibitem[{{Ribaudo} {et~al.}(2011){Ribaudo}, {Lehner}, {Howk}, {Werk}, {Tripp},
  {Prochaska}, {Meiring}, \& {Tumlinson}}]{ribaudo2011b}
{Ribaudo}, J., {Lehner}, N., {Howk}, J.~C., {et~al.} 2011, \apj, 743, 207

\bibitem[{{Rodr{\'{\i}}guez-Puebla} {et~al.}(2017){Rodr{\'{\i}}guez-Puebla},
  {Primack}, {Avila-Reese}, \& {Faber}}]{rodriguez-puebla2017}
{Rodr{\'{\i}}guez-Puebla}, A., {Primack}, J.~R., {Avila-Reese}, V., \& {Faber},
  S.~M. 2017, \mnras, 470, 651

\bibitem[{{Rose} {et~al.}(2011){Rose}, {Tadhunter}, {Holt}, {Ramos Almeida}, \&
  {Littlefair}}]{rose2011}
{Rose}, M., {Tadhunter}, C.~N., {Holt}, J., {Ramos Almeida}, C., \&
  {Littlefair}, S.~P. 2011, \mnras, 414, 3360

\bibitem[{{Ross} {et~al.}(2020){Ross}, {Bautista}, {Tojeiro}, {Alam}, {Bailey},
  {Burtin}, {Comparat}, {Dawson}, {de Mattia}, {du Mas des Bourboux},
  {Gil-Mar{\'\i}n}, {Hou}, {Kong}, {Lyke}, {Mohammad}, {Moustakas}, {Mueller},
  {Myers}, {Percival}, {Raichoor}, {Rezaie}, {Seo}, {Smith}, {Tinker},
  {Zarrouk}, {Zhao}, {Zhao}, {Bizyaev}, {Brinkmann}, {Brownstein}, {Rosell},
  {Chabanier}, {Choi}, {Chuang}, {Cruz-Gonzalez}, {de la Macorra}, {de la
  Torre}, {Escoffier}, {Fromenteau}, {Higley}, {Jullo}, {Kneib}, {McLane},
  {Mu{\~n}oz-Guti{\'e}rrez}, {Neveux}, {Newman}, {Nitschelm},
  {Palanque-Delabrouille}, {Paviot}, {Pullen}, {Rossi}, {Ruhlmann-Kleider},
  {Schneider}, {Maga{\~n}a}, {Vivek}, \& {Zhang}}]{ross2020}
{Ross}, A.~J., {Bautista}, J., {Tojeiro}, R., {et~al.} 2020, \mnras, 498, 2354

\bibitem[{{Rudie} {et~al.}(2012){Rudie}, {Steidel}, {Trainor}, {Rakic},
  {Bogosavljevi{\'c}}, {Pettini}, {Reddy}, {Shapley}, {Erb}, \&
  {Law}}]{rudie2012}
{Rudie}, G.~C., {Steidel}, C.~C., {Trainor}, R.~F., {et~al.} 2012, \apj, 750,
  67

\bibitem[{{Salim} {et~al.}(2007){Salim}, {Rich}, {Charlot}, {Brinchmann},
  {Johnson}, {Schiminovich}, {Seibert}, {Mallery}, {Heckman}, {Forster},
  {Friedman}, {Martin}, {Morrissey}, {Neff}, {Small}, {Wyder}, {Bianchi},
  {Donas}, {Lee}, {Madore}, {Milliard}, {Szalay}, {Welsh}, \& {Yi}}]{salim2007}
{Salim}, S., {Rich}, R.~M., {Charlot}, S., {et~al.} 2007, \apjs, 173, 267

\bibitem[{{Savage} {et~al.}(2010){Savage}, {Narayanan}, {Wakker}, {Stocke},
  {Keeney}, {Shull}, {Sembach}, {Yao}, \& {Green}}]{savage2010}
{Savage}, B.~D., {Narayanan}, A., {Wakker}, B.~P., {et~al.} 2010, \apj, 719,
  1526

\bibitem[{{Schaye}(2001)}]{schaye2001}
{Schaye}, J. 2001, \apj, 559, 507

\bibitem[{{Schaye} {et~al.}(2003){Schaye}, {Aguirre}, {Kim}, {Theuns}, {Rauch},
  \& {Sargent}}]{schaye2003}
{Schaye}, J., {Aguirre}, A., {Kim}, T.-S., {et~al.} 2003, \apj, 596, 768

\bibitem[{{Schaye} {et~al.}(2010){Schaye}, {Dalla Vecchia}, {Booth}, {Wiersma},
  {Theuns}, {Haas}, {Bertone}, {Duffy}, {McCarthy}, \& {van de
  Voort}}]{schaye2010}
{Schaye}, J., {Dalla Vecchia}, C., {Booth}, C.~M., {et~al.} 2010, \mnras, 402,
  1536

\bibitem[{{Schaye} {et~al.}(2015){Schaye}, {Crain}, {Bower}, {Furlong},
  {Schaller}, {Theuns}, {Dalla Vecchia}, {Frenk}, {McCarthy}, {Helly},
  {Jenkins}, {Rosas-Guevara}, {White}, {Baes}, {Booth}, {Camps}, {Navarro},
  {Qu}, {Rahmati}, {Sawala}, {Thomas}, \& {Trayford}}]{schaye2015}
{Schaye}, J., {Crain}, R.~A., {Bower}, R.~G., {et~al.} 2015, \mnras, 446, 521

\bibitem[{{Schlafly} \& {Finkbeiner}(2011)}]{schlafly2011}
{Schlafly}, E.~F., \& {Finkbeiner}, D.~P. 2011, \apj, 737, 103

\bibitem[{{Schlafly} {et~al.}(2019){Schlafly}, {Meisner}, \&
  {Green}}]{schlafly2019}
{Schlafly}, E.~F., {Meisner}, A.~M., \& {Green}, G.~M. 2019, \apjs, 240, 30

\bibitem[{{Schlegel} {et~al.}(1998){Schlegel}, {Finkbeiner}, \&
  {Davis}}]{schlegel1998}
{Schlegel}, D.~J., {Finkbeiner}, D.~P., \& {Davis}, M. 1998, \apj, 500, 525

\bibitem[{{Schreiber} {et~al.}(2015){Schreiber}, {Pannella}, {Elbaz},
  {B{\'e}thermin}, {Inami}, {Dickinson}, {Magnelli}, {Wang}, {Aussel}, {Daddi},
  {Juneau}, {Shu}, {Sargent}, {Buat}, {Faber}, {Ferguson}, {Giavalisco},
  {Koekemoer}, {Magdis}, {Morrison}, {Papovich}, {Santini}, \&
  {Scott}}]{schreiber2015}
{Schreiber}, C., {Pannella}, M., {Elbaz}, D., {et~al.} 2015, \aap, 575, A74

\bibitem[{{Schroetter} {et~al.}(2016){Schroetter}, {Bouch{\'e}}, {Wendt},
  {Contini}, {Finley}, {Pell{\'o}}, {Bacon}, {Cantalupo}, {Marino}, {Richard},
  {Lilly}, {Schaye}, {Soto}, {Steinmetz}, {Straka}, \&
  {Wisotzki}}]{schroetter2016}
{Schroetter}, I., {Bouch{\'e}}, N., {Wendt}, M., {et~al.} 2016, \apj, 833, 39

\bibitem[{{Schroetter} {et~al.}(2019){Schroetter}, {Bouch{\'e}}, {Zabl},
  {Contini}, {Wendt}, {Schaye}, {Mitchell}, {Muzahid}, {Marino}, {Bacon},
  {Lilly}, {Richard}, \& {Wisotzki}}]{schroetter2019}
{Schroetter}, I., {Bouch{\'e}}, N.~F., {Zabl}, J., {et~al.} 2019, \mnras, 490,
  4368

\bibitem[{{Shull} {et~al.}(2014){Shull}, {Danforth}, \& {Tilton}}]{shull2014}
{Shull}, J.~M., {Danforth}, C.~W., \& {Tilton}, E.~M. 2014, \apj, 796, 49

\bibitem[{{Skrutskie} {et~al.}(2006){Skrutskie}, {Cutri}, {Stiening},
  {Weinberg}, {Schneider}, {Carpenter}, {Beichman}, {Capps}, {Chester},
  {Elias}, {Huchra}, {Liebert}, {Lonsdale}, {Monet}, {Price}, {Seitzer},
  {Jarrett}, {Kirkpatrick}, {Gizis}, {Howard}, {Evans}, {Fowler}, {Fullmer},
  {Hurt}, {Light}, {Kopan}, {Marsh}, {McCallon}, {Tam}, {Van Dyk}, \&
  {Wheelock}}]{skrutskie2006}
{Skrutskie}, M.~F., {Cutri}, R.~M., {Stiening}, R., {et~al.} 2006, \aj, 131,
  1163

\bibitem[{{Smithsonian Astrophysical Observatory}(2000)}]{sao2000}
{Smithsonian Astrophysical Observatory}. 2000, {SAOImage DS9: A utility for
  displaying astronomical images in the X11 window environment}, , ,
  ascl:0003.002

\bibitem[{{Som} {et~al.}(2015){Som}, {Kulkarni}, {Meiring}, {York},
  {P{\'e}roux}, {Lauroesch}, {Aller}, \& {Khare}}]{som2015}
{Som}, D., {Kulkarni}, V.~P., {Meiring}, J., {et~al.} 2015, \apj, 806, 25

\bibitem[{{Spinrad} {et~al.}(1993){Spinrad}, {Filippenko}, {Yee}, {Ellingson},
  {Blades}, {Bahcall}, {Jannuzi}, {Bechtold}, \& {Dobrzycki}}]{spinrad1993}
{Spinrad}, H., {Filippenko}, A.~V., {Yee}, H.~K., {et~al.} 1993, \aj, 106, 1

\bibitem[{{Steidel} \& {Dickinson}(1992)}]{steidel1992}
{Steidel}, C.~C., \& {Dickinson}, M. 1992, \apj, 394, 81

\bibitem[{{STScI Development Team}(2012)}]{stsci2012}
{STScI Development Team}. 2012, {DrizzlePac: HST image software}, , ,
  ascl:1212.011

\bibitem[{{STScI Development Team}(2018)}]{stsci2018}
---. 2018, {synphot: Synthetic photometry using Astropy}, , , ascl:1811.001

\bibitem[{{Suresh} {et~al.}(2019){Suresh}, {Nelson}, {Genel}, {Rubin}, \&
  {Hernquist}}]{suresh2019}
{Suresh}, J., {Nelson}, D., {Genel}, S., {Rubin}, K. H.~R., \& {Hernquist}, L.
  2019, \mnras, 483, 4040

\bibitem[{{Tacconi} {et~al.}(2010){Tacconi}, {Genzel}, {Neri}, {Cox}, {Cooper},
  {Shapiro}, {Bolatto}, {Bouch{\'e}}, {Bournaud}, {Burkert}, {Combes},
  {Comerford}, {Davis}, {Schreiber}, {Garcia-Burillo}, {Gracia-Carpio}, {Lutz},
  {Naab}, {Omont}, {Shapley}, {Sternberg}, \& {Weiner}}]{tacconi2010}
{Tacconi}, L.~J., {Genzel}, R., {Neri}, R., {et~al.} 2010, \nat, 463, 781

\bibitem[{{Tejos} {et~al.}(2014){Tejos}, {Morris}, {Finn}, {Crighton},
  {Bechtold}, {Jannuzi}, {Schaye}, {Theuns}, {Altay}, {Le F{\`e}vre},
  {Ryan-Weber}, \& {Dav{\'e}}}]{tejos2014}
{Tejos}, N., {Morris}, S.~L., {Finn}, C.~W., {et~al.} 2014, \mnras, 437, 2017

\bibitem[{{The Astropy Collaboration} {et~al.}(2018){The Astropy
  Collaboration}, Price-Whelan, Sip{\H o}cz, G{\"u}nther, Lim, Crawford,
  Conseil, Shupe, Craig, Dencheva, Ginsburg, VanderPlas, Bradley,
  P{\'e}rez-Su{\'a}rez, de~Val-Borro, Contributors), Aldcroft, Cruz,
  Robitaille, Tollerud, Committee), Ardelean, Babej, Bach, Bachetti, Bakanov,
  Bamford, Barentsen, Barmby, Baumbach, Berry, Biscani, Boquien, Bostroem,
  Bouma, Brammer, Bray, Breytenbach, Buddelmeijer, Burke, Calderone,
  Rodr{\'\i}guez, Cara, Cardoso, Cheedella, Copin, Corrales, Crichton,
  D'Avella, Deil, Depagne, Dietrich, Donath, Droettboom, Earl, Erben, Fabbro,
  Ferreira, Finethy, Fox, Garrison, Gibbons, Goldstein, Gommers, Greco,
  Greenfield, Groener, Grollier, Hagen, Hirst, Homeier, Horton, Hosseinzadeh,
  Hu, Hunkeler, Ivezi{\'c}, Jain, Jenness, Kanarek, Kendrew, Kern, Kerzendorf,
  Khvalko, King, Kirkby, Kulkarni, Kumar, Lee, Lenz, Littlefair, Ma, Macleod,
  Mastropietro, McCully, Montagnac, Morris, Mueller, Mumford, Muna, Murphy,
  Nelson, Nguyen, Ninan, N{\"o}the, Ogaz, Oh, Parejko, Parley, Pascual, Patil,
  Patil, Plunkett, Prochaska, Rastogi, Janga, Sabater, Sakurikar, Seifert,
  Sherbert, Sherwood-Taylor, Shih, Sick, Silbiger, Singanamalla, Singer,
  Sladen, Sooley, Sornarajah, Streicher, Teuben, Thomas, Tremblay, Turner,
  Terr{\'o}n, van Kerkwijk, de~la Vega, Watkins, Weaver, Whitmore, Woillez,
  Zabalza, \& Contributors)}]{astropy2018}
{The Astropy Collaboration}, Price-Whelan, A.~M., Sip{\H o}cz, B.~M., {et~al.}
  2018, The Astronomical Journal, 156, 123

\bibitem[{{Thom} {et~al.}(2012){Thom}, {Tumlinson}, {Werk}, {Prochaska},
  {Oppenheimer}, {Peeples}, {Tripp}, {Katz}, {O'Meara}, {Ford}, {Dav{\'e}},
  {Sembach}, \& {Weinberg}}]{thom2012}
{Thom}, C., {Tumlinson}, J., {Werk}, J.~K., {et~al.} 2012, \apjl, 758, L41

\bibitem[{{Tripp} {et~al.}(2011){Tripp}, {Meiring}, {Prochaska}, {Willmer},
  {Howk}, {Werk}, {Jenkins}, {Bowen}, {Lehner}, {Sembach}, {Thom}, \&
  {Tumlinson}}]{tripp2011}
{Tripp}, T.~M., {Meiring}, J.~D., {Prochaska}, J.~X., {et~al.} 2011, Science,
  334, 952

\bibitem[{{Tumlinson} {et~al.}(2011){Tumlinson}, {Thom}, {Werk}, {Prochaska},
  {Tripp}, {Weinberg}, {Peeples}, {O'Meara}, {Oppenheimer}, {Meiring}, {Katz},
  {Dav{\'e}}, {Ford}, \& {Sembach}}]{tumlinson2011}
{Tumlinson}, J., {Thom}, C., {Werk}, J.~K., {et~al.} 2011, Science, 334, 948

\bibitem[{{Tumlinson} {et~al.}(2013){Tumlinson}, {Thom}, {Werk}, {Prochaska},
  {Tripp}, {Katz}, {Dav{\'e}}, {Oppenheimer}, {Meiring}, {Ford}, {O'Meara},
  {Peeples}, {Sembach}, \& {Weinberg}}]{tumlinson2013}
---. 2013, \apj, 777, 59

\bibitem[{{van de Voort} \& {Schaye}(2012)}]{vs2012}
{van de Voort}, F., \& {Schaye}, J. 2012, \mnras, 423, 2991

\bibitem[{{van de Voort} {et~al.}(2012){van de Voort}, {Schaye}, {Altay}, \&
  {Theuns}}]{vandevoort2012}
{van de Voort}, F., {Schaye}, J., {Altay}, G., \& {Theuns}, T. 2012, \mnras,
  421, 2809

\bibitem[{{van der Walt} {et~al.}(2011){van der Walt}, {Colbert}, \&
  {Varoquaux}}]{vanderwalt2011}
{van der Walt}, S., {Colbert}, S.~C., \& {Varoquaux}, G. 2011, Computing in
  Science and Engineering, 13, 22

\bibitem[{{Vazdekis} {et~al.}(2010){Vazdekis}, {S{\'a}nchez-Bl{\'a}zquez},
  {Falc{\'o}n-Barroso}, {Cenarro}, {Beasley}, {Cardiel}, {Gorgas}, \&
  {Peletier}}]{vazdekis2010}
{Vazdekis}, A., {S{\'a}nchez-Bl{\'a}zquez}, P., {Falc{\'o}n-Barroso}, J.,
  {et~al.} 2010, \mnras, 404, 1639

\bibitem[{Virtanen {et~al.}(2020)Virtanen, Gommers, Oliphant, Haberland, Reddy,
  Cournapeau, Burovski, Peterson, Weckesser, Bright, {van der Walt}, Brett,
  Wilson, Millman, Mayorov, Nelson, Jones, Kern, Larson, Carey, Polat, Feng,
  Moore, {VanderPlas}, Laxalde, Perktold, Cimrman, Henriksen, Quintero, Harris,
  Archibald, Ribeiro, Pedregosa, {van Mulbregt}, \& {SciPy 1.0
  Contributors}}]{scipy2020}
Virtanen, P., Gommers, R., Oliphant, T.~E., {et~al.} 2020, Nature Methods, 17,
  261

\bibitem[{{Wechsler} \& {Tinker}(2018)}]{wt2018}
{Wechsler}, R.~H., \& {Tinker}, J.~L. 2018, \araa, 56, 435

\bibitem[{{Weilbacher} {et~al.}(2020){Weilbacher}, {Palsa}, {Streicher},
  {Bacon}, {Urrutia}, {Wisotzki}, {Conseil}, {Husemann}, {Jarno}, {Kelz},
  {P{\'e}contal-Rousset}, {Richard}, {Roth}, {Selman}, \&
  {Vernet}}]{weilbacher2020}
{Weilbacher}, P.~M., {Palsa}, R., {Streicher}, O., {et~al.} 2020, \aap, 641,
  A28

\bibitem[{{Wendt} {et~al.}(2021){Wendt}, {Bouch{\'e}}, {Zabl}, {Schroetter}, \&
  {Muzahid}}]{wendt2021}
{Wendt}, M., {Bouch{\'e}}, N.~F., {Zabl}, J., {Schroetter}, I., \& {Muzahid},
  S. 2021, \mnras, 502, 3733

\bibitem[{{Werk} {et~al.}(2012){Werk}, {Prochaska}, {Thom}, {Tumlinson},
  {Tripp}, {O'Meara}, \& {Meiring}}]{werk2012}
{Werk}, J.~K., {Prochaska}, J.~X., {Thom}, C., {et~al.} 2012, \apjs, 198, 3

\bibitem[{{Werk} {et~al.}(2014){Werk}, {Prochaska}, {Tumlinson}, {Peeples},
  {Tripp}, {Fox}, {Lehner}, {Thom}, {O'Meara}, {Ford}, {Bordoloi}, {Katz},
  {Tejos}, {Oppenheimer}, {Dav{\'e}}, \& {Weinberg}}]{werk2014}
{Werk}, J.~K., {Prochaska}, J.~X., {Tumlinson}, J., {et~al.} 2014, \apj, 792, 8

\bibitem[{{Weymann} {et~al.}(2001){Weymann}, {Vogel}, {Veilleux}, \&
  {Epps}}]{weymann2001}
{Weymann}, R.~J., {Vogel}, S.~N., {Veilleux}, S., \& {Epps}, H.~W. 2001, \apj,
  561, 559

\bibitem[{{Wiersma} {et~al.}(2009){Wiersma}, {Schaye}, \&
  {Smith}}]{wiersma2009}
{Wiersma}, R.~P.~C., {Schaye}, J., \& {Smith}, B.~D. 2009, \mnras, 393, 99

\bibitem[{{Wiersma} {et~al.}(2011){Wiersma}, {Schaye}, \&
  {Theuns}}]{wiersma2011}
{Wiersma}, R. P.~C., {Schaye}, J., \& {Theuns}, T. 2011, \mnras, 415, 353

\bibitem[{{Wilde} {et~al.}(2021){Wilde}, {Werk}, {Burchett}, {Prochaska},
  {Tchernyshyov}, {Tripp}, {Tejos}, {Lehner}, {Bordoloi}, {O'Meara}, \&
  {Tumlinson}}]{wilde2021}
{Wilde}, M.~C., {Werk}, J.~K., {Burchett}, J.~N., {et~al.} 2021, \apj, 912, 9

\bibitem[{{Wisotzki} {et~al.}(2016){Wisotzki}, {Bacon}, {Blaizot},
  {Brinchmann}, {Herenz}, {Schaye}, {Bouch{\'e}}, {Cantalupo}, {Contini},
  {Carollo}, {Caruana}, {Courbot}, {Emsellem}, {Kamann}, {Kerutt}, {Leclercq},
  {Lilly}, {Patr{\'\i}cio}, {Sandin}, {Steinmetz}, {Straka}, {Urrutia},
  {Verhamme}, {Weilbacher}, \& {Wendt}}]{wisotzki2016}
{Wisotzki}, L., {Bacon}, R., {Blaizot}, J., {et~al.} 2016, \aap, 587, A98

\bibitem[{{Wotta} {et~al.}(2019){Wotta}, {Lehner}, {Howk}, {O'Meara},
  {Oppenheimer}, \& {Cooksey}}]{wotta2019}
{Wotta}, C.~B., {Lehner}, N., {Howk}, J.~C., {et~al.} 2019, \apj, 872, 81

\bibitem[{{Wotta} {et~al.}(2016){Wotta}, {Lehner}, {Howk}, {O'Meara}, \&
  {Prochaska}}]{wotta2016}
{Wotta}, C.~B., {Lehner}, N., {Howk}, J.~C., {O'Meara}, J.~M., \& {Prochaska},
  J.~X. 2016, \apj, 831, 95

\bibitem[{{York} {et~al.}(2000){York}, {Adelman}, {Anderson}, {Anderson},
  {Annis}, {Bahcall}, {Bakken}, {Barkhouser}, {Bastian}, {Berman}, {Boroski},
  {Bracker}, {Briegel}, {Briggs}, {Brinkmann}, {Brunner}, {Burles}, {Carey},
  {Carr}, {Castander}, {Chen}, {Colestock}, {Connolly}, {Crocker}, {Csabai},
  {Czarapata}, {Davis}, {Doi}, {Dombeck}, {Eisenstein}, {Ellman}, {Elms},
  {Evans}, {Fan}, {Federwitz}, {Fiscelli}, {Friedman}, {Frieman}, {Fukugita},
  {Gillespie}, {Gunn}, {Gurbani}, {de Haas}, {Haldeman}, {Harris}, {Hayes},
  {Heckman}, {Hennessy}, {Hindsley}, {Holm}, {Holmgren}, {Huang}, {Hull},
  {Husby}, {Ichikawa}, {Ichikawa}, {Ivezi{\'c}}, {Kent}, {Kim}, {Kinney},
  {Klaene}, {Kleinman}, {Kleinman}, {Knapp}, {Korienek}, {Kron}, {Kunszt},
  {Lamb}, {Lee}, {Leger}, {Limmongkol}, {Lindenmeyer}, {Long}, {Loomis},
  {Loveday}, {Lucinio}, {Lupton}, {MacKinnon}, {Mannery}, {Mantsch}, {Margon},
  {McGehee}, {McKay}, {Meiksin}, {Merelli}, {Monet}, {Munn}, {Narayanan},
  {Nash}, {Neilsen}, {Neswold}, {Newberg}, {Nichol}, {Nicinski}, {Nonino},
  {Okada}, {Okamura}, {Ostriker}, {Owen}, {Pauls}, {Peoples}, {Peterson},
  {Petravick}, {Pier}, {Pope}, {Pordes}, {Prosapio}, {Rechenmacher}, {Quinn},
  {Richards}, {Richmond}, {Rivetta}, {Rockosi}, {Ruthmansdorfer}, {Sandford},
  {Schlegel}, {Schneider}, {Sekiguchi}, {Sergey}, {Shimasaku}, {Siegmund},
  {Smee}, {Smith}, {Snedden}, {Stone}, {Stoughton}, {Strauss}, {Stubbs},
  {SubbaRao}, {Szalay}, {Szapudi}, {Szokoly}, {Thakar}, {Tremonti}, {Tucker},
  {Uomoto}, {Vanden Berk}, {Vogeley}, {Waddell}, {Wang}, {Watanabe},
  {Weinberg}, {Yanny}, {Yasuda}, \& {SDSS Collaboration}}]{york2000}
{York}, D.~G., {Adelman}, J., {Anderson}, John~E., J., {et~al.} 2000, \aj, 120,
  1579

\bibitem[{{Zabl} {et~al.}(2019){Zabl}, {Bouch{\'e}}, {Schroetter}, {Wendt},
  {Finley}, {Schaye}, {Conseil}, {Contini}, {Marino}, {Mitchell}, {Muzahid},
  {Pezzulli}, \& {Wisotzki}}]{zabl2019}
{Zabl}, J., {Bouch{\'e}}, N.~F., {Schroetter}, I., {et~al.} 2019, \mnras, 485,
  1961

\bibitem[{{Zahedy} {et~al.}(2019){Zahedy}, {Chen}, {Johnson}, {Pierce},
  {Rauch}, {Huang}, {Weiner}, \& {Gauthier}}]{zahedy2019}
{Zahedy}, F.~S., {Chen}, H.-W., {Johnson}, S.~D., {et~al.} 2019, \mnras, 484,
  2257

\bibitem[{{Zahedy} {et~al.}(2021){Zahedy}, {Chen}, {Cooper}, {Boettcher},
  {Johnson}, {Rudie}, {Chen}, {Cantalupo}, {Cooksey}, {Faucher-Gigu{\`e}re},
  {Greene}, {Lopez}, {Mulchaey}, {Penton}, {Petitjean}, {Putman}, {Rafelski},
  {Rauch}, {Schaye}, {Simcoe}, \& {Walth}}]{zahedy2021}
{Zahedy}, F.~S., {Chen}, H.-W., {Cooper}, T.~M., {et~al.} 2021, \mnras, 506,
  877

\end{thebibliography}

\appendix

\makeatletter
\renewcommand{\thefigure}{A\@arabic\c@figure}
\renewcommand{\thetable}{A\@arabic\c@table}
\makeatother
\setcounter{table}{0}
\setcounter{figure}{0}

\section{KCWI and MUSE IFU Observations}\label{sec:appa}

In this appendix we list the information for the cut candidate galaxies in Table~\ref{tab:galinfo_other} and the galaxy location information for all candidate galaxies in Table~\ref{tab:galinfoapp}. We display the rest of the \hst\ images with the IFU pointings and identified candidate galaxies in Figures~\ref{fig:PG1338}--\ref{fig:PG1522}.

\begin{deluxetable*}{lcccccccccccc}
\tabletypesize{\small}
\tabcolsep=2pt
\tablecaption{Cut Candidate Galaxy Information \label{tab:galinfo_other}}
\tablehead{\colhead{SE\#} & \colhead{$\rho$} & \colhead{$|\Delta v|$} & \colhead{$\log M_{\star}$} & \colhead{$\log M_{\rm h}$} & \colhead{$R_{\rm vir}$} & \colhead{$v_{\rm esc}$} & \colhead{$M_{F814W}$} & \colhead{SFR} & \colhead{$\log {\rm sSFR}$} & \colhead{[O/H]\tablenotemark{a}} & \colhead{$\Phi$} & \colhead{$i$\tablenotemark{b}} \\
\colhead{} &\colhead{(kpc)} &\colhead{(\kms)} &\colhead{[\msun]} &\colhead{[\msun]} &\colhead{(kpc)} &\colhead{(\kms)} &\colhead{(mag)} &\colhead{(\msun\ yr$^{-1}$)} &\colhead{[yr$^{-1}$]} &\colhead{} &\colhead{(deg)} &\colhead{(deg)}}
\startdata
\hline
\multicolumn{13}{c}{HE1003+0149, \hspace{0.5mm} $z_{\rm abs}$= 0.418522, \hspace{0.5mm} \logNHI=$16.89 \pm 0.04$, \hspace{0.5mm} [X/H]=$-1.08 \pm 0.06$} \\
\hline
997 & 126 & $211 \pm 1$ & 8.7 & 11.1 & 91 & 140 & $-$17.5 & $0.09 \pm 0.01$ & $-$9.76 & $-0.31^{+0.19}_{-0.20}$ & $82.4 \pm 2.6$ & 73 \\
\hline
\multicolumn{13}{c}{J1435+3604, \hspace{0.5mm} $z_{\rm abs}$=0.387594, \hspace{0.5mm} \logNHI=$16.18 \pm 0.06$, \hspace{0.5mm} [X/H]=$-1.12 \pm 0.11$} \\
\hline
1063 & 197 & $109 \pm 22$ & 9.3 & 11.4 & 115 & 175 & $-$17.9 & $0.21 \pm 0.02$ & $-$10.02 & \nodata & $5.2 \pm 2.0$ & 69 \\
\hline
\multicolumn{13}{c}{J1419+4207, \hspace{0.5mm} $z_{\rm abs}$=0.425592,  \hspace{0.5mm} \logNHI=$16.17 \pm 0.05$, \hspace{0.5mm} [X/H]=$-1.38 \pm 0.20$} \\
\hline
1066 & 78 & $664 \pm 19$ & 9.6 & 11.5 & 124 & 196 & $-$19.4 & $0.69 \pm 0.04$ & $-$9.75 & \nodata & $42.2 \pm 0.5$ & 61 \\
\hline
\multicolumn{13}{c}{HE1003+0149, \hspace{0.5mm} $z_{\rm abs}$=0.836989, \hspace{0.5mm} \logNHI=$16.52 \pm 0.02$, \hspace{0.5mm} [X/H]=$-1.51 \pm 0.09$} \\
\hline
1206 & 279 & $70 \pm 8$ & 10.1 & 11.8 & 131 & 271 & $-$20.7 & $12.67 \pm 0.37$ & $-$8.98 & $0.08 \pm 0.18$ & \nodata & \nodata \\
1229\tablenotemark{c} & 189 & $103 \pm 2$ & 9.3 & 11.4 & 97 & 199 & $-$17.6 & $0.20 \pm 0.03$ & $-$10.04 & $-0.06^{+0.19}_{-0.28}$ & $66.23 \pm 0.01$ & 52 \\
\hline
\multicolumn{13}{c}{HE1003+0149, \hspace{0.5mm} $z_{\rm abs}$=0.839400, \hspace{0.5mm} \logNHI=$16.13 \pm 0.01$, \hspace{0.5mm} [X/H] $<-1.74$} \\
\hline
1001\tablenotemark{d} & 157 & $564 \pm 2$ & 10.4 & 12.0 & 155 & 321 & $-$18.8 & $1.03 \pm 0.03$ & $-$10.39 & $-0.16 \pm 0.19$ & $56.9 \pm 4.1$ & 59 \\
1206 & 279 & $463 \pm 8$ & 10.1 & 11.8 & 131 & 271 & $-$20.7 & $12.67 \pm 0.37$ & $-$8.98 & $0.08 \pm 0.18$ & \nodata & \nodata \\
1229\tablenotemark{c} & 189 & $495 \pm 2$ & 9.3 & 11.4 & 97 & 199 & $-$17.6 & $0.20 \pm 0.03$ & $-$10.04 & $-0.06^{+0.19}_{-0.28}$ & $66.23 \pm 0.01$ & 52 \\
\hline
\multicolumn{13}{c}{J1435+3604, \hspace{0.5mm} $z_{\rm abs}$=0.372981, \hspace{0.5mm} \logNHI=$16.68 \pm 0.05$, \hspace{0.5mm} [X/H]=$-1.98^{+0.17}_{-0.15}$} \\
\hline
955 & 194 & $48 \pm 44$ & 9.1 & 11.3 & 104 & 157 & $-$17.7 & $0.16 \pm 0.02$ & $-$9.94 & \nodata & \nodata & 35 \\
1042 & 172 & $273 \pm 4$ & 9.1 & 11.2 & 102 & 154 & $-$19.2 & $1.69 \pm 0.04$ & $-$8.87 & \nodata & $80.8 \pm 0.8$ & 72 \\
\hline
\multicolumn{13}{c}{HE1003+0149, \hspace{0.5mm} $z_{\rm abs}$=0.837390, \hspace{0.5mm} \logNHI=$16.36 \pm 0.02$, \hspace{0.5mm} [X/H]=$-2.19^{+0.15}_{-0.18}$} \\
\hline
1206 & 279 & $135 \pm 8$ & 10.1 & 11.8 & 131 & 271 & $-$20.7 & $12.67 \pm 0.37$ & $-$8.98 & $0.08 \pm 0.18$ & \nodata & \nodata \\
1229\tablenotemark{c} & 189 & $168 \pm 2$ & 9.3 & 11.4 & 97 & 199 & $-$17.6 & $0.20 \pm 0.03$ & $-$10.04 & $-0.06^{+0.19}_{-0.28}$ & $66.23 \pm 0.01$ & 52 \\
\hline
\multicolumn{13}{c}{PKS0552$-$640, \hspace{0.5mm} $z_{\rm abs}$=0.345149, \hspace{0.5mm} \logNHI=$17.02 \pm 0.03$, \hspace{0.5mm} [X/H]=$-2.83^{+0.50}_{-0.56}$} \\
\hline
6085 & 142 & $394 \pm 1$ & 9.4 & 11.4 & 115 & 171 & $-$19.1 & $0.09 \pm 0.01$ & $-$10.45 & $0.09 \pm 0.18$ & $69.5 \pm 2.3$ & 47 \\
\hline
\multicolumn{13}{c}{PG1522+101, \hspace{0.5mm} $z_{\rm abs}$=0.728885, \hspace{0.5mm} \logNHI=$16.63 \pm 0.05$, \hspace{0.5mm} [X/H]=$-2.92 \pm 0.05$} \\
\hline
934 & 176 & $23 \pm 1$ & 9.7 & 11.6 & 115 & 220 & $-$20.5 & $7.02 \pm 0.05$ & $-$8.82 & $-0.16 \pm 0.18$ & $17.2 \pm 1.7$ & 61 \\
\enddata
\tablecomments{The above absorber column density and metallicity values are taken from the COS CGM Compendium \citep{lehner2018,lehner2019}. The error bars on the median absorber and galaxy metallicity represent the 68\% confidence interval. The candidate galaxies of the HE1003+0149 absorbers at $z_{\rm abs}$=0.836989, $z_{\rm abs}$=0.837390, and $z_{\rm abs}$=0.839400 are the same.}\tablenotetext{a}{The galaxy metallicity reported here is [O/H] = $\epsilon$(O) $-$ 8.69.}\tablenotetext{b}{The inclination values have errors around 5 to 10 degrees.}\tablenotetext{c}{The stellar mass of this galaxy was calculated using synthetic magnitudes.}\tablenotetext{d}{This mass estimate is in conflict with the value reported in \citet{narayanan2021} and is likely due to the differences in the assumed star formation history. We independently ran {\tt kcorrect} and {\tt eazy-py} (\citealt{brammer2008}, \url{https://github.com/gbrammer/eazy-py}) with photometry from the Hyper Suprime-Cam Subaru Strategic Program Data Release 3 \citep{aihara2022} and obtained a mass estimate of \logMstar\ = 9.8 and \logMstar\ = 9.9, respectively. Our measurement is consistent within the expected errors using this method.}
\end{deluxetable*}

\clearpage

\startlongtable
\begin{deluxetable*}{lcccc}
\tabletypesize{\small}
\tablecaption{Galaxy Location \label{tab:galinfoapp}}
\tablehead{\colhead{QSO} & \colhead{SE\#} & \colhead{RA} & \colhead{DEC} & \colhead{$z_{\rm gal}$}}
\startdata
HE0153-4520 & 966 & 01:55:12.88 & $-$45:06:34.94 & 0.225549 \\
PKS0405-123 & 3207 & 04:07:48.31 & $-$12:11:02.35 & 0.16696 \\
PKS0405-123 & 1822 & 04:07:51.15 & $-$12:11:37.46 & 0.16696 \\
HE0439-5254 & 1309 & 04:40:10.52 & $-$52:48:10.81 & 0.614700 \\
HE0439-5254 & 1083 & 04:40:12.74 & $-$52:48:20.26 & 0.614818 \\
HE0439-5254 & 1009 & 04:40:13.96 & $-$52:48:10.15 & 0.61508 \\
PKS0552-640 & 6085 & 05:52:20.30 & $-$64:02:05.19 & 0.343382 \\
PKS0552-640 & 6264 & 05:52:22.96 & $-$64:02:24.04 & 0.34391 \\
PKS0552-640 & 5939 & 05:52:23.14 & $-$64:02:36.68 & 0.34376 \\
PKS0552-640 & 78334 & 05:52:26.66 & $-$64:02:17.29 & 0.34481 \\
HE1003+0149 & 974 & 10:05:34.01 & +01:34:57.24 & 0.418784 \\
HE1003+0149 & 997 & 10:05:34.12 & +01:35:00.11 & 0.417522 \\
HE1003+0149 & 1001 & 10:05:34.19 & +01:34:57.91 & 0.83594 \\
HE1003+0149 & 1229 & 10:05:36.76 & +01:34:53.65 & 0.83636 \\
HE1003+0149 & 1206 & 10:05:37.26 & +01:34:26.65 & 0.83656 \\
PG1338+416 & 1118 & 13:41:01.39 & +41:22:56.66 & 0.34926 \\
J1419+4207 & 1003 & 14:19:10.96 & +42:07:55.67 & 0.2894 \\
J1419+4207 & 1066 & 14:19:11.41 & +42:07:47.82 & 0.42875 \\
J1419+4207 & 1163 & 14:19:12.21 & +42:07:36.77 & 0.2896 \\
J1435+3604 & 1060 & 14:35:13.01 & +36:04:25.67 & 0.3875 \\
J1435+3604 & 1042 & 14:35:14.01 & +36:04:25.23 & 0.37423 \\
J1435+3604 & 1063 & 14:35:14.16 & +36:04:19.99 & 0.3881 \\
J1435+3604 & 955 & 14:35:14.54 & +36:04:37.93 & 0.3732 \\
PG1522+101 & 1151 & 15:24:22.90 & +09:58:15.57 & 0.52007 \\
PG1522+101 & 934 & 15:24:26.14 & +09:58:33.11 & 0.728752 \\
PG1522+101 & 1077 & 15:24:26.16 & +09:58:17.63 & 0.5187 \\
J1619+3342 & 1006 & 16:19:15.95 & +33:42:35.11 & 0.4718 \\
J1619+3342 & 946 & 16:19:16.03 & +33:42:48.72 & 0.47052 \\
J1619+3342 & 1028 & 16:19:16.10 & +33:42:32.04 & 0.4710 \\
J1619+3342 & 903 & 16:19:16.54 & +33:42:38.36 & 0.47160 \\
J1619+3342 & 1180 & 16:19:17.26 & +33:42:33.35 & 0.47220 \\
J1619+3342 & 1179 & 16:19:17.69 & +33:42:33.86 & 0.47140 \\
J1619+3342 & 1137 & 16:19:18.38 & +33:42:35.78 & 0.4704
\enddata
\tablecomments{The candidate galaxies of the HE1003+0149 absorbers at $z_{\rm abs}$=0.836989, $z_{\rm abs}$=0.837390, and $z_{\rm abs}$=0.839400 are the same, so we only include them once in this table. The redshift significance comes from the {\tt REDROCK} errors. These values are reported as errors on $|\Delta v|$ in Tables~\ref{tab:galinfo} and \ref{tab:galinfo_other}.}
\end{deluxetable*}


\begin{figure}
    \epsscale{1.17}
    \plotone{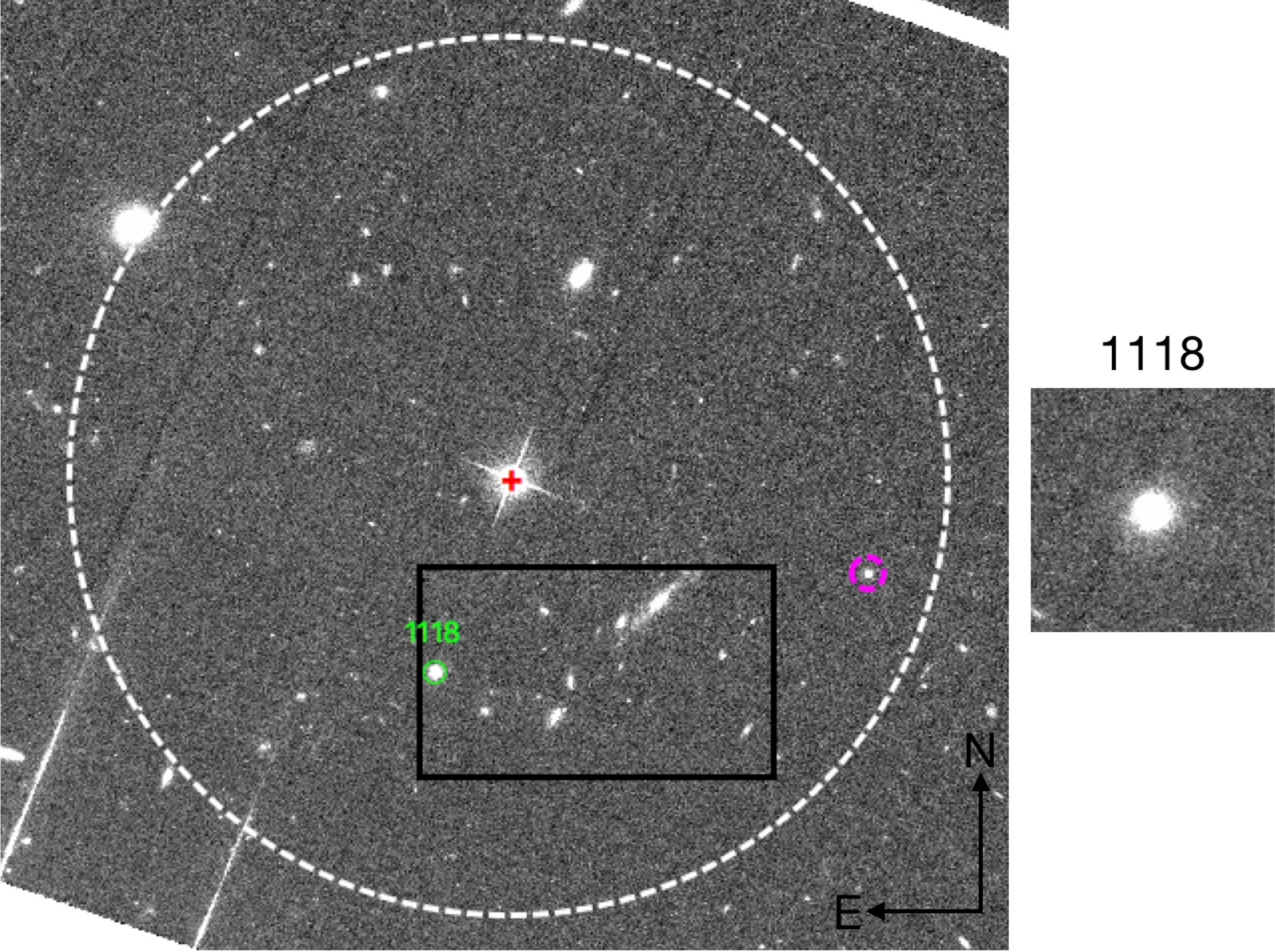}
    \caption{\hst/ACS image with outline of Keck/KCWI pointing for the field of PG1338+416. North is up, and East is to the left. The QSO is marked with the red cross, stars are indicated with magenta circles, and the white dashed circle has a radius of 200 kpc at the redshift of the absorber. The black box designates the KCWI FOV. The candidate galaxy of the absorber is marked by the green circle and labelled with the SE\#.} \label{fig:PG1338}
\end{figure}

\begin{figure}
    \epsscale{1.17}
    \plotone{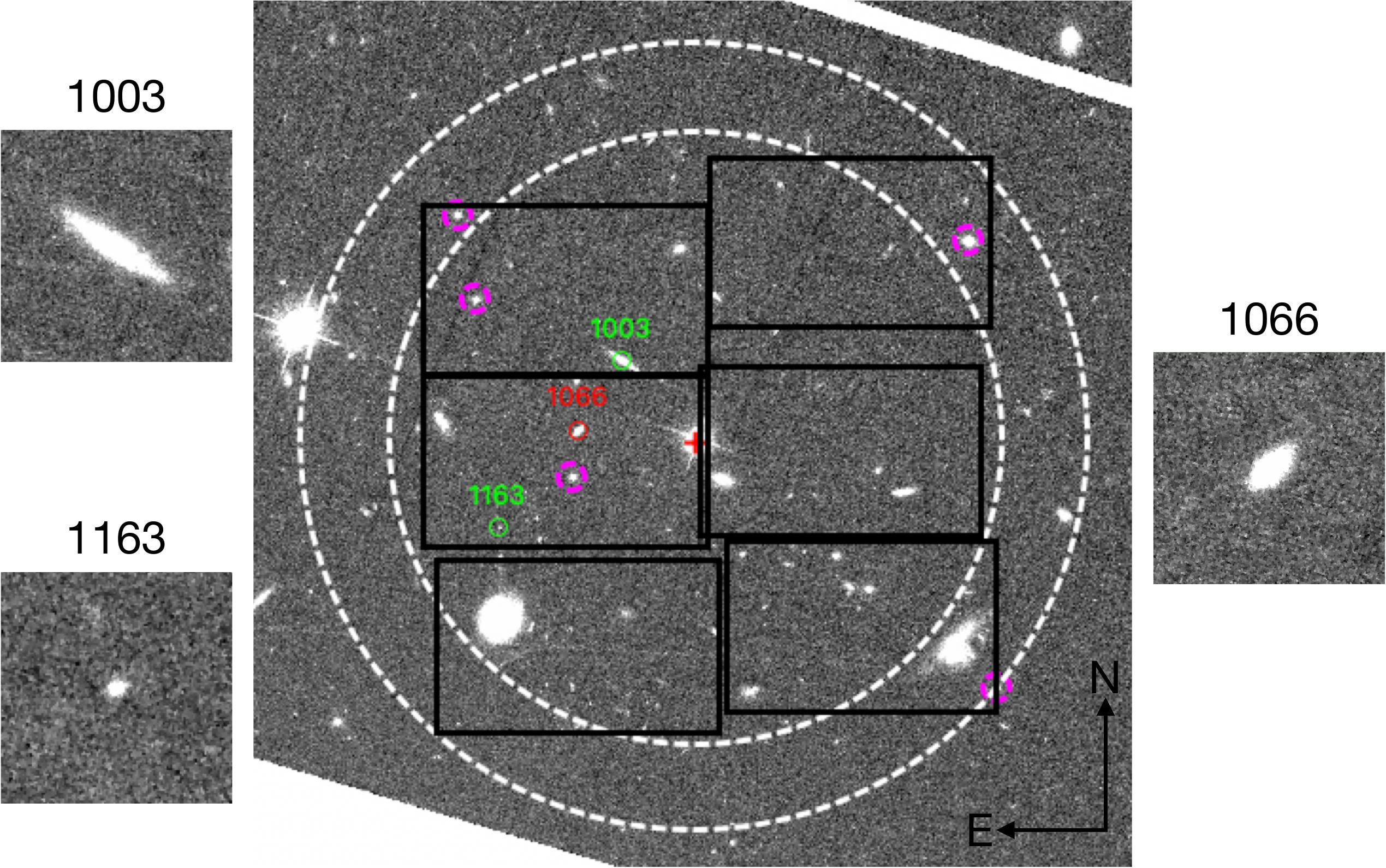}
    \caption{Same as Figure~\ref{fig:PG1338}, but for the field of J1419+4207. The candidate galaxies of the lowest redshift absorber are marked by the green circles and labelled with their SE\#. The candidate galaxy of the higher redshift absorber is marked by the red circle and labelled with the SE\#.} \label{fig:J1419}
\end{figure}

\begin{figure}
    \epsscale{1.17}
    \plotone{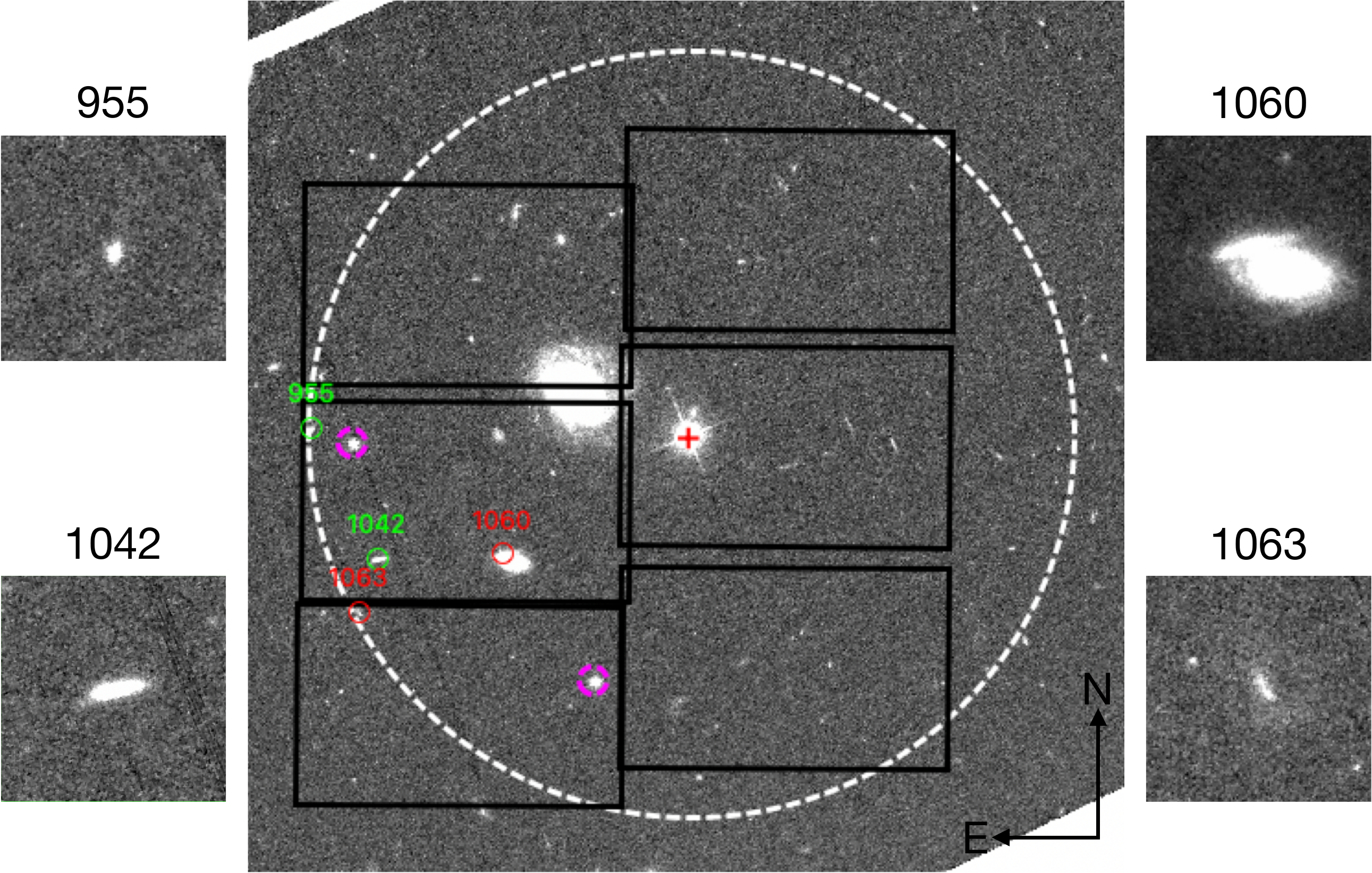}
    \caption{Same as Figure~\ref{fig:PG1338}, but for the field of J1435+3604. The white dashed circle has a radius of 200 kpc at \z=0.38. The candidate galaxies of the lowest redshift absorber are marked by the green circles and labelled with their SE\#. The candidate galaxies of the higher redshift absorber are marked by the red circles and labelled with their SE\#. Galaxy 1060 is blended with a lower redshift galaxy.} \label{fig:J1435}
\end{figure}

\begin{figure}
    \epsscale{1.17}
    \plotone{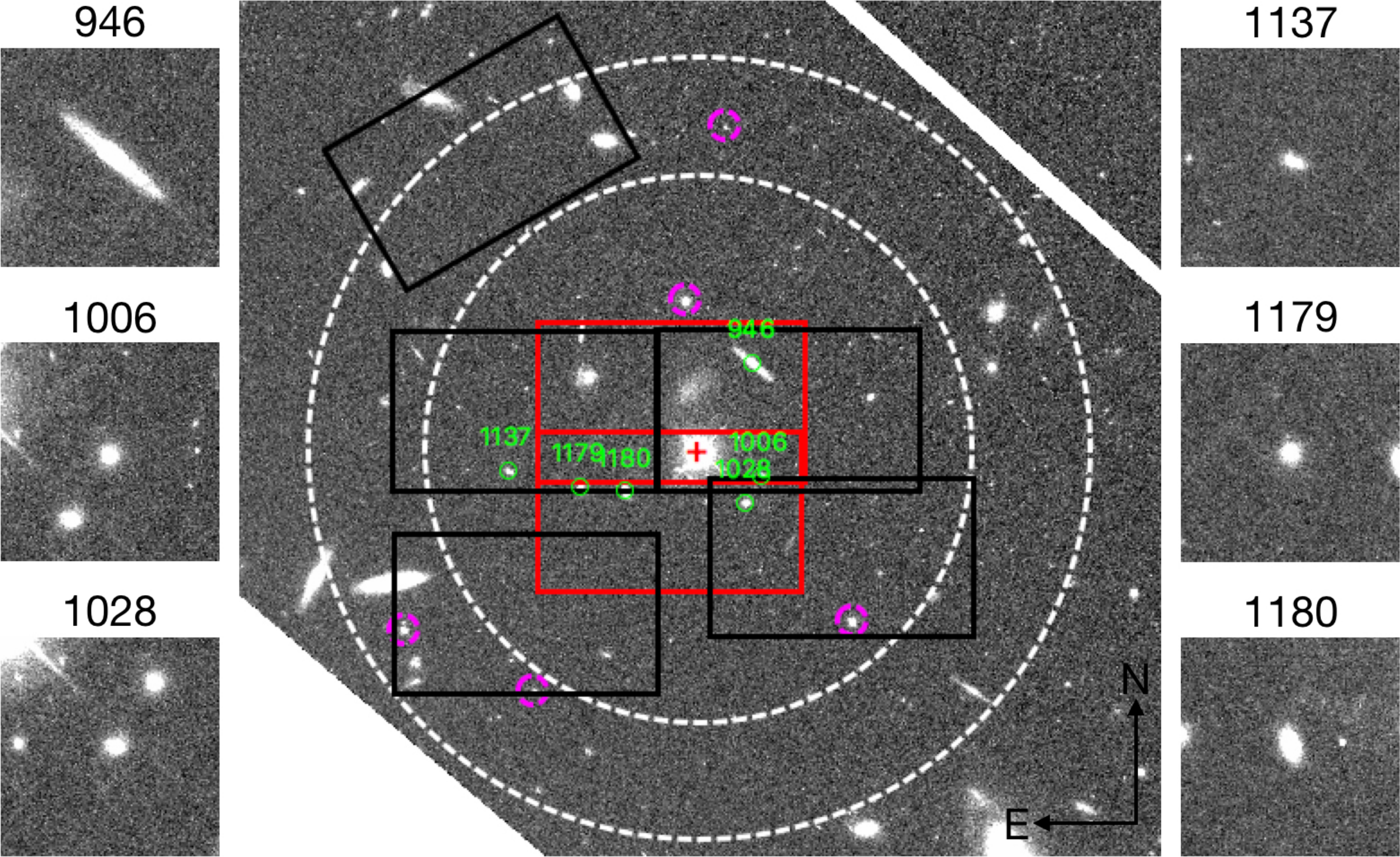}
    \caption{Same as Figure~\ref{fig:PG1338}, but for the field of J1619+3342. The candidate galaxies of the high redshift absorber are marked by the green circles and labelled with their SE\#. The high redshift absorber is a proximate absorber located in the QSO galaxy group. No candidate galaxies were found for the lower redshift absorber. The black pointings were observed for 900 seconds. The red pointings were observed for 1200\,seconds.} \label{fig:J1619}
\end{figure}


\begin{figure}
    \epsscale{1.17}
    \plotone{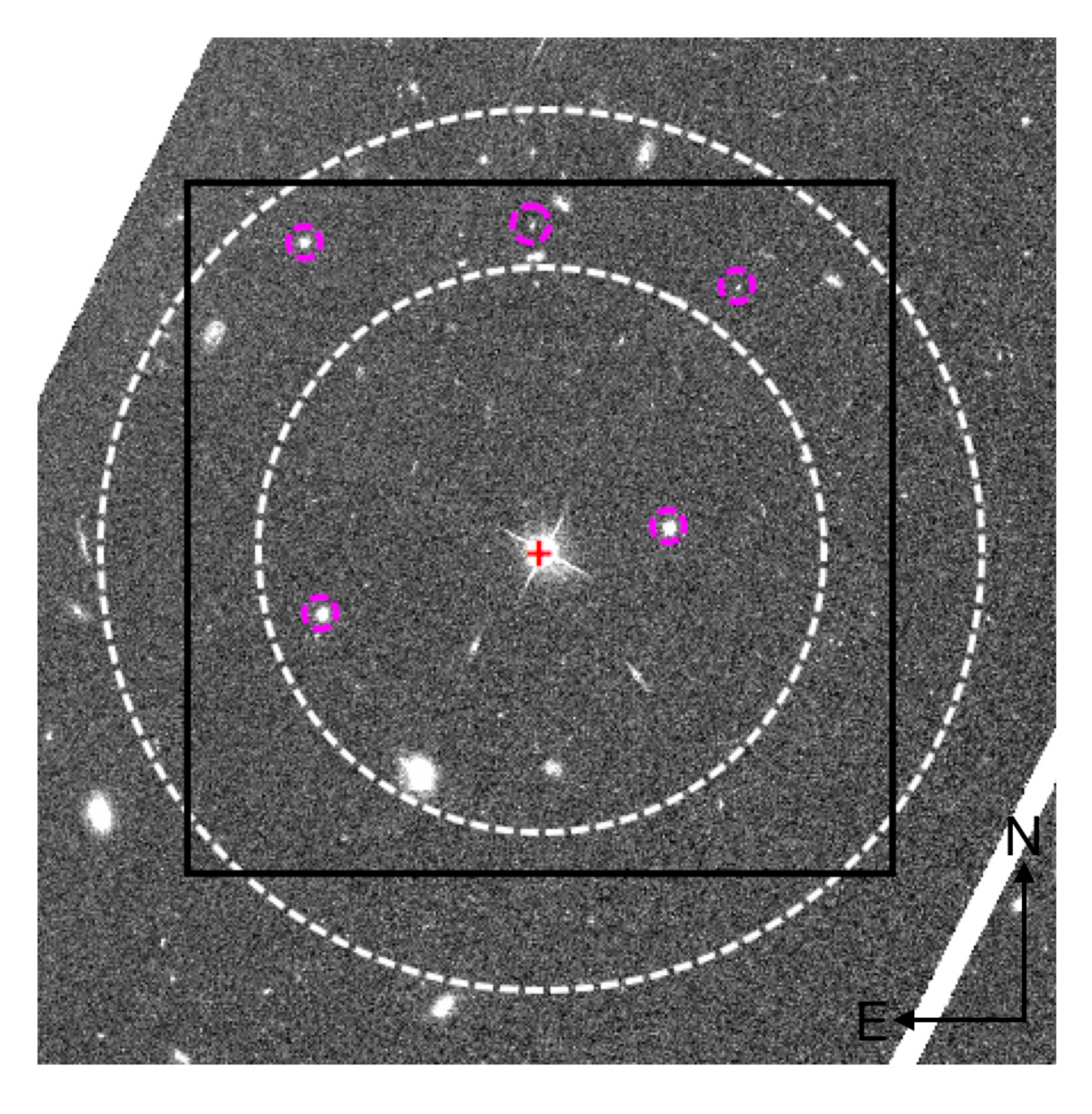}
    \caption{\hst/ACS image with outline of VLT/MUSE pointing for the field of PHL1377. North is up, and East is to the left. The QSO is marked with the red cross, stars are indicated with magenta circles, and the white dashed circles have a radius of 200 kpc at the redshifts of the absorbers. The black box designates the MUSE FOV. No candidate galaxies were found for either absorber.} \label{fig:PHL1377}
\end{figure}

\begin{figure}
    \epsscale{1.17}
    \plotone{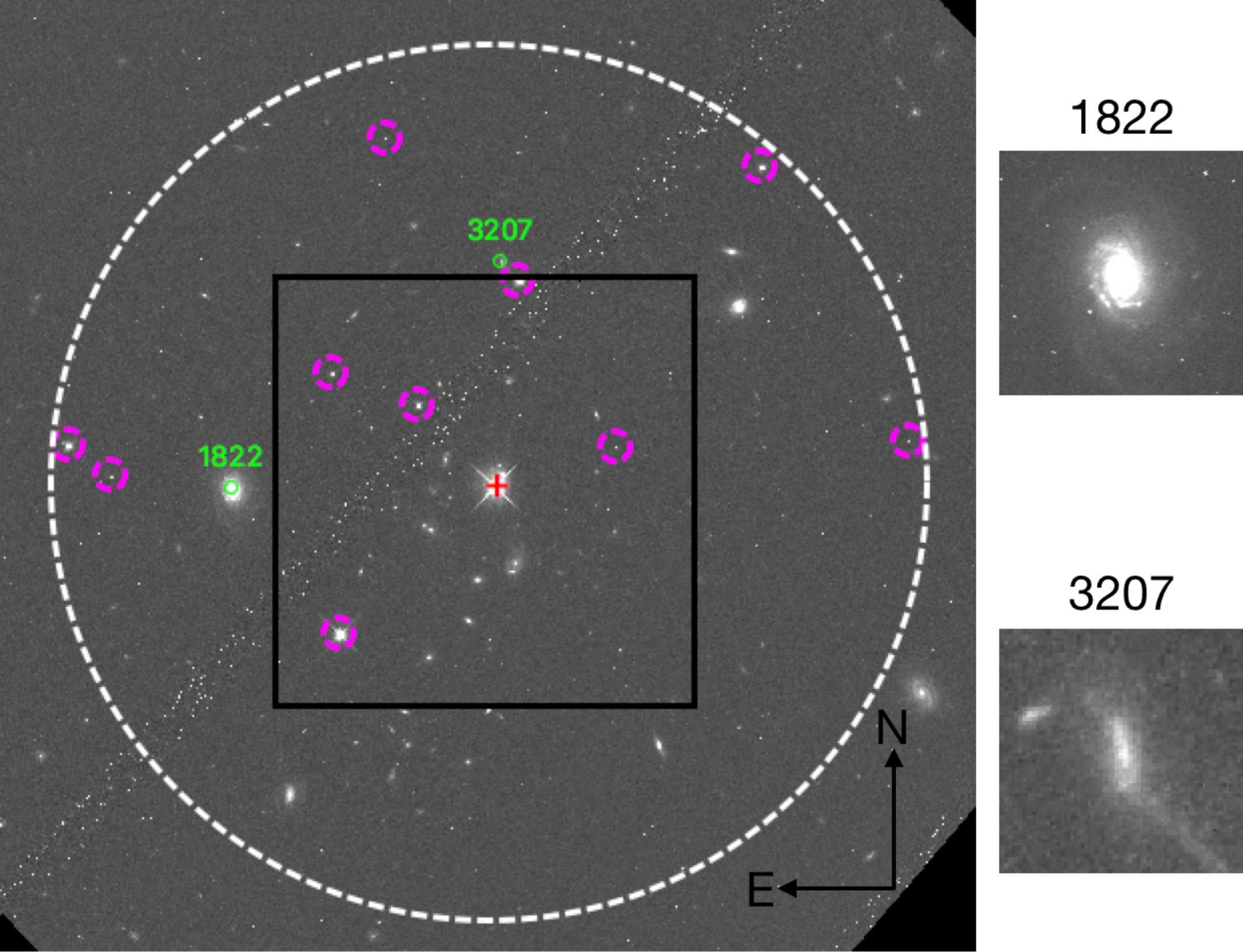}
    \caption{Same as Figure~\ref{fig:PHL1377}, but for the field of PKS0405$-$123. No candidate galaxies were found in the MUSE FOV, but the candidate galaxies for this absorber have been previously identified in \citet{spinrad1993}, \citet{prochaska2006}, \citet{chen2009}, \citet{savage2010}, and \citet{johnson2013}.} \label{fig:PKS0405}
\end{figure}

\begin{figure}
    \epsscale{1.17}
    \plotone{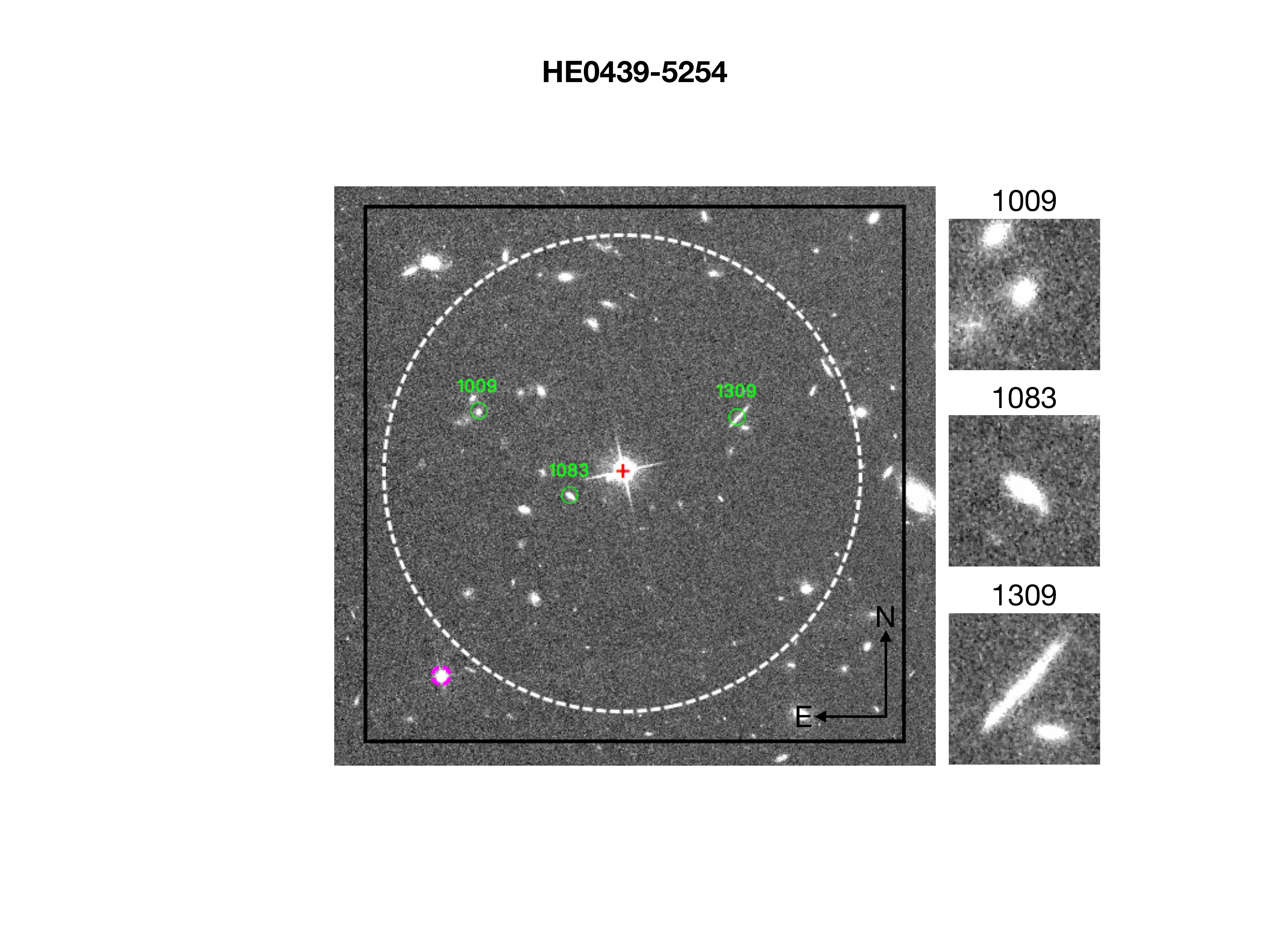}
    \caption{Same as Figure~\ref{fig:PHL1377}, but for the field of HE0439$-$5254. The candidate galaxies of the absorber are marked by the green circles and labelled with their SE\#.} \label{fig:HE0439}
\end{figure}

\begin{figure}
    \epsscale{1.17}
    \plotone{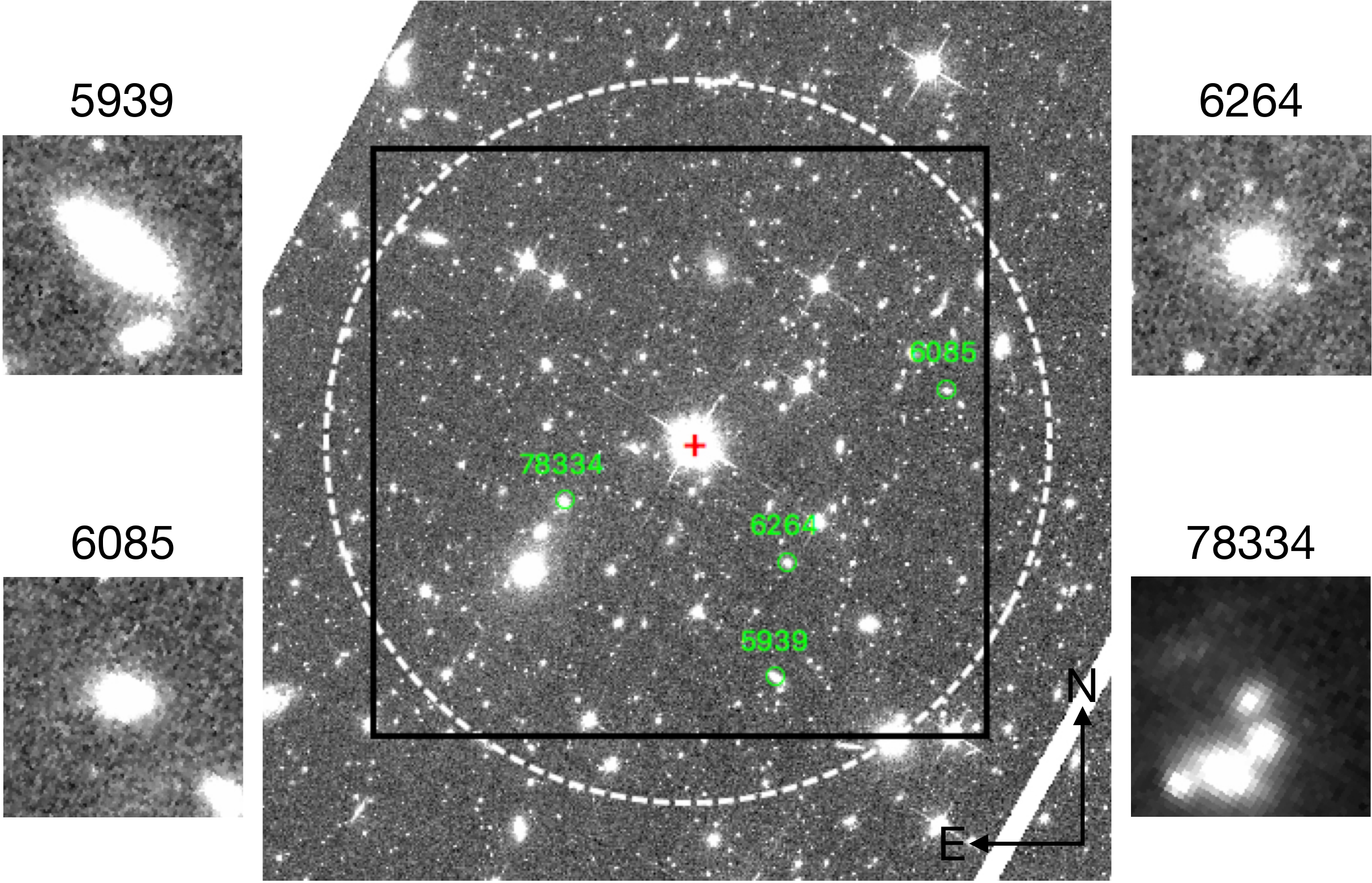}
    \caption{Same as Figure~\ref{fig:PHL1377}, but for the field of PKS0552$-$640. Most of the objects in this field are stars, so we do not mark them individually. The candidate galaxies of the absorber are marked by the green circles and labelled with their SE\#.} \label{fig:PKS0552}
\end{figure}

\begin{figure}
    \epsscale{1.17}
    \plotone{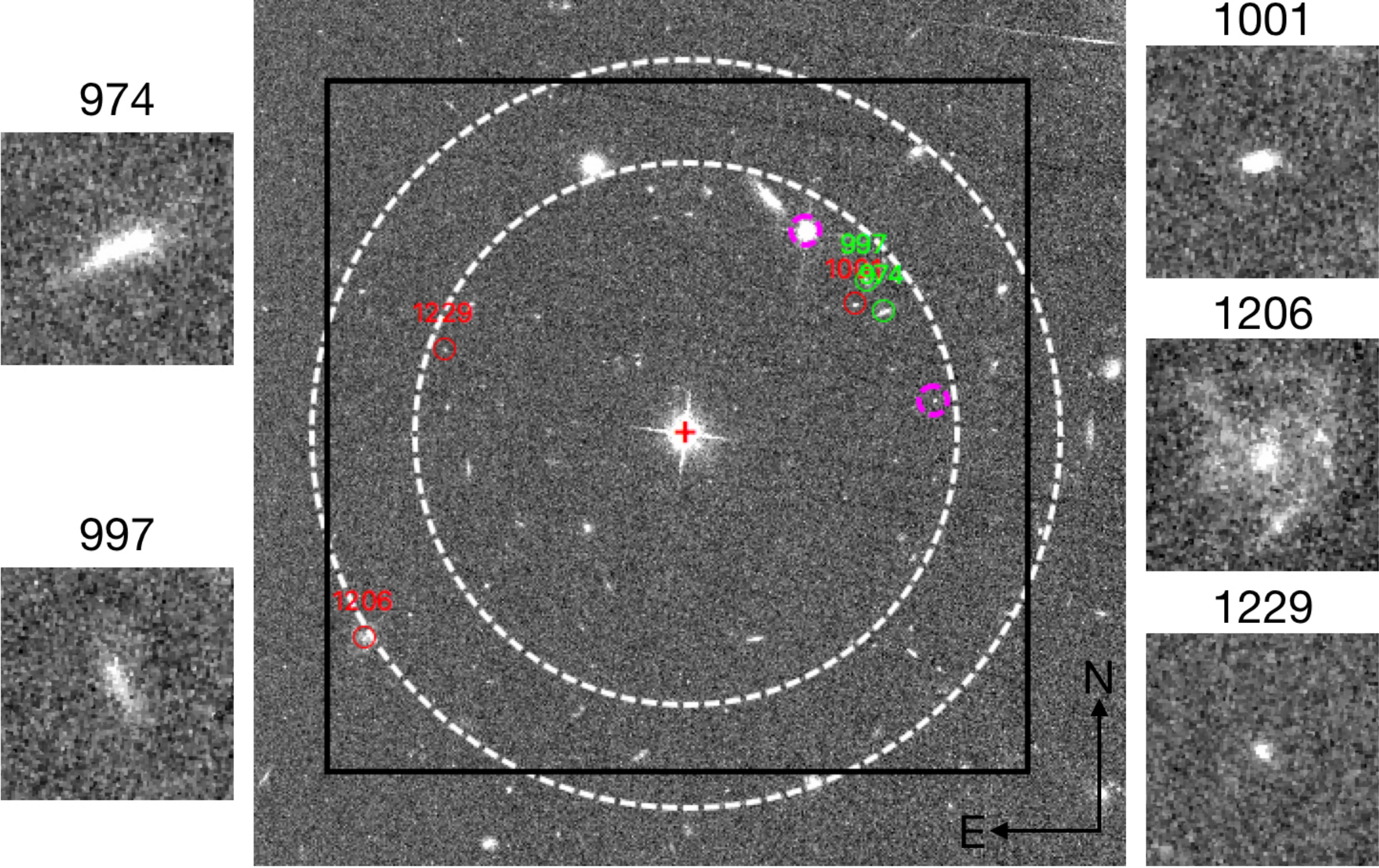}
    \caption{Same as Figure~\ref{fig:PHL1377}, but for the field of HE1003+0149. The white dashed circles have a radius of 200 kpc at the redshift of the low redshift absorber and at \z=0.837 to represent the three high redshift absorbers. The candidate galaxies of the low redshift absorber are marked by the green circles and labelled with their SE\#. The candidate galaxies of the three high redshift absorbers are marked by the red circles and labelled with their SE\#.} \label{fig:HE1003}
\end{figure}

\begin{figure}
    \epsscale{1.17}
    \plotone{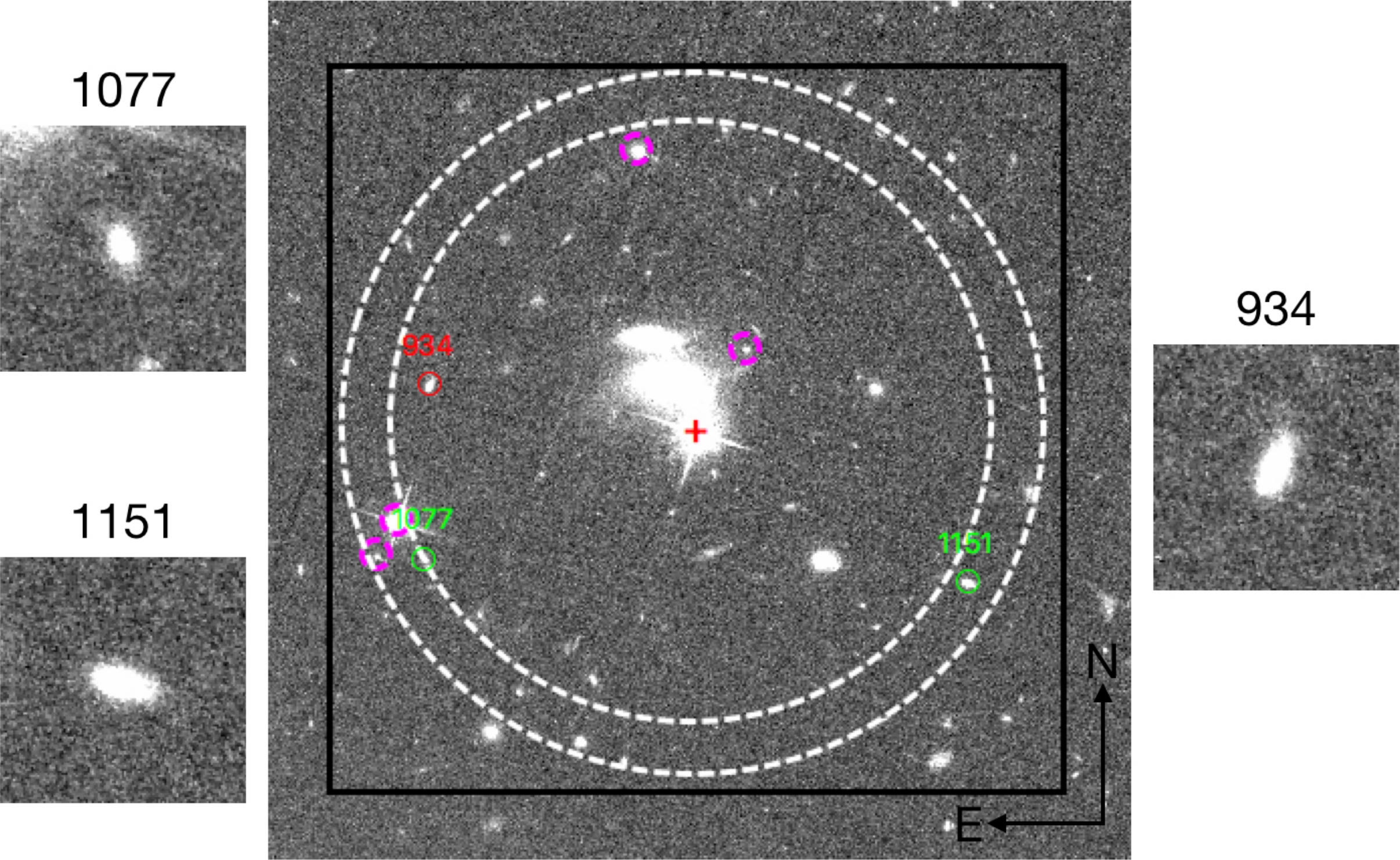}
    \caption{Same as Figure~\ref{fig:PHL1377}, but for the field of PG1522+101. The candidate galaxies of the low redshift absorber are marked by the green circles and labelled with their SE\#. The candidate galaxy of the higher redshift absorber is marked by the red circle and labelled with the SE\#.} \label{fig:PG1522}
\end{figure}

\clearpage

\section{Galaxy Characteristics for the Entire Candidate Sample}\label{sec:appb}

Here we duplicate the galaxy property plots in the main paper and include the entire candidate galaxy sample in Figure~\ref{fig:galpropsall}.

\begin{figure*}
    \epsscale{1.2}
    \plotone{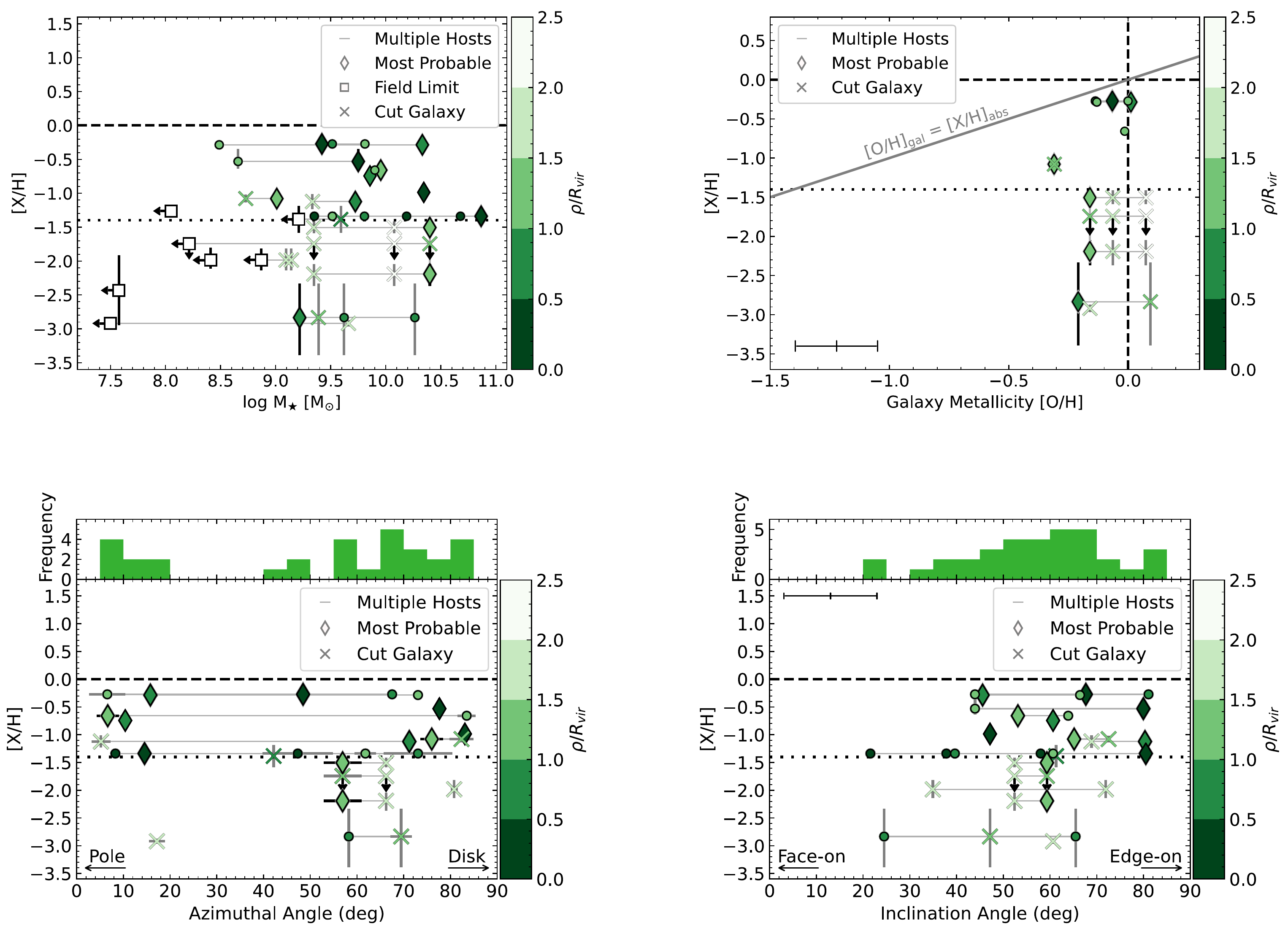}
    \caption{Absorber metallicity versus galaxy stellar mass (top left; compare to Figure~\ref{fig:metstmass}), absorber metallicity versus galaxy metallicity (top right; compare to Figure~\ref{fig:metgalmet}), absorber metallicity versus galaxy azimuthal angle (bottom left; compare to Figure~\ref{fig:metaz}), and absorber metallicity versus galaxy inclination angle (bottom right; compare to Figure~\ref{fig:metincl}) for the full candidate galaxy sample. See the main text for a description of the figures.} \label{fig:galpropsall}
\end{figure*}

\listofchanges

\end{document}